\documentclass[aps,prb,groupedaddress,showpacs
]{revtex4} 
\usepackage{graphicx,graphics,amsfonts,amssymb,color,bm,verbatim}

\def\be{\begin{equation}} 
\def\ee{\end{equation}} 
\def\ba{\begin{eqnarray}} 
\def\ea{\end{eqnarray}} 
\def\bc{\begin{center}} 
\def\ec{\end{center}} 
\def\sgn{{\rm sgn}} 

\def\p{\partial}

\begin{document} 

\title{Quantum theory of the third-order nonlinear electrodynamic effects in graphene} 

\author{S. A. Mikhailov} 
\email[Electronic mail: ]{sergey.mikhailov@physik.uni-augsburg.de} 

\affiliation{Institute of Physics, University of Augsburg, D-86135 Augsburg, Germany} 

\date{\today} 

\begin{abstract}
The linear energy dispersion of graphene electrons leads to a strongly nonlinear electromagnetic response of this material. We develop a general quantum theory of the third-order nonlinear local dynamic conductivity of graphene $\sigma_{\alpha\beta\gamma\delta}(\omega_1,\omega_2,\omega_3)$, which describes its nonlinear response to a uniform electromagnetic field. The derived analytical formulas describe a large number of different nonlinear phenomena such as the third harmonic generation, the four wave mixing, the saturable absorption, the second harmonic generation stimulated by a dc electric current, etc., which may be used in different terahertz and optoelectronic devices. \end{abstract} 


 
\maketitle 

\tableofcontents

\section{Introduction}

Electrodynamic properties of graphene have attracted considerable attention in recent years \cite{Bonaccorso10,Hartmann14,Glazov14,Roadmap15}. Microwave and optical response, plasma and cyclotron resonances, nonlinear electromagnetic properties have been investigated both theoretically and experimentally in many research groups in the world \cite{Henriksen08,Gusynin07b,Falkovsky07a,Mikhailov07d,Wunsch06,Hwang07,Bostwick07,Ryzhii07,Kuzmenko08,Nair08,Peres08,Li08,Mak08}. Special interest to these topics is generated by the opportunity to use many of these fundamental phenomena for microwave-, terahertz- and optoelectronic applications \cite{Sun10,Hartmann14,Roadmap15}. 

The most distinctive feature of graphene is the linear energy dispersion of its quasi-particles, electrons and holes, $ E_\pm({\bm p})=\pm v_F|{\bm p}|$. It was theoretically predicted \cite{Mikhailov07e} and then experimentally confirmed\cite{Dragoman10,Hendry10}, that due to this property graphene should demonstrate strongly nonlinear electromagnetic behavior. Physically this can be understood as follows \cite{Mikhailov07e}. Assume that a particle with the linear spectrum is placed in the uniform external electric field ${\bm E}(t)={\bm E}_0\cos\omega t$. According to Newton equations of motion its momentum will oscillate as ${\bm p}(t)\propto \sin\omega t$, while the velocity ${\bm v}(t)= \p  E/\p {\bm p}\propto{\bm p}(t)/|{\bm p}(t)|$, as well as the current ${\bm j}(t)$, which are \textit{not} proportional to the momentum, will be strongly nonlinear functions of ${\bm p}(t)$ and will therefore contain higher frequency harmonics,
\be 
 j(t)\propto v(t)\propto \sgn(\sin\omega t)\propto \sin\omega t+\frac 13 \sin 3\omega t +\dots.
\ee
In other words, since graphene electrons have no mass and can move in any direction only with the velocity $v_F\approx 10^8$ cm/s, at the return points they have to instantaneously change their velocity from $+10^8$ cm/s to $-10^8$ cm/s which is accompanied by a strong acceleration and leads to emission of electromagnetic radiation. 
The nonlinear electrodynamic phenomena in graphene have been further studied in Refs.  \cite{Mikhailov08a,Mikhailov08b,Mikhailov09a,Mikhailov11b,Mikhailov11c,Mikhailov12b,Mikhailov14c,Smirnova14,Yao14,Peres14,Cheng14a,Cheng14b,Cheng15,Cheng15b} (theory) and in Refs. \cite{Dean09,Dean10,Wu11,Gu12,Zhang12,Bykov12,Kumar13,Hong13,An14,Lin14,Popa10,Popa11} (experiment). 

In the first theoretical publications the nonlinear (third-order) electromagnetic response of graphene was studied in the quasiclassical approximation, using the Boltzmann kinetic approach. Such a theory takes into account only the {\em intra}-band oscillations of the graphene electrons, which is valid at low (microwave/terahertz) frequencies $\hbar\omega\lesssim 2E_F$, where $\omega$ is the radiation frequency, $E_F\equiv |\mu|$ is the Fermi energy and $\mu$ is the chemical potential. At higher (infrared, optical) frequencies the {\em inter}-band quantum transitions between the electron and hole bands should be taken into account and a quantum theory is needed. Recently, a quantum theory of the third-order nonlinear electrodynamic response of graphene to a \textit{uniform} external electric field has been proposed in several publications \cite{Cheng14a,Cheng14b,Cheng15,Mikhailov14c,Semnani15}. For example, in Ref. \cite{Cheng14a} the authors calculated the third-order conductivity $\sigma^{(3)}_{\alpha\beta\gamma\delta} (\omega_1,\omega_2,\omega_3)$ in the relaxation-free approximation neglecting all scattering effects. In Ref. \cite{Mikhailov14c} the special case of the third-harmonic generation $\omega_1=\omega_2=\omega_3=\omega$ has been studied, with the relaxation effects described by a single phenomenological relaxation time. In Ref. \cite{Cheng15} the authors generalized their results \cite{Cheng14a} by including the impurities and phonon scattering effects having introduced two phenomenological relaxation rates $\Gamma_i$ and $\Gamma_e$ corresponding to the intra- and inter-band electronic transitions. In addition, they took into account the finite temperature effects. In the paper \cite{Cheng15} the authors analytically solved the semiconductor Bloch equations for the optical graphene response with the electromagnetic interaction described in the length gauge ($\hat H'=-e{\bm r}\cdot {\bm E}$, Ref. \cite{Aversa95}). In Ref. \cite{Cheng15b} they confirmed their analytical results by numerical calculations. 

Although quite a number of  theories have been already proposed, their results are still contradictory, apparently due to extreme complexity of the problem. For example, the largest contribution to the third harmonic generation effect was predicted at $\hbar\omega\simeq 2E_F$ in Ref. \cite{Mikhailov14c} but at $\hbar\omega\simeq 2E_F/3$ in Refs. \cite{Cheng14a,Cheng15}. Analyzing the formulas of Ref. \cite{Cheng15} we have found a number of mistakes (for example, analytical results of Ref. \cite{Cheng15} did not reproduce the graphs of that paper) and contradictions with Ref. \cite{Cheng14a}. (After this manuscript has been submitted for publication, the authors of Ref. \cite{Cheng15} published Erratum, Ref. \cite{Cheng16err}, in which a misprint in the main formulas of Ref. \cite{Cheng15} was corrected.) The development of the theory of the third-order nonlinear electrodynamic phenomena in graphene cannot therefore by considered as completed and additional independent calculations are required.

In this paper we develop an analytical theory of the third-order nonlinear effects in graphene and calculate the local third-order conductivity $\sigma^{(3)}_{\alpha\beta\gamma\delta} (\omega_1,\omega_2,\omega_3)$. To solve the problem we apply a different from Refs. \cite{Cheng14a,Cheng15} approach, using the density matrix formalism and describing the \textit{uniform} electromagnetic field by a space-dependent scalar potential $\phi({\bf r},t)\propto \exp(i{\bf q\cdot r})$, where the wave-vector ${\bf q}$ is first assumed to be small but finite, with the limit ${\bf q\to 0}$ being taken in final formulas. This way we avoid formally divergent matrix elements of the perturbation with the electron Bloch wave-functions. We study a few specific physical effects, including the third-harmonic generation and the saturable absorption effects, and show that the output power emitted by graphene-based nonlinear devices resonantly depends on the input frequencies, with the resonance position governed by the electron density in graphene. This allows one to control the device operation by the gate voltage which makes the discussed effects very interesting for high-frequency electronic and optoelectronic applications. Our results also help to remove the discrepancies between the theories proposed so far.

The paper is organized as follows. In Section \ref{sec:spectrum} we introduce the tight-binding spectrum and wave functions of graphene electrons and calculate matrix elements of some operators which are needed in the rest of the paper. In Section \ref{sec:linear} the linear response of graphene \cite{Gusynin07b,Falkovsky07a,Mikhailov07d} is discussed. The results for the first-order conductivity $\sigma^{(1)}_{\alpha\beta} (\omega)$ are presented in Eqs. (\ref{sigma1}) -- (\ref{inter-cond}) and in Figure \ref{fig:sigma-1}. They are used then in Section \ref{Kerr}.

Section \ref{sec:nonlin} contains the main results of our work. Here, as well as  in Appendixes \ref{app:deriv-current} and \ref{app:conduct}, we give a detailed derivation of the third conductivity $\sigma^{(3)}_{\alpha\beta\gamma\delta} (\omega_1,\omega_2,\omega_3)$. A summary of results for this function can be found in Eqs. (\ref{sigma3resultSymm}) -- (\ref{FuncS03}). 
In Section \ref{sec:mono} we analyze our results in the case of an external monochromatic radiation with the frequency $\omega$ and consider two physical effects: the third harmonic generation and the saturable absorption (Kerr effect). Finally, in Section \ref{sec:conclusion} we summarize all obtained results and discuss the main conclusions from this work. Some technical details are collected in Appendixes.

\section{Electronic spectrum, wave functions and matrix elements\label{sec:spectrum}} 

\subsection{Tight binding energy and wave functions\label{sec:tbenergy}}

\begin{figure} 
\includegraphics[width=0.44\textwidth]{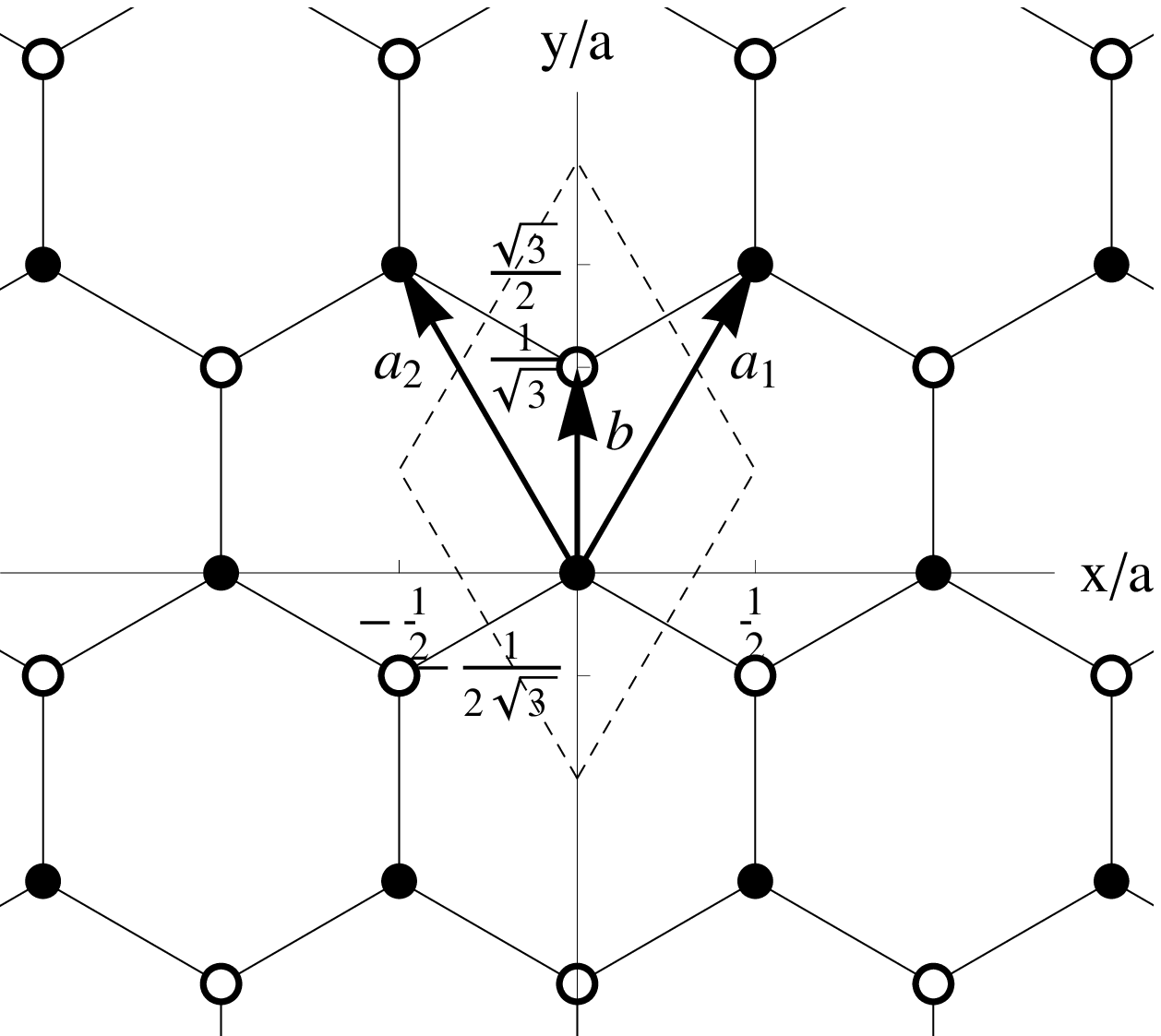} \hfill 
\includegraphics[width=0.5\textwidth]{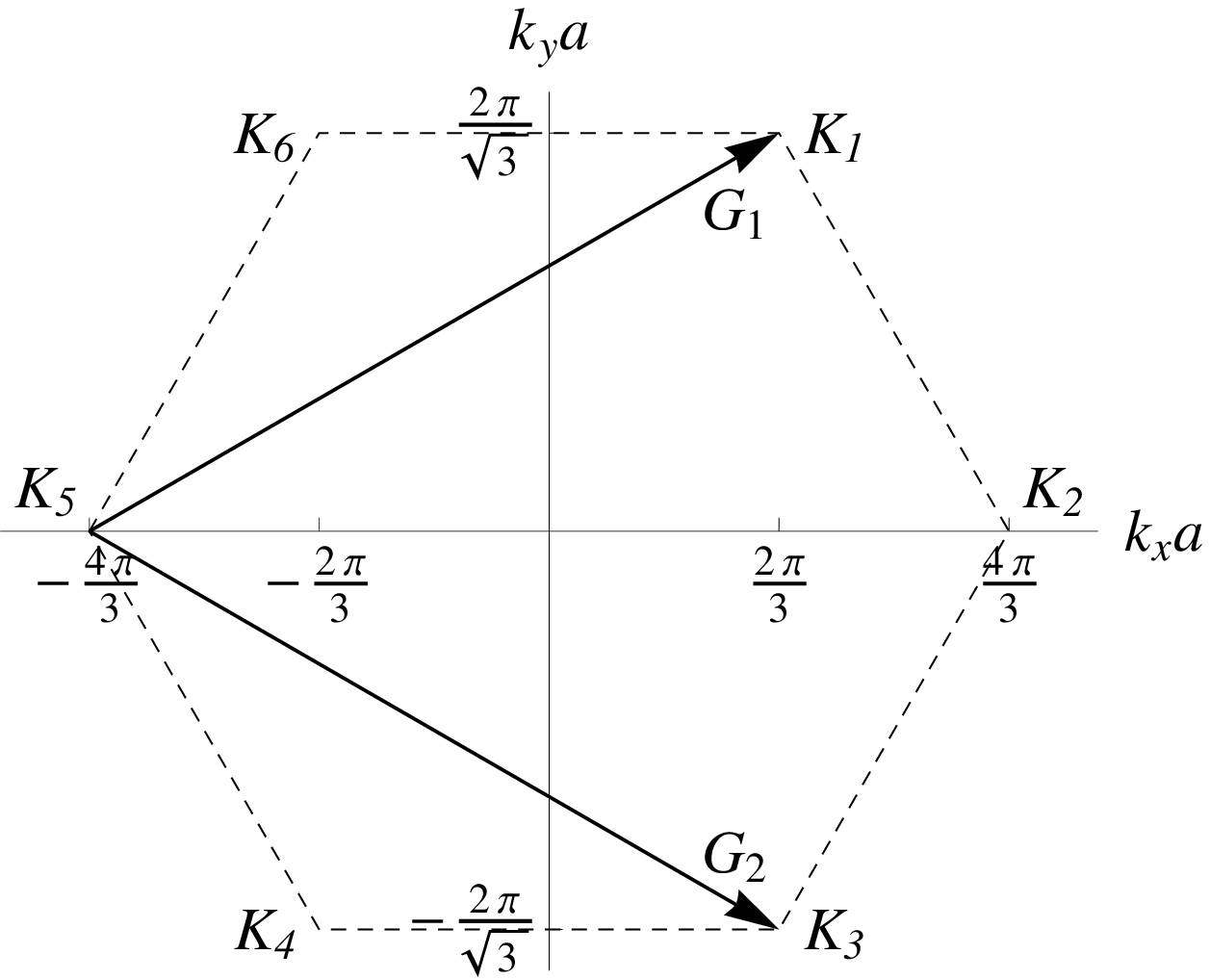} 
\caption{\label{lattice} Left: The honey-comb lattice of graphene; ${\bf a}_1=a(1/2,\sqrt{3}/2)$, ${\bf a}_2=a(-1/2,\sqrt{3}/2)$, ${\bf b}=a(0,1/\sqrt{3})$, $a$ is the lattice constant. Right: The Brillouin zone of graphene; ${\bf G}_1=(2\pi/a)(1,1/\sqrt{3})$, ${\bf G}_2=(2\pi/a)(1,-1/\sqrt{3})$, $|\textbf{G}_1|=|\textbf{G}_2|=G=4\pi/(\sqrt{3}a)$; ${\bf K}_1=-{\bf K}_4=(2\pi/a)(1/3,1/\sqrt{3})$, ${\bf K}_2=-{\bf K}_5=(2\pi/a)(2/3,0)$,
${\bf K}_3=-{\bf K}_6=(2\pi/a)(1/3,-1/\sqrt{3})$. 
Dashed lines show the boundaries of the elementary cell in the direct and reciprocal space. 
} 
\end{figure}

The carbon atoms in graphene occupy a 2D plane ($z=0$) and are arranged in a honey-comb lattice, Fig. \ref{lattice},left, composed out of two triangular sublattices shifted by a vector ${\bf b}$ with respect to each other. All points of the first sublattice (black circles in Fig. \ref{lattice}a) are given by the vectors $n_1{\bf a}_1+n_2{\bf a}_2$ and those of the second sublattice (open circles) by $n_1{\bf a}_1+n_2{\bf a}_2+{\bf b}$, where $n_1$ and $n_2$ are integers. The basis vectors ${\bf a}_1$, ${\bf a}_2$ and the vector ${\bf b}$ are chosen as shown in Fig. \ref{lattice}a, ${\bf a}_1=a(1/2,\sqrt{3}/2)$, ${\bf a}_2=a(-1/2,\sqrt{3}/2)$, ${\bf b}=a(0,1/\sqrt{3})$, where $a=|{\bf a}_1|=|{\bf a}_2|=2.46$  \AA \ is the lattice constant. 

The carbon atom has four electrons on its outer shell. Three of them, so called $\sigma$-electrons, form chemical bonds with their neighbors. The fourth ($\pi$-) electron moves in the periodic potential of the honey-comb lattice,
\be 
U({\bf r},z)=\sum_{\bf a}\left[U_a({\bf r-a},z)+U_a({\bf r-a-b},z)\right],
\label{potential}
\ee 
where $U_a$ is the atomic potential, ${\bf r}=(x,y)$ is a two-dimensional vector, and the sum is taken over all ${\bf a}$ vectors of the lattice. We will describe its motion within the tight-binding approximation. Within this approach the single-particle Hamiltonian is
\be 
\hat H_0=\frac{{\bf \hat p}^2}{2m}+\frac{\hat p_z^2}{2m}+U({\bf r},z),
\label{Hamilt0}
\ee 
where ${\bf \hat p}=-i\hbar (\p_x,\p_y)$ and $m$ is the free electron mass. Following the standard procedure of the tight binding approximation and assuming only the nearest-neighbors interaction we find the energy \cite{Wallace47}
\be
E_{l{\bf k}}=(-1)^lt |{\cal Z}_{\bf k}|, 
\label{energy}
\ee
and the wave functions of the Hamiltonian (\ref{Hamilt0})
\be 
\Psi_{l{\bf k}}({\bf r},z)=|l{\bf k}\rangle =\frac 1{\sqrt{S}}e^{i{\bf k\cdot r}}u_{l{\bf k}}({\bf r},z) ,\label{wavefunction} 
\ee 
where
\be 
u_{l{\bf k}}({\bf r},z)=|l{\bf k})= \sqrt{\frac A2}\sum_{\bf a} e^{-i{\bf k\cdot (r-a)}}
\left[(-1)^l\zeta_{\bf k}\psi_a 
({\bf r-a},z)+ \psi_a 
({\bf r-a-b},z) 
\right] 
\label{blochfactor} 
\ee 
is the Bloch factor, 
$l=1,2$ is the band index, ${\bf k}=(k_x,k_y)$, $t$ is the transfer integral (in graphene $t\approx 3$ eV), $S$ and $A$ are the areas of the sample and of the elementary cell, respectively, and $\psi_a$ is the atomic wave function. The functions ${\cal Z}_{\bf k}$ and $\zeta_{\bf k}$ in (\ref{energy}), (\ref{blochfactor}) are defined as 
\be 
{\cal Z}_{\bf k}=1 
+e^{i{\bf k}\cdot{\bf a}_1} 
+e^{i{\bf k}\cdot{\bf a}_2}=1+2\cos(k_xa/2)e^{i\sqrt{3}k_ya/2}, 
\label{calS}
\ee
\be 
\zeta_{\bf k}={\cal Z}_{\bf k}/|{\cal Z}_{\bf k}|=e^{i\Phi_{\bf k}},
\label{zeta}
\ee
where $\Phi_{\bf k}$ is the phase of the complex function ${\cal Z}_{\bf k}$. 
They satisfy the relations
\be
{\cal Z}_{\bf -k}={\cal Z}^\star_{\bf k}, \ \ {\cal Z}_{\bf k+G}={\cal Z}_{\bf k},\ \ 
\zeta_{\bf -k}=\zeta^\star_{\bf k}, \ \ \zeta_{\bf k+G}=\zeta_{\bf k},
\label{Szetaprop}
\ee
where ${\bf G}$ are the 2D reciprocal lattice vectors which can be chosen as shown in Fig. \ref{lattice},right, ${\bf G}_1=(2\pi/ a)(1,1/\sqrt{3})$, ${\bf G}_2=(2\pi/ a)(1,-1/\sqrt{3})$. The energy $E_{l{\bf k}}$ and the wave functions $|l{\bf k}\rangle$ are periodic in ${\bf k}$-space, $E_{l,{\bf k+G}}=E_{l{\bf k}}$ and $|l,{\bf k+G}\rangle =|l{\bf k}\rangle$. The functions (\ref{wavefunction}) and (\ref{blochfactor}) are normalized as follows
\be 
\langle l'{\bf k}'|l{\bf k}\rangle\equiv \int_{S}d{\bf r}\int_{-\infty}^\infty dz\Psi^\star_{l'{\bf k}'}({\bf r},z)\Psi_{l{\bf k}}({\bf r},z)=\delta_{{\bf k}{\bf k}'}\delta_{ll'},\label{norm<>}
\ee 
\be 
(l'{\bf k}|l{\bf k})\equiv \frac 1A\int_Ad{\bf r}\int_{-\infty}^\infty dz u^\star_{l'{\bf k}}({\bf r},z) u_{l{\bf k}}({\bf r},z)
=\delta_{ll'}.\label{norm()}
\ee 

Near the Dirac points, at the corners of the hexagon shaped Brillouin zone ${\bf k=K}_j$ ($j=1,..,6$ is the valley index, Fig. \ref{lattice}b), the function ${\cal Z}_{\bf k}$ assumes the form 
\be
{\cal Z}_{\bf k}^j\approx -\frac{\sqrt{3}a}2  
\left[(-1)^j \tilde k^j_x + i \tilde k^j_y\right], \ \ {\bf \tilde k}_j={\bf k-K}_j,\label{SnearDirac}
\ee
where ${\bf K}_1=-{\bf K}_4=(2\pi/a)(1/3,1/\sqrt{3})$, ${\bf K}_2=-{\bf K}_5=(2\pi/a)(2/3,0)$, ${\bf K}_3=-{\bf K}_6=(2\pi/a)(1/3,-1/\sqrt{3})$. The energy (\ref{energy}) and the function $\zeta_{\bf k}$ are then (at $|{\bf \tilde k}_j|a\ll 1$)
\be
E_{l{\bf k}}^j= 
(-1)^l\frac{t\sqrt{3}a}{2} |{\bf \tilde k}_j|
\equiv (-1)^lv_F\hbar |{\bf \tilde k}_j|,\label{energynearDirac}
\ee
\be
\zeta_{{\bf k}j}=-
\frac
{(-1)^j \tilde k^j_x + i \tilde  k^j_y}
{|{\bf \tilde k}_j|}=(-1)^{j+1} e^{(-1)^j i\varphi_{\bf k}},
\label{ZetanearDirac}
\ee
where the effective (Fermi) velocity is $v_F=t\sqrt{3}a/2\hbar\approx 10^8$ cm/s, and $\varphi_{\bf k}$ is the polar angle of the vector ${\bf \tilde k}$, ${\bf \tilde k}=(\tilde k_x,\tilde k_y)=\tilde k(\cos \varphi_{\bf k},\sin \varphi_{\bf k})$.

In typical graphene structures the chemical potential $\mu$ lies in the vicinity of Dirac points, $|\mu|\ll t$. In this paper we will assume that the energy of the photon is also small as compared to the overlap integral $t$, $\hbar\omega\ll t$. Under these conditions the energy and the wave-functions of electrons can be described by formulas (\ref{energynearDirac}) and (\ref{ZetanearDirac}). Possible resonances originated from quantum transitions in other (than ${\bf K}_j$) points of the Brillouin zone, as well as the trigonal warping effects are not considered in this paper.

\subsection{Matrix elements}

To calculate the system response in subsequent chapters we will need some matrix elements with the tight-binding functions (\ref{wavefunction}). 

\subsubsection{Matrix element of the exponential function}

Consider the matrix element $\langle l{\bf k}|e^{i{\bf q}\cdot{\bf r}}|l'{\bf k}'\rangle$ of the exponential function $e^{i{\bf q}\cdot{\bf r}}$. Assuming as above that the neighboring wave functions do not overlap we get
\be
\langle l{\bf k}|e^{i{\bf q}\cdot{\bf r}}|l'{\bf k}'\rangle= 
\delta_{{\bf k},{\bf k'+q}} \frac 12 
\left[
1+(-1)^{l'+l} \zeta^\star_{\bf k}\zeta_{{\bf k-q}}
\right] .
\label{MEexp}
\ee
The matrix element is nonzero only at ${\bf k}={\bf k'+q}$. 
Eq. (\ref{MEexp}) is valid in the whole Brillouin zone and at arbitrary values of ${\bf q}$. Since we aim to calculate the local (${\bf q=0}$) conductivity $\sigma^{(3)}_{\alpha\beta\gamma\delta} (\omega_1,\omega_2,\omega_3)$ we will need these matrix elements at ${\bf q\to 0}$, however, the terms linear in $q$ must be kept. Then the matrix element (\ref{MEexp}) assumes the form
\be
\langle l{\bf k}|e^{i{\bf q}\cdot{\bf r}}|l'{\bf k}'\rangle_{\bf q\to 0}
=
\delta_{{\bf k},{\bf k'+q}} 
\left(
\delta_{l'l}-\frac {(-1)^{l'+l}}2q_\alpha\eta^\alpha_{\bf k}
\right)
\label{MEexpexpansion}
\ee
where 
\be 
\eta^\alpha_{\bf k} = 
\frac{\p \zeta_{{\bf k}}}{\p  k_\alpha}\zeta^\star_{\bf k}
= 
i\frac{\p \Phi_{\bf k}}{\p  k_\alpha}
\label{eta}.
\ee

Near the Dirac points $j=1$ and $j=2$ the function $\zeta_{{\bf k}}$ is given by Eq. (\ref{ZetanearDirac}). Substituting this in Eq. (\ref{eta}) we get 
\be 
\eta^\alpha_{{\bf k}j} = \mp i \frac{\p \varphi_{\bf k}}{\p  k_\alpha} = \pm  \frac {i}{k^2}\epsilon_{\alpha\beta}k_\beta,\label{etanearDirac}
\ee
where the upper (lower) sign corresponds to $j=1$ ($j=2$) and the two-dimensional Levi-Civita symbol is defined in (\ref{Levi}), see Appendix 
\ref{app:usefulformulas}. Notice that the more general formula (\ref{eta}) is valid in the whole Brillouin zone.  

\subsubsection{Matrix element of the velocity}

Another quantity that we will need below is the diagonal in ${\bf k}$ (${\bf k}={\bf k'}$) matrix element of the velocity operator $\langle l{\bf k}| {\bf \hat v}|l'{\bf k}\rangle$, where ${\bf \hat v}={\bf \hat p}/m$ (here $m$ is the free electron mass). To find it, consider the matrix element $\langle l{\bf k}| [\hat H_0,e^{i{\bf q\cdot r}}]|l'{\bf k}'\rangle$. On the one hand it equals
\be 
\langle l{\bf k}| [\hat H_0,e^{i{\bf q\cdot r}}]|l'{\bf k}'\rangle=
(E_{l{\bf k}}-E_{l'{\bf k}'})\langle l{\bf k}| e^{i{\bf q\cdot r}}|l'{\bf k}'\rangle.
\ee
On the other hand
\be 
[\hat H_0,e^{i{\bf q\cdot r}}]=\frac 1{2m}[{\bf \hat p}^2,e^{i{\bf q\cdot r}}]
=\frac {\hbar{\bf q}} {2}\cdot \left({\bf \hat v}e^{i{\bf q\cdot r}} + e^{i{\bf q\cdot r}}{\bf \hat v}
\right),
\ee
so that 
\be 
\frac {\hbar{\bf q}} {2}\cdot \langle l{\bf k}| {\bf \hat v}e^{i{\bf q\cdot r}} + e^{i{\bf q\cdot r}}{\bf \hat v}|l'{\bf k}'\rangle=(E_{l{\bf k}}-E_{l'{\bf k}'})\langle l{\bf k}| e^{i{\bf q\cdot r}}|l'{\bf k}'\rangle.
\label{derivVelocity1}
\ee
In the limit of small ${\bf q}$ this gives
\be 
\frac {\hbar q_\alpha} {2} 
\langle l{\bf k}| 
\{ \hat v_\alpha,e^{i{\bf q\cdot r}} \}_+|l'{\bf k-q}\rangle=(E_{l{\bf k}}-E_{l'{\bf k-q}})
\left(
\delta_{l'l}-\frac {(-1)^{l'+l}}2q_\alpha\eta^\alpha_{\bf k}
\right)
\label{derivVelocity2}
\ee
where $\{\hat a,\hat b\}_+=\hat a\hat b+\hat b\hat a$ is the anticommutator. 
Now consider two cases, $l= l'$ and $l\neq l'$. For the diagonal and off-diagonal matrix elements we get, keeping only the lowest terms in ${\bf q}$:
\be 
q_\alpha \left(\hbar  \langle l{\bf k}| \hat  v_\alpha |l{\bf k}\rangle-\frac{\p E_{l{\bf k}}}{\p k_\alpha}\right)=0,\label{v1}
\ee
\be 
q_\alpha \Big( 2\hbar  \langle l{\bf k}| \hat v_\alpha |l'{\bf k}\rangle-(E_{l{\bf k}}-E_{l'{\bf k}})  
 \eta^\alpha_{\bf k}\Big)=0, \ \ l\neq l' .\label{v2}
\ee
Since the direction of the vector ${\bf q}$ in both formulas (\ref{v1}) and (\ref{v2}) is arbitrary, the terms in parenthesis must vanish. Thus we get for the intra- and inter-band matrix elements of the velocity
\be 
\langle l{\bf k}| \hat  v_\alpha |l{\bf k}\rangle=\frac 1\hbar \frac{\p E_{l{\bf k}}}{\p k_\alpha},\ \ \ l=l',
\label{velocity-on}
\ee
\be 
\langle l{\bf k}| \hat v_\alpha |l'{\bf k}\rangle=\frac{E_{l{\bf k}}-E_{l'{\bf k}}}{2\hbar}   \eta^\alpha_{\bf k},\ \ \ l\neq l'.
\label{velocity-off}
\ee
The formulas (\ref{velocity-on}) and (\ref{velocity-off}) are valid not only near the Dirac points but in the whole Brillouin zone. Near the Dirac point one can use Eq. (\ref{energynearDirac}) for the energy and Eq. (\ref{etanearDirac}) for the $\eta^\alpha_{\bf k}$-function.

\section{Linear response of graphene\label{sec:linear}}

\subsection{Preliminary remarks} 

Our ultimate goal is to calculate the current response ${\bf j}(t)$ of a uniform graphene layer to an external \textit{uniform} electric field ${\bf E}(t)$ parallel to it. The field is assumed to have an arbitrary time dependence and be presented by Fourier expansion
\be 
{\bf E}(t)=\int_{-\infty}^\infty d\omega {\bf E}_\omega e^{-i\omega t}.
\label{fieldFourier}
\ee
The current response ${\bf j}(t)$ will have to be calculated in up to the third order in the electric field ${\bf E}$.

The external ac electric field can be included in the Hamiltonian of the system in two ways, using a vector or a scalar potential $\phi({\bf r},t)$. We choose the second option and write the Hamiltonian in the form
\be 
\hat H=\hat H_0+\hat H_1=\hat H_0-e\phi({\bf r},t),
\ee
where the potential can be expanded in the Fourier integral over time, 
\be
 \phi({\bf r},t)=\int_{-\infty}^\infty d\omega \phi_{\omega}({\bf r})e^{-i\omega t}. \label{elpotential}
\ee
Notice that at the moment we assume $\phi({\bf r},t)$ to be space-dependent,
\be 
\phi_{\omega}({\bf r})=\sum_{\bf q}\phi_{{\bf q}\omega}e^{i{\bf q}\cdot {\bf r}},
\label{phi_expand_coord}
\ee
where the sums are taken over discretized ${\bf q}$ values, and the transition from the sum to the integral is performed using the usual rule
\be 
\sum_{\bf q}(\dots)\to \frac S{(2\pi)^2}\int\int (\dots) d{\bf q}.
\ee 
This has to be done since the field is proportional to the gradient of the potential, ${\bf E}({\bf r},t)=-\nabla \phi({\bf r},t)$, and its Fourier component is ${\bf E}_{{\bf q}\omega}=-i{\bf q}\phi_{{\bf q}\omega}$. In order to calculate the local (${\bf q\to 0}$) conductivities $\sigma^{(1)}$ and $\sigma^{(3)}$ we have, first, to find the current response at a finite wave-vector and then take the limit ${\bf q\to 0}$ in the final formulas. The factor ${\bf q}\phi_{{\bf q}\omega}$ should be kept finite in the taking-the-limit procedure. 

In this Section we discuss this process in sufficient detail for the linear response problem. The nonlinear response is then treated in Section \ref{sec:nonlin}. 

\subsection{Density matrix}

The system response to the potential (\ref{elpotential}) is described by Liouville equation for the density matrix $\hat\rho$
\be
\frac{\p\hat\rho}{\p t}=\frac 1{i\hbar}[\hat H_0+\hat H_1,\hat\rho]-\gamma(\hat\rho-\hat\rho_0),\label{liou}
\ee
where $\gamma$ is the phenomenological relaxation rate due to the scattering of electrons by impurities, phonons and other lattice imperfections (for simplicity, we restrict our consideration by one parameter $\gamma$). The unperturbed Hamiltonian $\hat H_0$ and the equilibrium density matrix $\hat\rho_0$ satisfy the eigen-value equations
\be 
\hat H_0|\lambda\rangle=E_{\lambda}|\lambda\rangle, \ \ \  
\hat \rho_0|\lambda\rangle=f_\lambda|\lambda\rangle, 
\ee 
where $|\lambda\rangle=|l{\bf k}\sigma\rangle$ (we have added the spin variable $\sigma$) and $f_\lambda=f_{l{\bf k}}$ is the Fermi distribution function
\be 
f_{l{\bf k}}=\left(1+\exp\frac{E_{l{\bf k}}-\mu}T\right)^{-1}.\label{fermi}
\ee
Expanding the density matrix in powers of the electric field, $\hat\rho=\hat\rho_0+\hat\rho_1+\hat\rho_2+\hat\rho_3+\dots$, $\hat\rho_n\propto E^n$, and solving the Liouville equation in the first order we get
\be 
\langle\lambda|\rho_1|\lambda'\rangle_t=\int_{-\infty}^\infty d\omega 
\frac {f_{\lambda'}-f_{\lambda}}{E_{\lambda'}-E_{\lambda}+\hbar(\omega+i\gamma)} 
\langle\lambda|h_{\omega}|\lambda'\rangle e^{-i\omega t},
\label{rho1sol}
\ee
where $h_{\omega}\equiv h_{\omega}({\bf r})=-e\phi_{\omega}({\bf r})$ and the subscript $t$ indicates that the density matrix elements $\langle\lambda|\rho_1|\lambda'\rangle_t$ depend on time [the integrand in (\ref{rho1sol}), apart from the exponential function, will then be denoted as $\langle\lambda|\rho_1|\lambda'\rangle_\omega$].  

\subsection{Current density} 

The two-dimensional current density is calculated according to the formula
\be 
{\bf j}({\bf r}_0,t)=
-\frac e2\sum_{\lambda\lambda'} 
\langle\lambda'|\{{\bf \hat v}\delta({\bf r}_0-{\bf r})\}_+|\lambda\rangle \langle \lambda| \rho|\lambda'\rangle_t. \label{current}
\ee
In the first order we get for the $\omega$-Fourier component of the current
\be 
{\bf j}_\omega({\bf r}_0)=
\frac {e^2}2
\sum_{\lambda\lambda'} 
\langle\lambda'|\{{\bf \hat v}\delta({\bf r}_0-{\bf r})\}_+|\lambda\rangle 
\frac {f_{\lambda'}-f_{\lambda}}{E_{\lambda'}-E_{\lambda}+\hbar(\omega+i\gamma)} 
\langle\lambda|\phi_{\omega}({\bf r})|\lambda'\rangle .
\ee
Now we expand the potential in the Fourier integral in ${\bf q}$, Eq. (\ref{phi_expand_coord}), and use the Dirac function representation 
\be 
\delta({\bf r}_0-{\bf r})=\frac 1S\sum_{\bf \tilde q}e^{i{\bf \tilde q}\cdot({\bf r}_0-{\bf r})}.\label{Delta-representation}
\ee
Then we get 
\ba 
{\bf j}_{{\bf \tilde q}\omega}&=&
\frac {e^2}{2S}\sum_{\tilde {\bf q}}\phi_{{\bf q}\omega}
\sum_{\lambda\lambda'} 
\langle\lambda'|\{{\bf \hat v},e^{-i{\bf \tilde q}\cdot {\bf r}}\}_+|\lambda\rangle 
\frac {f_{\lambda'}-f_{\lambda}}{E_{\lambda'}-E_{\lambda}+\hbar(\omega + i\gamma)} 
\langle\lambda|e^{i{\bf q\cdot r}}|\lambda'\rangle .
\label{currentGeneral}
\ea
After the substitution $|\lambda\rangle=|l{\bf k}\sigma\rangle$ the matrix elements in this expression give the factors $\delta_{\bf k',k-\tilde q}\delta_{\bf k,k'+q}\propto\delta_{\bf q,\tilde q}$ so that the sum over $\tilde {\bf q}$ disappears.

The formula (\ref{currentGeneral}) is general and valid at arbitrary ${\bf q}$. In the limit ${\bf q}\to{\bf 0}$ we can simplify it, substituting the matrix element expansion (\ref{MEexpexpansion}), thus getting the intra-band (first term) and inter-band (second term) contributions to the current: 
\be 
j^\alpha_{\omega}=
\left[
\frac {ie^2g_s}{\hbar(\omega + i\gamma) S}\sum_{l{\bf k} } 
\langle l{\bf k}| \hat v_\alpha|l{\bf k}\rangle 
\left(-\frac{\p f_{l{\bf k}}}{\p k_\beta}\right)
+
\frac {ie^2g_s}{2S}
\sum_{l{\bf k}}  
\langle \bar l{\bf k}| \hat v_\alpha |l{\bf k}\rangle 
\frac {f_{\bar l{\bf k}}-f_{l{\bf k}}}{E_{\bar l{\bf k}}-E_{l{\bf k}}+\hbar(\omega + i\gamma)} 
\eta^\beta_{\bf k} \right]E_{\omega}^\beta.\label{curr}
\ee
Here $g_s=2$ is the spin degeneracy factor, 
\be 
E_{\omega}^\beta=\lim_{\bf q\to 0}E_{{\bf q}\omega}^\beta=\lim_{\bf q\to 0}\Big(-iq_\beta \phi_{{\bf q}\omega}\Big),
\ee
and we have introduced a designation $\bar l$ which means \textit{not} $l$, i.e.
\be 
\bar l=
\left\{
\begin{array}{ll}
2, &\textrm{  if }l=1, \\
1, &\textrm{  if }l=2.\\
\end{array}
\right.\label{notl}
\ee 
Notice that the sums in (\ref{curr}) are taken over the whole Brillouin zone of graphene, therefore the valley degeneracy factor $g_v=2$ has not yet appeared in this expression. 

\subsection{First-order conductivity}

\begin{figure}
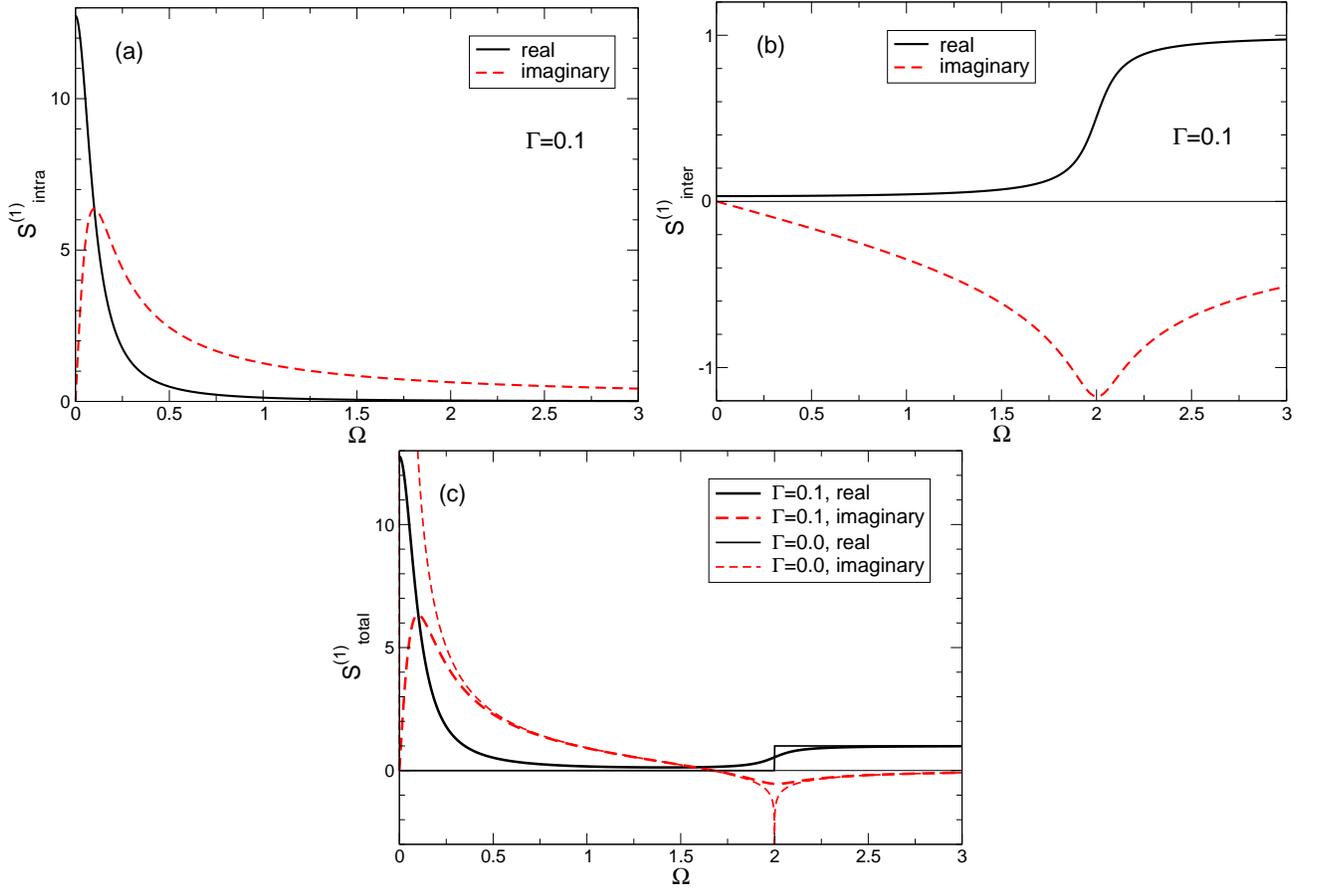

\includegraphics[width=8.5cm]{fig2a.eps}
\includegraphics[width=8.5cm]{fig2b.eps}
\includegraphics[width=8.5cm]{fig2c.eps}
\caption{\label{fig:sigma-1} (a) The intra-band (\ref{drude}), (b) inter-band (\ref{inter-cond}), and (c) total conductivity of a single graphene layer as a function of the frequency $\Omega =\hbar\omega/E_F$ at $\Gamma=\hbar\gamma/E_F=0.1$. In (c) the conductivity in the relaxation-free limit ($\Gamma=0$, thin curves) is also shown.}
\end{figure}

The value in the parenthesis in (\ref{curr}) is the linear (first-order) local conductivity of graphene
\be 
\sigma^{(1)}_{\alpha\beta}(\omega)=\frac {ie^2g_s}{\hbar(\omega + i\gamma) S}\sum_{l{\bf k} } 
\langle l{\bf k}| \hat v_\alpha|l{\bf k}\rangle 
\left(-\frac{\p f_{l{\bf k}}}{\p k_\beta}\right)
+
\frac {ie^2g_s}{2S}
\sum_{l{\bf k}}  
\langle \bar l{\bf k}| \hat v_\alpha |l{\bf k}\rangle 
\frac {f_{\bar l{\bf k}}-f_{l{\bf k}}}{E_{\bar l{\bf k}}-E_{l{\bf k}}+\hbar(\omega + i\gamma)} 
\eta^\beta_{\bf k} .\label{conductiv1}
\ee
This quantity has been calculated in a number of publications, see e.g. \cite{Falkovsky07a,Gusynin07b,Mikhailov07d}, therefore we do not discuss here further details of the integral calculations in (\ref{conductiv1}). Assuming that the temperature is zero, $T=0$, and taking into account that the main contribution to the sums over ${\bf k}$ is given by the vicinity of Dirac points we get 
\be 
\sigma^{(1)}_{\alpha\beta}(\omega)=
\sigma^{(1)}_0
{\cal S}^{(1)}_{\alpha\beta}(\Omega),\label{sigma1}
\ee 
where
\be 
\sigma^{(1)}_0=\frac{e^2g_sg_v}{16\hbar}=\frac{e^2}{4\hbar}
\label{universal}
\ee
is the universal optical conductivity \cite{Kuzmenko08,Nair08} and the dimensionless function 
\be 
{\cal S}^{(1)}_{\alpha\beta}(\Omega)=
\delta_{\alpha\beta} 
\left({\cal S}^{(1)}_{\rm intra}(\Omega)+{\cal S}^{(1)}_{\rm inter}(\Omega)\right),\label{sigma1dimless}
\ee 
consists of two, intra-band and inter-band, contributions:
\be
{\cal S}^{(1)}_{\rm intra}(\Omega)=\frac 4\pi
\frac{i}{\Omega+i\Gamma}, 
\label{drude}
\ee
\be 
{\cal S}^{(1)}_{\rm inter}(\Omega)=
\frac {i}{\pi}
\ln\frac{2-(\Omega + i \Gamma)}{2+(\Omega + i \Gamma)}.\label{inter-cond}
\ee
We have introduced here two dimensionless parameters 
\be 
\Omega=\frac{\hbar\omega}{|\mu|}, \ \ \ \Gamma=\frac{\hbar\gamma}{|\mu|}.
\label{dimlessOmegaGamma}
\ee
The intra-band contribution (\ref{drude}) has the Drude form; the inter-band conductivity (\ref{inter-cond}) has a step-like (logarithmic) singularity at $\hbar\omega\simeq 2E_F$ in its real (imaginary) part. At large frequencies $\hbar\omega\gg 2E_F$ the conductivity assumes the universal value (\ref{universal}). 
Figure \ref{fig:sigma-1} shows the intra-band, inter-band and the total conductivity of graphene at $\Gamma=0.1$.

\section{Third-order response of graphene\label{sec:nonlin}}

\subsection{Density matrix and  third-order current}

Now we need to calculate the next term in the expansion of the electric current in powers of the electric field. The second-order current response $\propto {\bf E}^2(t)$ vanishes due to the central symmetry of graphene. In the third order in the perturbation (\ref{elpotential}) the solution of Liouville equation (\ref{liou}) can be written in the form 

\ba 
\langle\lambda|\rho_3|\lambda'\rangle_t&=&
\int_{-\infty}^\infty d\omega_1 
\int_{-\infty}^\infty d\omega_2  
\int_{-\infty}^\infty d\omega_3 
\frac {e^{-i(\omega_1 +\omega_2+\omega_3 ) t}}
{E_{\lambda'}-E_{\lambda}+\hbar(\omega_1 +\omega_2+\omega_3) +i\hbar \gamma}
\sum_{\lambda''\lambda'''}
\langle\lambda| h_{\omega_3} |\lambda'''\rangle
\langle\lambda'''|h_{\omega_2}|\lambda''\rangle 
\langle\lambda''| h_{\omega_1} |\lambda'\rangle 
\nonumber \\ &\times&
\Bigg[
\frac{1}
{E_{\lambda'}-E_{\lambda'''} + \hbar (\omega_1 +\omega_2) +i\hbar\gamma }
\Bigg(
\frac {f_{\lambda'}-f_{\lambda''}}
{E_{\lambda'}-E_{\lambda''}+\hbar\omega_1+i\hbar\gamma} 
-
\frac {f_{\lambda''}-f_{\lambda'''}}
{E_{\lambda''}-E_{\lambda'''}+\hbar\omega_2+i\hbar\gamma} 
\Bigg)
\nonumber \\ &-&
\frac{1}
{E_{\lambda''}-E_{\lambda} + \hbar (\omega_2 +\omega_3) +i\hbar\gamma }
\Bigg(
\frac {f_{\lambda''}-f_{\lambda'''}}
{E_{\lambda''}-E_{\lambda'''}+\hbar\omega_2+i\hbar\gamma} 
-
\frac {f_{\lambda'''}-f_{\lambda}}
{E_{\lambda'''}-E_{\lambda}+\hbar\omega_3+i\hbar\gamma} 
\Bigg)
\Bigg] .\label{rho3t}
\ea
In order to find the current density in the third-order we should substitute Eq. (\ref{rho3t}) in the current definition (\ref{current}). Using the delta-function representation (\ref{Delta-representation}) and the Fourier expansion of the potential (\ref{phi_expand_coord}) we get for the third-order current:
\ba 
&&{\bf j}^{(3)}({\bf r}_0,t)=
\int_{-\infty}^\infty d\omega_1 
\int_{-\infty}^\infty d\omega_2  
\int_{-\infty}^\infty d\omega_3 
\frac {e^4}{2S}\sum_{{\bf \tilde q}{\bf q}_1{\bf q}_2{\bf q}_3} 
\phi_{{\bf q}_1\omega_1}\phi_{{\bf q}_2\omega_2}\phi_{{\bf q}_3\omega_3}
\sum_{\lambda\lambda'} 
\frac {\langle\lambda'|\{{\bf \hat v},e^{-i{\bf \tilde q}\cdot{\bf r}}\}_+|\lambda\rangle e^{i{\bf \tilde q}\cdot{\bf r}_0-i(\omega_1 +\omega_2+\omega_3) t}}
{E_{\lambda'}-E_{\lambda}+\hbar(\omega_1 +\omega_2+\omega_3) +i\hbar \gamma}
\nonumber \\ &\times&
\Bigg[
\sum_{\lambda''\lambda'''}
\frac{\langle\lambda| e^{i{\bf q}_3\cdot{\bf r}_3}|\lambda'''\rangle
\langle\lambda'''|e^{i{\bf q}_2\cdot{\bf r}_2} |\lambda''\rangle 
\langle\lambda''| e^{i{\bf q}_1\cdot{\bf r}_1}  |\lambda'\rangle }
{E_{\lambda'}-E_{\lambda'''} + \hbar (\omega_1 +\omega_2) +i\hbar\gamma }
\Bigg(
\frac {f_{\lambda'}-f_{\lambda''}}
{E_{\lambda'}-E_{\lambda''}+\hbar\omega_1+i\hbar\gamma} 
-
\frac {f_{\lambda''}-f_{\lambda'''}}
{E_{\lambda''}-E_{\lambda'''}+\hbar\omega_2+i\hbar\gamma} 
\Bigg)
\nonumber \\ &-&
\sum_{\lambda''\lambda'''}
\frac{\langle\lambda| e^{i{\bf q}_3\cdot{\bf r}_3}|\lambda'''\rangle
\langle\lambda'''|e^{i{\bf q}_2\cdot{\bf r}_2} |\lambda''\rangle 
\langle\lambda''| e^{i{\bf q}_1\cdot{\bf r}_1}  |\lambda'\rangle }
{E_{\lambda''}-E_{\lambda} + \hbar (\omega_2 +\omega_3) +i\hbar\gamma }
\Bigg(
\frac {f_{\lambda''}-f_{\lambda'''}}
{E_{\lambda''}-E_{\lambda'''}+\hbar\omega_2+i\hbar\gamma} 
-
\frac {f_{\lambda'''}-f_{\lambda}}
{E_{\lambda'''}-E_{\lambda}+\hbar\omega_3+i\hbar\gamma} 
\Bigg)
\Bigg] .
\label{current3}
\ea
The current density (\ref{current3}) contains the product of three matrix elements of the type (\ref{MEexp}),
\be 
\langle\lambda| e^{i{\bf q}_3\cdot{\bf r}_3}|\lambda'''\rangle
\langle\lambda'''|e^{i{\bf q}_2\cdot{\bf r}_2} |\lambda''\rangle 
\langle\lambda''| e^{i{\bf q}_1\cdot{\bf r}_1}  |\lambda'\rangle .
\label{3me}
\ee
Each of these factors is the sum of the intra- and inter-band contributions, Eq. (\ref{MEexpexpansion}). Expanding the product (\ref{3me}) we obtain a total of eight summands:
\begin{enumerate}
\item One term containing the product of three intraband contributions; we will label it as $(3/0)$ term (three intra- and no inter-band factors). This term is purely classical; the corresponding contribution to the current can be obtained by solving the classical (Boltzmann) kinetic equation.
\item Three terms containing the product of two intraband and one interband contributions; we will label them as $(2/1)$ term (two intra- and one inter-band factors). 
\item Three terms containing the product of one intraband and two interband contributions; they will be labeled as $(1/2)$ term. 
\item And one term containing the product of three interband contributions; this is a purely quantum contribution; it will be labeled as $(0/3)$ term.
\end{enumerate}
In order to find all these contributions we have to calculate the sums over $\lambda=(\sigma l{\bf k})$, $\lambda'=(\sigma' l'{\bf k}')$, $\lambda''=(\sigma'' l''{\bf k}'')$, and $\lambda'''=(\sigma''' l'''{\bf k}''')$ in Eq. (\ref{current3}). Perfoming the summations over all spin indexes and all but one factors $l{\bf k}$ we reduce the expression for the third-order current to the following form (for details see Appendix \ref{app:deriv-current}):
\be 
j_\alpha^{(3)}(t)=
\underbrace{j_{\alpha}^{(3)}(t)}_{(3/0)}+
\underbrace{j_\alpha^{(3)}(t)}_{(2/1)}+
\underbrace{j_\alpha^{(3)}(t)}_{(1/2)}+
\underbrace{j_\alpha^{(3)}(t)}_{(0/3)},
\label{three-currents}
\ee
where
\ba 
\underbrace{j_{\alpha}^{(3)}(t)}_{(3/0)} &=&
\int_{-\infty}^\infty d\omega_1 
\int_{-\infty}^\infty d\omega_2  
\int_{-\infty}^\infty d\omega_3 E^{\beta}_{\omega_1}E^{\gamma}_{\omega_2}E^{\delta}_{\omega_3}
e^{-i(\omega_1 +\omega_2+\omega_3) t}
\nonumber \\ &\times&
\frac 1{(\omega_1 +\omega_2+\omega_3 +i\gamma) 
(\omega_1 +\omega_2 +i\gamma)(\omega_1+i\gamma)}
\frac {ie^4g_s}{\hbar^3 S} 
\sum_{ l  {\bf k}} 
\langle l {\bf k}| \hat v_\alpha| l  {\bf k}\rangle
\frac{\p^3 f_{ l{\bf k}}}{\p k_\beta\p k_\gamma\p k_\delta} ,
\label{intra3}
\ea
\ba 
\underbrace{j_\alpha^{(3)}(t)}_{(2/1)}&=&
\int_{-\infty}^\infty d\omega_1 
\int_{-\infty}^\infty d\omega_2  
\int_{-\infty}^\infty d\omega_3 
E^{\beta}_{\omega_1}
E^\gamma_{\omega_2}
E^\delta_{\omega_3}
e^{-i(\omega_1 +\omega_2+\omega_3) t}
\nonumber \\ &\times&
\frac {-ie^4g_s}{2S}
\sum_{l {\bf k}} 
\frac {\langle \bar l{\bf k}| \hat v_\alpha|l{\bf k}\rangle }
{E_{\bar l{\bf k}}-E_{l{\bf k}}+\hbar(\omega_1 +\omega_2+\omega_3) +i\hbar \gamma}
\Bigg\{
\frac {1}{\hbar\omega_1+i\hbar\gamma} 
\frac{\eta^\delta_{{\bf k}}}
{\hbar (\omega_1 +\omega_2) +i\hbar\gamma }
\frac{\p^2 (f_{\bar l{\bf k}}- f_{ l{\bf k}})}
{\p k_\beta\p k_\gamma}
\nonumber \\ &+&
\frac {1}{\hbar\omega_1+i\hbar\gamma} 
\frac{\p}{\p k_\delta}
\Bigg(
\frac{\eta^\gamma_{{\bf k}}}
{E_{\bar l{\bf k}}-E_{ l{\bf k}} + \hbar (\omega_1 +\omega_2) +i\hbar\gamma }
\frac{\p (f_{\bar l{\bf k}}- f_{ l{\bf k}})}{\p k_\beta}
\Bigg)
\nonumber \\ &+&
\frac{\p }{\p k_\delta}
\Bigg[
\frac{1}
{E_{\bar l{\bf k}}-E_{ l{\bf k}} + \hbar (\omega_1 +\omega_2) +i\hbar\gamma }\frac{\p }{\p k_\gamma}
\left(\eta^\beta_{{\bf k}}
\frac {f_{ \bar l{\bf k}}-f_{ l{\bf k}}}
{E_{\bar l{\bf k}}-E_{ l{\bf k}}+\hbar\omega_1+i\hbar\gamma} \right)
\Bigg] 
\Bigg\}
\label{intra2inter},
\ea
\ba 
\underbrace{ j_\alpha^{(3)}(t)}_{(1/2)}&=&
\int_{-\infty}^\infty d\omega_1 
\int_{-\infty}^\infty d\omega_2  
\int_{-\infty}^\infty d\omega_3 
E^{\beta}_{\omega_1}
E^{\gamma}_{\omega_2}
E^{\delta}_{\omega_3}
e^{-i(\omega_1 +\omega_2+\omega_3) t}
\nonumber \\ &\times&
\frac {ie^4g_s}{2S}
\frac {1}{\hbar(\omega_1 +\omega_2+\omega_3) +i\hbar \gamma}
\sum_{l {\bf k}} 
\langle l{\bf k}| \hat v_\alpha|l{\bf k}\rangle 
\Bigg[
\frac{1}{\hbar (\omega_1 +\omega_2) +i\hbar\gamma }
\frac{\p }{\p k_\delta}
\Bigg(
\eta^\beta_{{\bf k}}\eta^\gamma_{{\bf k}}
\frac {f_{ l{\bf k}}-f_{ \bar l{\bf k}}}
{E_{l{\bf k}}-E_{ \bar l{\bf k}}+\hbar\omega_1+i\hbar\gamma}  \Bigg)
\nonumber \\ &+&
\frac{\eta^\delta_{{\bf k}} }
{E_{ l{\bf k}}-E_{ \bar l,{\bf k}} + \hbar (\omega_1 +\omega_2) +i\hbar\gamma }
\frac{\p}{\p k_\gamma}
\Bigg(\eta^\beta_{{\bf k}}
\frac {f_{ l{\bf k}}-f_{ \bar l{\bf k}}}
{E_{l{\bf k}}-E_{ \bar l{\bf k}}+\hbar\omega_1+i\hbar\gamma} 
\Bigg)
\nonumber \\ &+&
\frac{\eta^\gamma_{{\bf k}} \eta^\delta_{{\bf k}}}{\hbar\omega_1+i\hbar\gamma}
\frac{1}
{E_{ l{\bf k}}-E_{ \bar l{\bf k}} + \hbar (\omega_1 +\omega_2) +i\hbar\gamma }
\frac{\p (f_{ l{\bf k}}- f_{ \bar l{\bf k}})}{\p k_\beta}
\Bigg] ,
\label{intrainter2}
\ea
and 
\ba 
\underbrace{ j^{(3)}_\alpha(t)}_{(0/3)}&=&
\int_{-\infty}^\infty d\omega_1 
\int_{-\infty}^\infty d\omega_2  
\int_{-\infty}^\infty d\omega_3 
E^{\beta}_{\omega_1}
E^{\gamma}_{\omega_2}
E^{\delta}_{\omega_3}
e^{-i(\omega_1 +\omega_2+\omega_3) t}
\nonumber \\ &\times&
\frac {ie^4g_s}{4S}
\frac{1 }{\hbar (\omega_1 +\omega_2) +i\hbar\gamma }
\sum_{l {\bf k}} 
\frac {\langle \bar l{\bf k}| \hat v_\alpha|l{\bf k}\rangle }
{E_{ \bar l{\bf k}}-E_{l{\bf k}}+\hbar(\omega_1 +\omega_2+\omega_3) +i\hbar \gamma}
\eta^\beta_{{\bf k}}\eta^\gamma_{{\bf k}}\eta^\delta_{{\bf k}}
\nonumber \\ &\times&
\Bigg(
\frac {1}
{E_{\bar l{\bf k}}-E_{ l{\bf k}}+\hbar\omega_1+i\hbar\gamma} 
+
\frac {1}
{E_{ l{\bf k}}-E_{ \bar l{\bf k}}+\hbar\omega_1+i\hbar\gamma} 
\Bigg) (f_{ l{\bf k}}-f_{ \bar l{\bf k}})
\label{inter3}.
\ea

One sees a certain structure in formulas (\ref{intra3}) -- (\ref{inter3}). The term $(3/0)$ has three poles at low frequencies $\omega_j\simeq -i\gamma$, $j=1,2,3$, and contains the third-order derivative of the distribution function $f_{ l{\bf k}}$; all other terms have at maximum two poles at $\omega_j\simeq -i\gamma$. Hence, the term $(3/0)$ is the largest one at low (microwave, terahertz) frequencies. The second term $(2/1)$ contains the inter-band energy denominator $\sim [E_{\bar l{\bf k}}-E_{l{\bf k}}+\hbar(\omega_1 +\omega_2+\omega_3) +i\hbar \gamma]^{-1}$ and the second derivative of the distribution function. Since at low temperatures $T/E_F\simeq 0$ the first derivative $\p (f_{ l{\bf k}}- f_{ \bar l{\bf k}})/\p E$ is proportional to the delta function $\delta(E-E_F)$, one should expect that the term $(2/1)$ has a second-order pole at $\hbar(\omega_1 +\omega_2+\omega_3)\simeq 2E_F$ and hence is the largest one at high (mid- and near-infrared) frequencies. The third and fourth contributions $(1/2)$ and $(0/3)$ contain the first and zeroth derivatives of the Fermi functions and hence are expected to be smaller as compared to the first two terms. It is also worth noting that the $(3/0)$ and $(1/2)$ contributions contain the diagonal (intra-band) matrix elements of the velocity $\langle l {\bf k}| \hat v_\alpha| l  {\bf k} \rangle$, while the $(2/1)$ and $(0/3)$ contributions contain its off-diagonal (inter-band) matrix elements $\langle \bar l {\bf k}| \hat v_\alpha| l  {\bf k} \rangle$.

Now we define the third-order conductivity and calculate different contributions to it at low ($T=0$) temperatures.

\subsection{Third-order conductivity\label{sec:3rdcond}}

The third-order fourth-rank conductivity tensor of graphene $\sigma_{\alpha\beta\gamma\delta}^{(3)}(\omega_1,\omega_2,\omega_3)$ is defined as follows
\be
j_{\alpha}^{(3)}(t)=
\int_{-\infty}^\infty d\omega_1 
\int_{-\infty}^\infty d\omega_2  
\int_{-\infty}^\infty d\omega_3 
\sigma_{\alpha\beta\gamma\delta}^{(3)}(\omega_1,\omega_2,\omega_3)
E^\beta_{\omega_1} 
E^\gamma_{\omega_2} 
E^\delta_{\omega_3} 
e^{-i(\omega_1 +\omega_2+\omega_3) t}.
\label{sigma3}
\ee
From the definition (\ref{sigma3}) and the formulas (\ref{intra3}) -- (\ref{inter3}) one sees that, first, simultaneous permutations of the arguments $\omega_1$, $\omega_2$, $\omega_3$ and the indexes $\beta$, $\gamma$, $\delta$ does not change the third-order current. The conductivity tensor $\sigma_{\alpha\beta\gamma\delta}^{(3)}(\omega_1,\omega_2,\omega_3)$ can thus be always presented in the symmetric form
\ba 
\sigma_{\alpha\beta\gamma\delta}^{(3)} (\omega_1,\omega_2,\omega_3)&=&
\frac 1{3!}
\Big(
\tilde\sigma_{\alpha\beta\gamma\delta}^{(3)} (\omega_1,\omega_2,\omega_3)+
\tilde\sigma_{\alpha\beta\delta\gamma}^{(3)}(\omega_1,\omega_3,\omega_2)+
\tilde\sigma_{\alpha\gamma\beta\delta}^{(3)}(\omega_2,\omega_1,\omega_3)
\nonumber \\ &+&
\tilde\sigma_{\alpha\gamma\delta\beta}^{(3)}(\omega_2,\omega_3,\omega_1)+
\tilde\sigma_{\alpha\delta\beta\gamma}^{(3)}(\omega_3,\omega_1,\omega_2)+
\tilde\sigma_{\alpha\delta\gamma\beta}^{(3)} (\omega_3,\omega_2,\omega_1)
\Big),\label{sigma3resultSymm}
\ea
where the (unsymmetrical) tilde-conductivities $\tilde\sigma_{\alpha\beta\gamma\delta}^{(3)} (\omega_1,\omega_2,\omega_3)$ can be obtained directly from the expressions (\ref{intra3}) -- (\ref{inter3}).
Second, in accordance with the expressions (\ref{intra3}) -- (\ref{inter3}) the unsymmetrical conductivity $\tilde\sigma_{\alpha\beta\gamma\delta}^{(3)} (\omega_1,\omega_2,\omega_3)$ contains four contributions according to the terms $(3/0)$, $(2/1)$, $(1/2)$ and $(0/3)$. We write them in the form 
\be 
\tilde \sigma_{\alpha\beta\gamma\delta}^{(3)} (\omega_1,\omega_2,\omega_3)=
\sigma_0^{(3)}
{\cal S}^{(3)}_{\alpha\beta\gamma\delta}(\Omega_1,\Omega_2,\Omega_3),
\label{sigma3resultfull}
\ee
where 
\be 
\sigma_0^{(3)}=\frac {e^4g_sg_v\hbar v_F^2}{16\pi E_F^4}
\label{sigma^30}
\ee 
and 
\be 
{\cal S}^{(3)}_{\alpha\beta\gamma\delta}(\Omega_1,\Omega_2,\Omega_3)=
{\cal S}^{(3/0)}_{\alpha\beta\gamma\delta}(\Omega_1,\Omega_2,\Omega_3)+
{\cal S}^{(2/1)}_{\alpha\beta\gamma\delta}(\Omega_1,\Omega_2,\Omega_3)+
{\cal S}^{(1/2)}_{\alpha\beta\gamma\delta}(\Omega_1,\Omega_2,\Omega_3)+
{\cal S}^{(0/3)}_{\alpha\beta\gamma\delta}(\Omega_1,\Omega_2,\Omega_3).
\label{sigma3resultfulldimless}
\ee
The dimensionless functions ${\cal S}^{(m/n)}_{\alpha\beta\gamma\delta}(\Omega_1,\Omega_2,\Omega_3)$ depend on the dimensionless frequencies $\Omega_j$, $j=1,2,3$, and the dimensionless scattering parameter $\Gamma$ as defined in Eq. (\ref{dimlessOmegaGamma}). Introducing the designations 
\be 
\Delta_{\alpha\beta\gamma \delta}= \delta_{\alpha\beta}\delta_{\gamma \delta}+
\delta_{\alpha\gamma }\delta_{\beta\delta}+
\delta_{\alpha\delta}\delta_{\beta\gamma},\label{Delta3}
\ee
\be 
O_{1}=\frac{\Omega_1+i\Gamma}2, \ \ \ 
O_{12}=\frac{\Omega_1+\Omega_2+i\Gamma}2, \ \ \ 
O_{123}=\frac{\Omega_1+\Omega_2+\Omega_3+i\Gamma}2,\label{O-definitions}
\ee
and calculating the contributions ${\cal S}^{(m/n)}_{\alpha\beta\gamma\delta}(\Omega_1,\Omega_2,\Omega_3)$ (see Appendix \ref{app:conduct} for details) we get:
\be 
{\cal S}^{(3/0)}_{\alpha\beta\gamma\delta}(\Omega_1,\Omega_2,\Omega_3)=
\frac {i\Delta_{\alpha\beta\gamma \delta}  
}{ 8O_{123}O_{12}O_{1}},\label{FuncS30}
\ee
\ba 
{\cal S}^{(2/1)}_{\alpha\beta\gamma\delta}(\Omega_1,\Omega_2,\Omega_3) &=&
-\frac i4\Bigg(
\frac {\delta_{\alpha\delta}\delta_{\beta\gamma}+(  \Delta_{\alpha \beta \gamma \delta}/4)O_{123}}{O_1O_{12} (1+ O_{123})^2}
+ \frac{(\delta_{\alpha \beta}\delta_{\gamma\delta}
 -\Delta_{\alpha \beta \gamma \delta} /4)}{O_1( 1 + O_{12})(1+O_{123})}
+\frac {\delta_{\alpha\gamma} \delta_{\beta \delta} -
\Delta_{\alpha \beta \gamma \delta} /4}{O_1 ( 1 + O_{12})(1+O_{123})^2}
\nonumber \\ &&
-(\delta_{\alpha\beta} \delta_{\gamma \delta} -\Delta_{\alpha\beta\gamma \delta}/4) 
{\cal J}_1(O_1,O_{12},O_{123})
+\left( \delta_{\alpha \gamma} \delta_{\beta\delta} - 
 \Delta_{\alpha \beta \gamma \delta}/4 \right)
{\cal J}_2(O_1,O_{12},O_{123})\Bigg)
 \nonumber \\ &&-\Big((O_1,O_{12},O_{123})\to(-O_1,-O_{12},-O_{123})\Big)
,\label{FuncS21}
\ea
\ba 
{\cal S}^{(1/2)}_{\alpha\beta\gamma\delta}(\Omega_1,\Omega_2,\Omega_3) &=&
\frac {i }{4}
\Bigg(
\frac {\Delta_{\alpha \beta \gamma\delta}/4}{O_1^2O_{123}O_{12}}
\ln(1+O_1)
- \frac {\Delta_{\alpha \beta \gamma\delta}/4}{O_{123}O_{12}O_1}
+
\frac{(\delta_{\alpha \beta}  
\delta_{\gamma\delta}-
 \Delta_{\alpha \beta \gamma \delta}/4 
)} {O_{123}O_1 (1 + O_{12}) }
\nonumber \\ &-&
\frac {\delta_{\alpha\beta}\delta_{\gamma\delta} 
-\Delta_{\alpha \beta \gamma \delta}/4 }{O_{123}}{\cal J}_3(O_{1},O_{12})
+
\frac {\delta_{\alpha  \gamma}\delta_{\beta\delta} 
-\Delta_{\alpha \beta \gamma \delta}/4 }{O_{123}}{\cal J}_4(O_{1},O_{12})
\Bigg)
\nonumber \\ &&-\Big((O_1,O_{12},O_{123})\to(-O_1,-O_{12},-O_{123})\Big)
\label{FuncS12},
\ea 
\be
{\cal S}^{(0/3)}_{\alpha\beta\gamma\delta}(\Omega_1,\Omega_2,\Omega_3) =
\frac{i\Delta_{\alpha  \beta  \gamma \delta} }{32 O_{12} }
{\cal J}_5(O_1,O_{123}).\label{FuncS03}
\ee
In the third lines of Eqs. (\ref{FuncS21}) and (\ref{FuncS12}) one should subtract the corresponding expressions from the first two lines, with the arguments $(O_1,O_{12},O_{123})$ being replaced by $(-O_1,-O_{12},-O_{123})$. The integrals ${\cal J}_n$ in (\ref{FuncS30}) -- (\ref{FuncS03}) are defined as follows 
\be 
{\cal J}_1(a,b,c)=
\int_{1}^\infty  \frac {dx}{( x + a)(x +  b)(x + c)}\Bigg(
\frac{1}{x^2}
+\frac{1}{x (x + c)}
-\frac { 1  }{ (x + b)(x + c)}-
\frac { 2  }{ (x + c)^2}
\Bigg),\label{IntI1}
\ee
\be 
{\cal J}_2(a,b,c)=
\int_{1}^\infty  \frac{dx}
{x ( x + a)(x + b)^2(x + c)},\label{IntI2}
\ee
\be 
{\cal J}_3(a,b)=\int_{1}^\infty 
\frac {dx }{x^2 (x+a)( x + b)},\label{IntI3}
\ee
\be 
{\cal J}_4(a,b)=\int_{1}^\infty  
\frac {dx }{x(x+a) ( x + b)^2},\label{IntI4}
\ee
\be 
{\cal J}_5(a,b)=
\int_{1}^\infty \frac{dx}{x^2} 
\Bigg(\frac {1}{ x+a} 
-\frac {1}{ x-a} \Bigg) 
\Bigg(\frac { 1 }{ x+b}
-\frac { 1}{ x-b}\Bigg).
\label{IntI5}
\ee
Their explicit expressions are:
\ba 
{\cal J}_1(a,b,c)&=&
\frac{1}{a b c }
+\frac{-a b+2 a c+3 b c-4 c^2}{c (c-a)^2 (b-c)^2 (c+1)}
-\frac{1}{(a-b) (b-c)^2(b+1)}
-\frac{1}{(a-c) (b-c) (c+1)^2}
\nonumber \\ &-&
\frac{-a^3-2 a^2 c+3 a b c+a c^2-b c^2}{a^2 (a-b)^2 (a-c)^3}\ln(a+1)
-\frac{-2 a b+a c+3 b^2-b c}{b^2 (a-b)^2 (b-c)^2 }\ln(b+1)
\nonumber \\ &-&
\frac{a^2+a b-4 a c-3 b c+5 c^2}{c (a-c)^3 (c-b)^2}\ln(c+1),
\label{IntI1abc}
\ea
\ba 
{\cal J}_2(a,b,c)&=&
\frac{1}{b(a-b) (b-c) (b+1)}
-\frac{1}{a (a-b)^2 (a-c)}\ln(a+1)
\nonumber \\ &-&
\frac{2 a b-a c-3 b^2+2 b c}{b^2 (a-b)^2 (b-c)^2}\ln(b+1)
-\frac{1}{c (c-a) (b-c)^2 }\ln (c+1),
\label{IntI2abc}
\ea
\be 
{\cal J}_3(a,b)=
\frac{1}{a b }
+\frac{1}{a^2 (a-b) }\ln(1+a)
-\frac{1}{b^2 (a-b) }\ln(1+b),
\label{IntI3ab}
\ee
\be 
{\cal J}_4(a,b)=
-\frac{1}{b (a-b) (1+b)}
+\frac{1}{a (a-b)^2}\ln (1+a)
+\frac{a-2 b}{b^2 (a-b)^2 }\ln (1+b),
\label{IntI4ab}
\ee
\be 
{\cal J}_5(a,b)=
\frac{4}{a b }
-\frac{2 b}{a^2 (a-b) (a+b) }\ln \frac{(1-a)}{(1+a)}
+\frac{2 a}{b^2(a-b) (a+b) }\ln \frac{(1-b)}{(1+b)}.
\label{IntI5ab}
\ee
The formulas (\ref{IntI1abc}) -- (\ref{IntI5ab}) are valid at arbitrary values of the arguments $a$, $b$ and $c$, but are not always convenient for numerical plotting of the results since the factors $(a-b)$, $(a-c)$, $(b-c)$ in the denominators may cause error messages ``division by zero''. It is easy to verify that in all limits $a\to b$, $a\to c$, $b\to c$, etc., the integrals ${\cal J}_n$ remain finite [this is seen directly from the definitions (\ref{IntI1}) -- (\ref{IntI5}); notice also that the arguments $a$, $b$, $c$ are complex at a finite $\Gamma$]. The explicit expressions for ${\cal J}_n$ in the special cases $a=b$, $a=c$, etc., are provided in Appendix \ref{specialJn}.

As we have discussed above, the contribution $(3/0)$ has three poles at low frequencies [the factors $1/O_{123}O_{12}O_{1}$ in Eq. (\ref{FuncS30})], and the contribution $(2/1)$ has the second-order pole at $\hbar(\omega_1 +\omega_2+\omega_3)\simeq \pm 2E_F$ [the factors $1/(1\pm O_{123})^2$ in Eq. (\ref{FuncS21}) and $1/(1\pm c)^2$ in Eq. (\ref{IntI1abc})]. In addition, the function ${\cal S}^{(3)}_{\alpha\beta\gamma\delta}$ has the first-order pole at the two-photon absorption frequency $\hbar(\omega_1 +\omega_2)\simeq \pm 2E_F$ [the factors $1/(1\pm O_{12})$ in Eqs. (\ref{FuncS21}) and (\ref{FuncS12}) and $1/(1\pm b)$ in Eqs. (\ref{IntI1abc}) -- (\ref{IntI2abc})], as well as (weaker) logarithmic singularities at $O_{123}\simeq \pm 1$, $O_{12}\simeq \pm 1$, and $O_{1}\simeq \pm 1$. 

The third-order conductivity tensor (\ref{sigma3resultSymm}) satisfies the relation
\be 
\left(\sigma^{(3)}_{\alpha\beta\gamma\delta} (-\omega_1,-\omega_2,-\omega_3)\right)^\star=\sigma^{(3)}_{\alpha\beta\gamma\delta} (\omega_1,\omega_2,\omega_3).
\ee
It has three independent components 
\be \sigma^{(3)}_{xyxy}=\sigma^{(3)}_{yxyx},\ \  \sigma^{(3)}_{xyyx}=\sigma^{(3)}_{yxxy},\ \  \sigma^{(3)}_{xxyy}=\sigma^{(3)}_{yyxx},\label{abcd-relation0}
\ee 
and the component $\sigma^{(3)}_{xxxx}=\sigma^{(3)}_{yyyy}$ which satisfies the equality
\be 
\sigma^{(3)}_{xxxx}=\sigma^{(3)}_{xxyy}+\sigma^{(3)}_{xyyx}+\sigma^{(3)}_{xyxy}.
\label{abcd-relation}
\ee
This agrees with the phenomenological relations for the third-order response function in an isotropic medium \cite{Boyd08}. 

The formulas obtained in this Section are the main result of this work. They show that the Fermi-energy related resonance in the third-order conductivity is the case at $\Omega_1+\Omega_2+\Omega_3\sim 2$ (which corresponds to $\hbar\omega=2E_F/3$ if all three frequency arguments are the same), as it was first pointed out in Ref. \cite{Cheng14a}. In our previous publication \cite{Mikhailov14c} not all contributions to the mixed current terms $(2/1)$, Eq. (\ref{intra2inter}), and $(1/2)$, Eq. (\ref{intrainter2}), have been taken into account; as a consequence, the main resonance in the third-harmonic-generation term was found at $\hbar\omega=2E_F$ instead of at $\hbar\omega=2E_F/3$ and the absolute value of the third-order conductivity was overestimated. In the relaxation-free limit $\Gamma\to 0$ the results (\ref{sigma3resultSymm})--(\ref{FuncS03})  mainly agree with those obtained in Ref. \cite{Cheng14a}, with the exception of a small vicinity of the pole at $|\Omega_1+\Omega_2+\Omega_3-2|\lesssim \Gamma$ where a certain discussion is needed (it can be found below in Section \ref{sec:THG}). The results (\ref{sigma3resultSymm})--(\ref{FuncS03}), which have been obtained by the method different from that used in Ref. \cite{Cheng15}, also agree with those of Ref. \cite{Cheng15} if to correct them according to the Erratum \cite{Cheng16err}. 

In the rest of the paper we discuss some consequences from the obtained results. In particular, the corrected results for the third-harmonic generation effect are considered in Section \ref{sec:THG}. 

\subsection{Asymptotes at low and high frequencies}

\begin{figure}
\includegraphics[width=8.5cm]{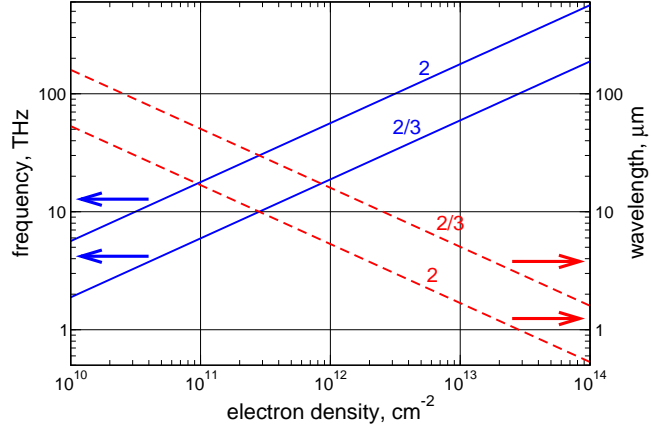}
\caption{The relations $\hbar\omega=2E_F=2\hbar v_F\sqrt{\pi n_s}$ (label $2$) and $\hbar\omega=2E_F/3$ (label $2/3$) plotted as the frequency $f=\omega/2\pi$ and the wavelength $\lambda=c/f$ versus the electron density $n_s$.\label{fig:freq-dens}}
\end{figure}

First, consider the system response at low frequencies, when all frequency parameters satisfy the conditions $|\Omega_j+i\Gamma|\ll 1$, or $\Omega_j\ll 1$ and $\Gamma=\hbar/E_F\tau\ll 1$. The condition $\Gamma\ll 1$ means $v_F\tau\sqrt{\pi n_s}\gg 1$, i.e., the mean-free path should exceed the average distance between electrons. At $\tau=(0.1-1)$ ps it means $n_s\gtrsim 10^{10}$ cm$^{-2}$. Since a typical graphene electron density exceeds $\sim 10^{11}$ cm$^{-2}$, the condition $\Gamma\ll 1$ is usually satisfied in real experiments. The second condition, $\hbar\omega\ll E_F$, is satisfied at up to mid-infrared frequencies, see Figure \ref{fig:freq-dens}. For example, at the density $n_s\simeq 10^{12}$ cm$^{-2}$ this condition corresponds to $f\ll 30$ THz. 

Under the conditions $|\Omega_j+i\Gamma|\ll 1$, $j=1,2,3$, the term $(3/0)$ gives the largest contribution to $\sigma^{(3)}_{\alpha\beta\gamma\delta} (\omega_1,\omega_2,\omega_3)$, and 
\be 
\sigma_{\alpha\beta\gamma\delta}^{(3)} (\omega_1,\omega_2,\omega_3)\approx 
\underbrace{\sigma_{\alpha\beta\gamma\delta}^{(3)} (\omega_1,\omega_2,\omega_3)}_{(3/0)}=\sigma_0^{(3)}
\frac {i\Delta_{\alpha\beta\gamma \delta}}{6 (\Omega_1+i\Gamma)(\Omega_2 +i\Gamma)(\Omega_3 +i\Gamma)},\hspace{5mm}|\Omega_j+i\Gamma|\ll 1.
\label{SmallOmegaResult}
\ee
This result was obtained within the quasi-classical approach in Ref. \cite{Mikhailov07e}, and then within the full quantum theory in Refs. \cite{Cheng14a,Cheng15}. It describes the low-frequency (microwave, terahertz) response of doped graphene. 

In the opposite limit, when all frequencies are large, $\Omega_j\gg 1$, the most important contributions to $\sigma^{(3)}_{\alpha\beta\gamma\delta} (\omega_1,\omega_2,\omega_3)$ come from the imaginary parts of logarithm functions in (\ref{FuncS21}) -- (\ref{FuncS03}), (\ref{IntI1abc}) -- (\ref{IntI5ab}), 
\be  
\lim_{\Omega_1\to \infty}\ln \frac{1+O_1}{1-O_1}= i\pi
\ee
(and similarly for $O_{12}$ and $O_{123}$). This gives
\be 
\sigma_{\alpha\beta\gamma\delta}^{(3)} (\omega_1,\omega_2,\omega_3)\approx 
\sigma_0^{(3)}
\frac{\pi\Delta_{\alpha  \beta  \gamma \delta} }
{3(\Omega_1+\Omega_2)(\Omega_2+\Omega_3)
(\Omega_3+\Omega_1)(\Omega_1+\Omega_2+\Omega_3)},\hspace{5mm}\Gamma\ll 1\ll|\Omega_j|.
\label{LargeOmegaResult}
\ee
The result (\ref{LargeOmegaResult}) agrees with the one obtained in Ref. \cite{Cheng14a}. It describes the optical response of intrinsic graphene.

\section{Third-order response to a monochromatic radiation\label{sec:mono}}

Now we consider the third-order response of an isolated, freely hanging in air  graphene layer to an external monochromatic radiation with the frequency $\omega$ and the intensity $I_\omega$ normally incident on the layer. Let the external electromagnetic radiation be linearly polarized in the $x$-direction and have only one frequency harmonic,
\be 
E_\beta(t)=E_0\delta_{x\beta}\cos\omega t.\label{extfield}
\ee
Then the Fourier component of the field is 
\be 
E^\beta_{\omega_1}=\frac {1}2 E_0\delta_{x\beta}\Big(\delta(\omega_1-\omega)+ \delta(\omega_1+\omega)\Big),
\ee
and the intensity of the incident wave reads 
\be 
I_\omega=\frac c{2\pi}|E_\omega|^2=\frac c{8\pi}E_0^2,
\ee
$E_\omega=E_0/2$. The electric field really acting on the graphene electrons [$\sim E_0/(1+2\pi\sigma^{(1)}/c)$] differs from (\ref{extfield}) due to the self-consistent screening effects (for their influence on the third-order response of graphene see, e.g., Ref. \cite{Savostianova15}). The difference is determined by the parameter $2\pi\sigma^{(1)}(\omega)/c$ which is small ($\lesssim (1-2)$\%) at infrared and optical frequencies. If to ignore this difference the third-order current (\ref{sigma3}) assumes the form
\be
j_\alpha^{(3)}(t)=
\frac {1}8E_0^3
\Bigg(
\sigma^{(3)}_{\alpha xxx}(\omega,\omega,\omega)e^{-i3\omega t} 
+\Big( \sigma^{(3)}_{\alpha xxx}(\omega,\omega,-\omega)
+\sigma^{(3)}_{\alpha xxx}(\omega,-\omega,\omega)+\sigma^{(3)}_{\alpha xxx}(-\omega,\omega,\omega)\Big)e^{-i\omega t} 
+ \textrm{c.c.}
\Bigg) ,\label{induced3WcurrentFull}
\ee
where c.c. means the complex conjugate. Since $\sigma^{(3)}_{y xxx}$ equals zero, the current is polarized in the same ($x$-) direction and has two contributions, one at the third harmonic $3\omega$ and another at the frequency of the incident wave $\omega$ (the Kerr effect). 

\subsection{Third harmonic generation\label{sec:THG}}

\begin{figure}
\includegraphics[width=8.5cm]{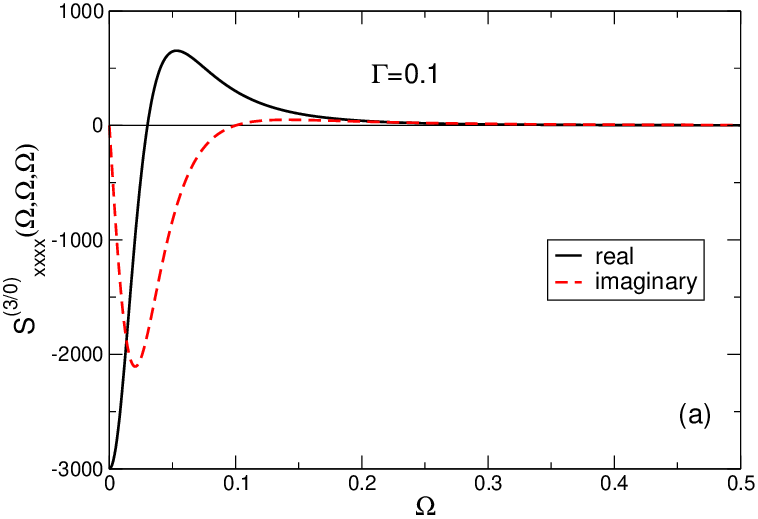}
\includegraphics[width=8.5cm]{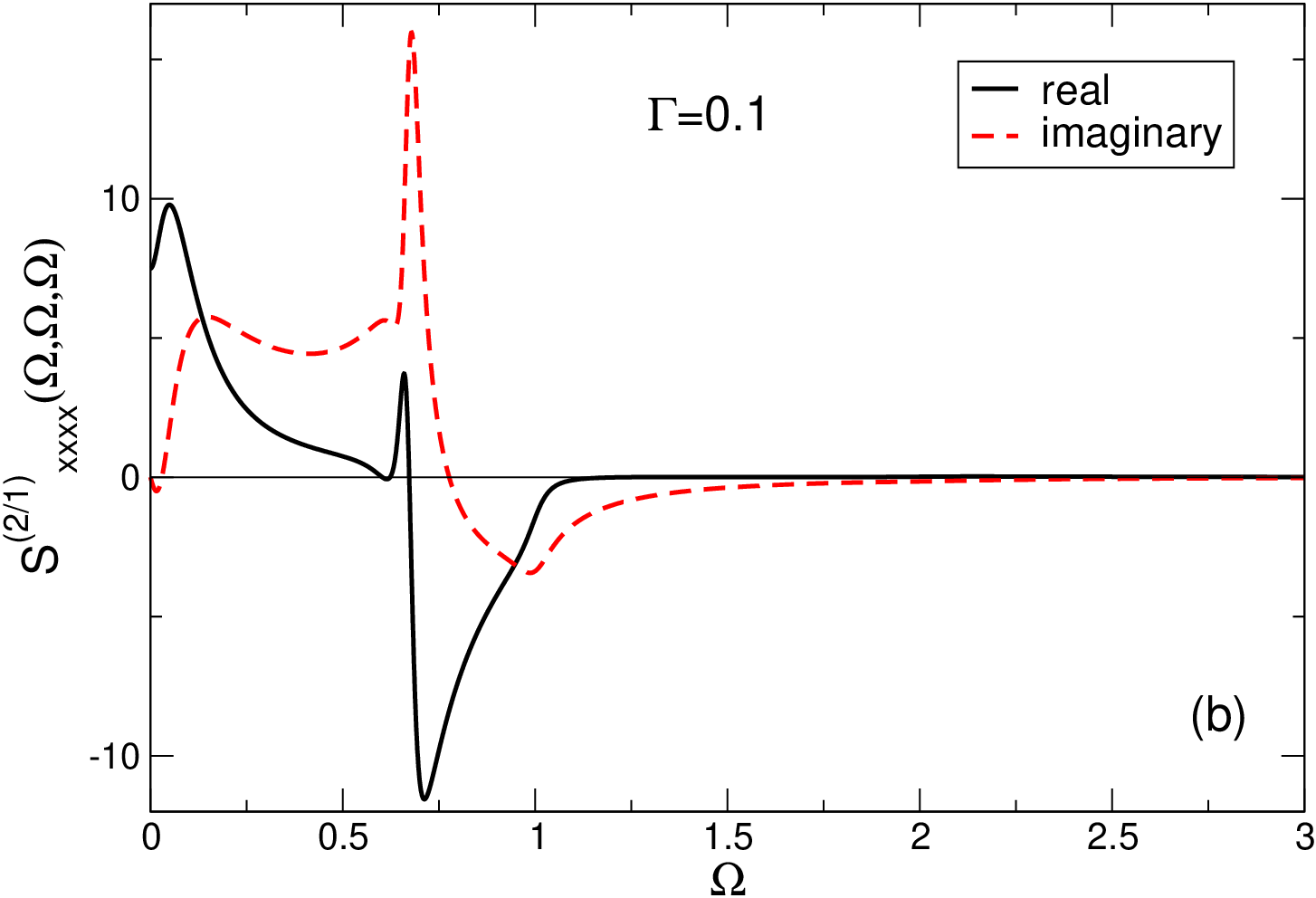}
\includegraphics[width=8.5cm]{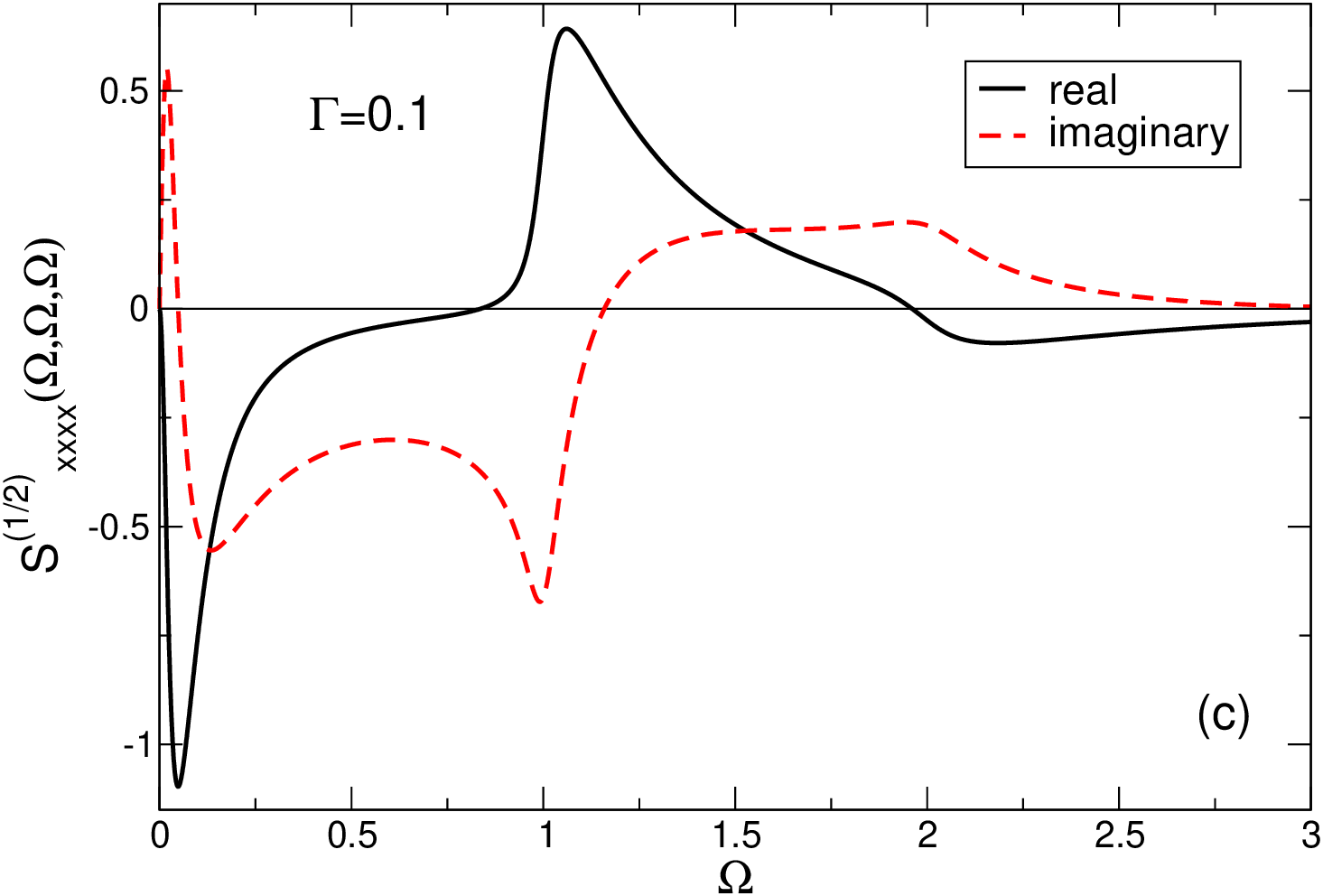}
\includegraphics[width=8.5cm]{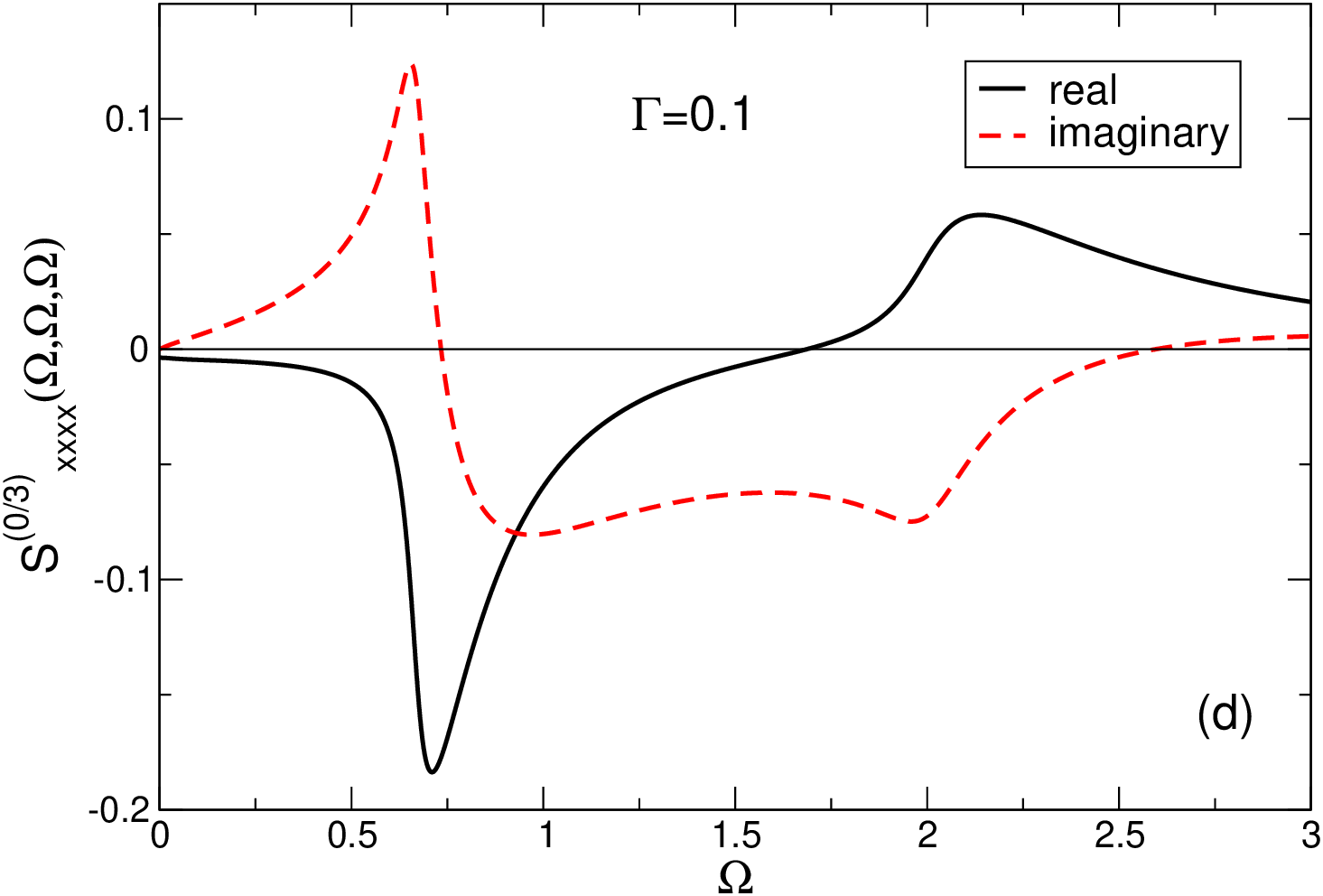}
\caption{Four contributions to the third order dimensionless conductivity ${\cal S}_{xxxx}^{(3)} (\omega,\omega,\omega)$: (a) $(3/0)$ contribution, (b) $(2/1)$ contribution, (c) $(1/2)$ contribution, (d) $(0/3)$ contribution. The effective relaxation rate is $\Gamma=0.1$. \label{fig:3Wcontribs}}
\end{figure}

\begin{figure}
\includegraphics[width=8.5cm]{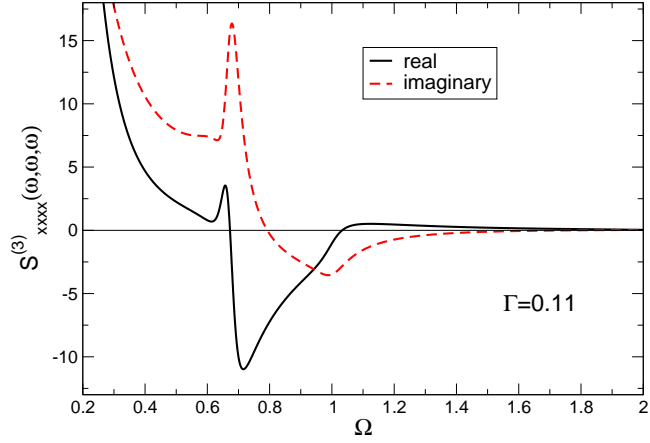}
\caption{The total third order conductivity $\sigma_{xxxx}^{(3)} (\omega,\omega,\omega)/\sigma_{0}^{(3)}$ under the conditions of Figure 3(b) of Ref. \cite{Cheng15}. The parameter $\Gamma=0.11$ corresponds to the numbers (the relaxation rate 33 meV, the Fermi energy 0.3 eV) used in Ref. \cite{Cheng15}.
\label{fig:3WSipe}}
\end{figure}

The third-harmonic current in Eq. (\ref{induced3WcurrentFull}) with the Fourier component $j^{(3)}_{3\omega}=\sigma^{(3)}_{xxxx}(\omega,\omega,\omega)(E_0/2)^3$ is determined by the conductivity $\sigma^{(3)}_{xxxx}(\omega,\omega,\omega)$. The Fourier components of the induced electric and magnetic fields at the frequency $3\omega$ are 
\be 
E_{3\omega}=H_{3\omega}=\frac {2\pi}c j^{(3)}_{3\omega}=\frac {2\pi}c \sigma^{(3)}_{xxxx}(\omega,\omega,\omega)E_\omega^3.
\ee
The intensity of the emitted third-harmonic wave (in the positive and negative $z$-directions) is then
\be 
I_{3\omega}=\frac c{2\pi}|E_{3\omega}|^2=\left(\frac {2\pi }{c}\right)^4
\left| \sigma^{(3)}_{xxxx}(\omega,\omega,\omega)\right|^2 I_\omega^3;
\ee
it can be written as
\be 
\frac{I_{3\omega}}{I_\omega}=
\left( \frac{e^4}{\hbar^2c^2}
 \frac {   I_\omega}{ \pi  n_s^2 \hbar v_F^2}\right)^2
\left|
{\cal S}^{(3)}_{xxxx}(\Omega,\Omega,\Omega)
\right|^2 ,\label{IdivI}
\ee
where ${\cal S}^{(3)}_{xxxx}(\Omega,\Omega,\Omega)$ is a sum of four contributions, Eq. (\ref{sigma3resultfulldimless}). These contributions are shown in Figure \ref{fig:3Wcontribs} (at $\Gamma=0.1$). The following features are seen:
\begin{enumerate}
\item The classical $(3/0)$ contribution has a large resonance at low frequencies $\Omega\lesssim\Gamma$. The contributions $(2/1)$ has a sharp resonance at $\Omega\simeq 2/3$ corresponding to the three-photon absorption edge and weaker singularities at $\Omega\simeq \Gamma$ and at $\Omega\simeq 1$; the latter corresponds to the two-photon absorption. The contribution $(1/2)$ has resonances at $\Omega\lesssim\Gamma$, $\Omega\simeq 1$ and $\Omega\simeq 2$, and the purely quantum contribution $(0/3)$ has weak (logarithmic) singularities at the frequencies of the single-photon ($\Omega\simeq 2$) and triple-photon ($\Omega\simeq 2/3$) inter-band transitions.
\item The absolute values of the different contributions to the third-order conductivity gradually decrease from the purely classical $(3/0)$ to the purely quantum $(0/3)$ contributions: at $\Gamma=0.1$ the scale of the $(3/0)$ term is about few thousands, of the $(2/1)$ term -- about 10, of the $(1/2)$ term about 1 and of the $(0/3)$ contribution -- about 0.1. The largest contribution at low (microwave, terahertz) frequencies is thus expected from the classical term $(3/0)$, while the largest (and resonant) contribution at higher (mid- and near-infrared) frequencies is expected at the resonance $\Omega=2/3$ ($\hbar\omega=2E_F/3$) from the $(2/1)$ contribution. 
\end{enumerate}
Figure \ref{fig:3WSipe} shows the total third-harmonic conductivity ${\cal S}^{(3)}_{xxxx}(\Omega,\Omega,\Omega)$ calculated with our formulas (\ref{intra3})--(\ref{inter3}) at parameters corresponding to Figure 3(b) from Ref. \cite{Cheng15}. The figures agree with each other.

\begin{figure}
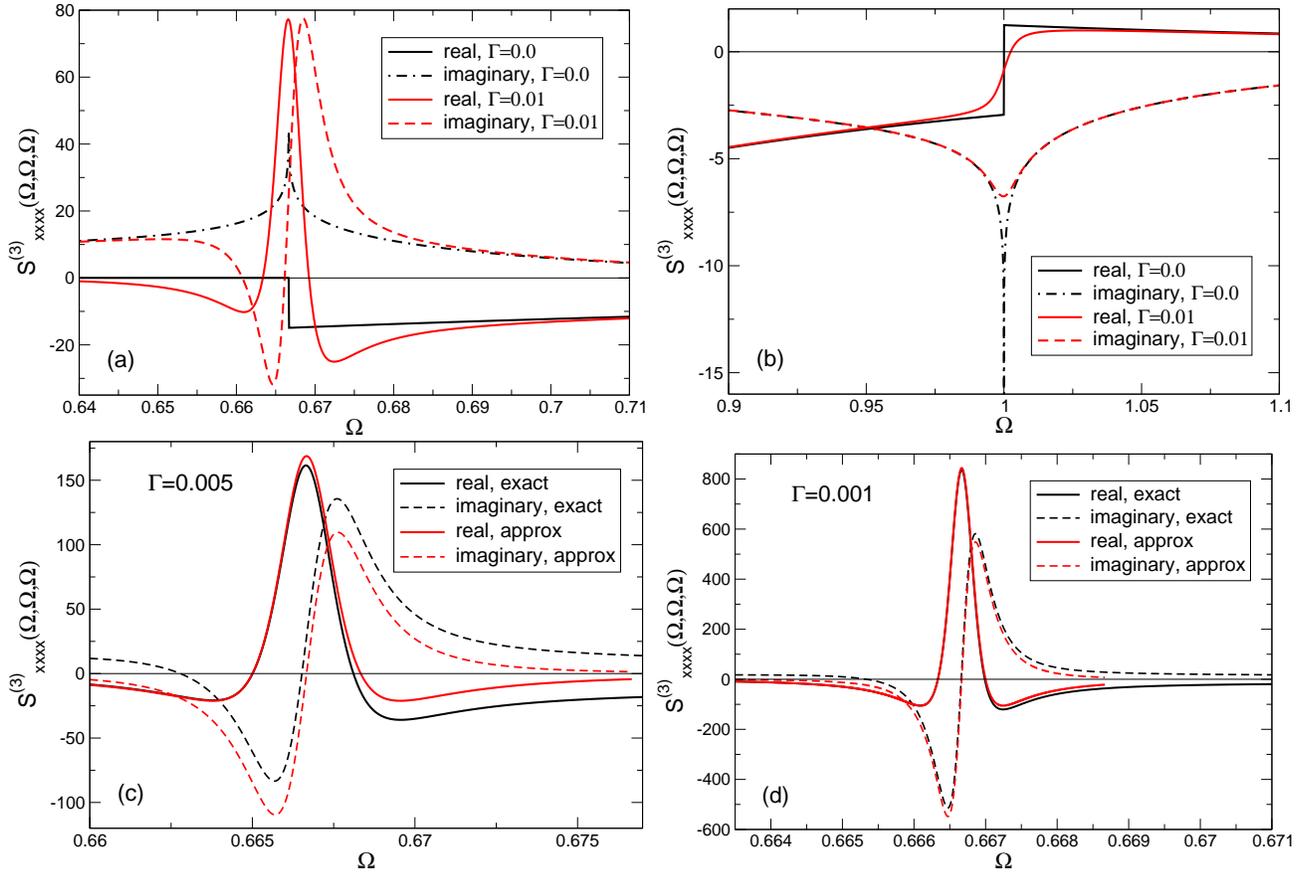

\includegraphics[width=8.5cm]{fig6a.eps}
\includegraphics[width=8.5cm]{fig6b.eps}
\includegraphics[width=8.5cm]{fig6c.eps}
\includegraphics[width=8.5cm]{fig6d.eps}
\caption{The total third order conductivity $\sigma_{xxxx}^{(3)} (\omega,\omega,\omega)/\sigma_{0}^{(3)}$ (a) near the triple-photon absorption edge $3\hbar\Omega\simeq 2E_F$, and (b) near the double-photon absorption edge $2\hbar\Omega\simeq 2E_F$, in the relaxation-free limit $\Gamma=0$ and at a finite $\Gamma=0.01$. In the panels (c) and (d) the exact solution (\ref{sigma3resultfulldimless}) is compared to the approximation (\ref{SwwwApprox23}) at (c) $\Gamma=0.005$ and $\Gamma=0.001$. The width of the resonance decreases, its height increases, and the accuracy of the approximation (\ref{SwwwApprox23}) improves when $\Gamma$ tends to zero.
\label{fig:3Wnear23}}
\end{figure}

The most interesting feature in Figures \ref{fig:3Wcontribs} and \ref{fig:3WSipe} is the case near the triple-photon absorption edge at $\Omega=2/3$ (it was not discussed in detail in Ref. \cite{Cheng15}). The behavior of the third-order conductivity near this point, at small values of $\Gamma$ and especially in the limit $\Gamma\to 0$, is rather nontrivial. To illustrate this we show in Figure \ref{fig:3Wnear23} the total dimensionless third-order conductivity ${\cal S}^{(3)}_{xxxx}(\Omega,\Omega,\Omega)$ (the sum of all four contributions) near the point $\Omega=2/3$ and, for comparison, near the point $\Omega=1$. The real and imaginary parts of this function are shown at $\Gamma=0$ (the relaxation-free limit, black curves) and at $\Gamma=0.01$ (red curves). Consider, first, the vicinity of the point $\Omega=1$ (the two-photon absorption edge), Figure \ref{fig:3Wnear23}(b). If $\Gamma=0$, the real part of ${\cal S}^{(3)}_{xxxx}(\Omega,\Omega,\Omega)$ has a step-like feature near the point $\Omega=1$, while the imaginary part -- a logarithmic singularity, similar to those which are the case in the linear response conductivity, Fig. \ref{fig:sigma-1}. At the finite value of $\Gamma=0.01$ these singularities are smeared out, as one would expect. Near the point $\Omega=2/3$ the behavior of the real and imaginary parts of ${\cal S}^{(3)}_{xxxx}(\Omega,\Omega,\Omega)$ is completely different. At $\Gamma=0$ one sees, again, the step-like and logarithmic singularities in the real and imaginary parts, respectively. At $\Gamma=0.01$, however, these singularities are not smeared out but an additional, much stronger, resonance appears near $\Omega=2/3$. To understand this behavior we have analyzed the general formulas (\ref{sigma3resultfulldimless}) -- (\ref{FuncS03}) and found that near the point $\Omega=\Omega_0=2/3$, apart from the logarithmic singularity, the function ${\cal S}^{(3)}_{xxxx}(\Omega,\Omega,\Omega)$ has the contribution
\be 
{\cal S}^{(3)}_{xxxx}(\Omega,\Omega,\Omega) \approx
-\frac {3}{32}
\frac {\Gamma}{( \Omega-\Omega_0+i\Gamma/3)^2},\ \ \textrm{ at }|\Omega-2/3|\lesssim \Gamma,\label{SwwwApprox23}
\ee
which results from the second order poles $\sim (1\pm O_{123})^{-2}$ in (\ref{FuncS21}) [and $\sim (1\pm c)^{-2}$ in (\ref{IntI1abc})]. How well this approximation reproduces the exact expression for ${\cal S}^{(3)}_{xxxx}(\Omega,\Omega,\Omega)$ is illustrated in Figures \ref{fig:3Wnear23}(c),(d). One sees that the smaller $\Gamma$, the closer is the approximation (\ref{SwwwApprox23}) to the exact expression, and the larger is the resonance. When $\Gamma$ decreases, the width of the resonance tends to zero while its height tends to infinity since exactly at $\Omega=2/3$  ${\cal S}^{(3)}_{xxxx}(\Omega,\Omega,\Omega) \propto 1/\Gamma$.  

In Ref. \cite{Cheng15} the authors argued that their results (obtained at finite values of the relaxation parameters) recover those of Ref. \cite{Cheng14a} in the limit $\Gamma_e,\Gamma_i\to 0$. This is however valid only at frequencies far from the resonance (\ref{SwwwApprox23}). As seen from the above discussion the resonant contribution to ${\cal S}^{(3)}_{xxxx}(\Omega,\Omega,\Omega)$, Eq. (\ref{SwwwApprox23}), could not be obtained in the relaxation-free theory \cite{Cheng14a}, since the relaxation parameter $\Gamma$ was set to zero from the very beginning. It should be emphasized that, since in real experimental systems the $\Gamma$ parameter is small ($\Gamma\ll 1$) but \textit{always finite}, the contribution (\ref{SwwwApprox23}) \textit{is dominant} around the frequency $\Omega\simeq 2/3$. The position of this resonance depends on the Fermi energy, therefore this resonance can be used for creation of efficient, resonant, voltage tunable optoelectronic devices operating in the mid- and near-infrared spectral range.

\begin{figure}
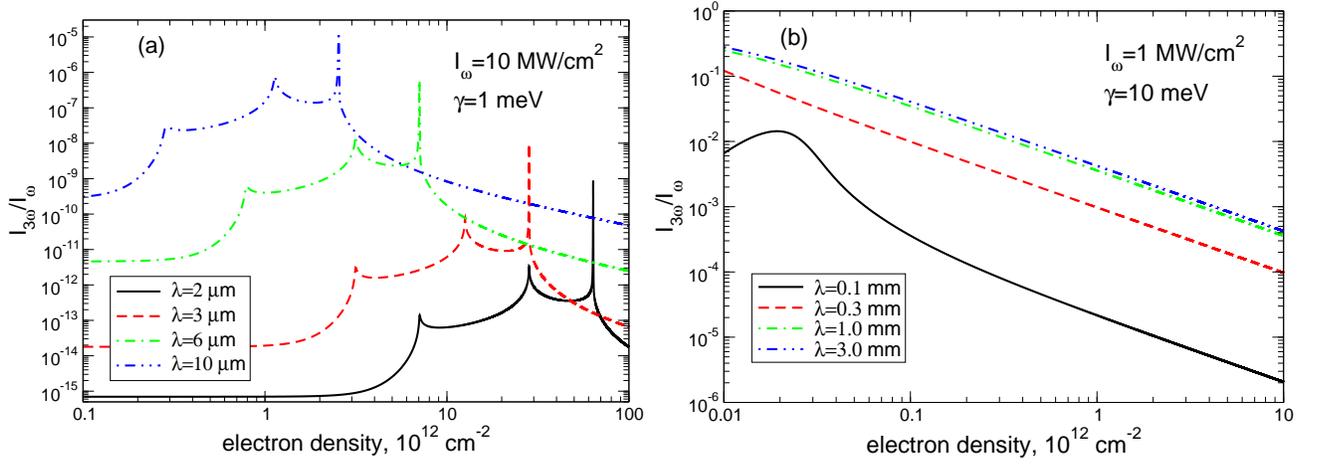

\includegraphics[width=8.5cm]{fig7a.eps}
\includegraphics[width=8.5cm]{fig7b.eps}
\caption{The up-conversion efficiency $I_{3\omega}/I_\omega$, Eq. (\ref{IdivI}), as a function of the electron density, in the range of (a) infrared and (b) terahertz frequencies. $\lambda$ is the wavelength of the \textit{incident} wave. The resonances in (a) correspond to $\Omega=2$, $1$, and $2/3$ (from left to right).
\label{fig:IdivI}}
\end{figure}

Now consider the up-conversion efficiency of a single graphene layer at low and high frequencies. In Figure \ref{fig:IdivI} we show the intensity $I_{3\omega}$ of the third harmonic, normalized to the intensity of the incident wave $I_\omega$, Eq. (\ref{IdivI}), as a function of the electron density. At infrared frequencies, Figure \ref{fig:IdivI}(a), the up-conversion efficiency is of order of $10^{-5}-10^{-9}$, dependent on the wavelength, at the chosen set of parameters ($I_\omega=10$ MW/cm$^2$, $\hbar\gamma=1$ meV) and resonantly increases around the points $\hbar\omega/E_F=2$, $1$, and $2/3$. These resonances are very narrow and sharp which can be used for a fine electrical tuning of the graphene based frequency multipliers. At terahertz frequencies, Figure \ref{fig:IdivI}(b), the up-conversion efficiency is substantially higher (for this graph we have chosen a ten times lower input power density $I_\omega=1$ MW/cm$^2$ and a ten times larger relaxation rate $\hbar\gamma=10$ meV) and smoothly depends on the electron density.

The results obtained in this Section are valid for an isolated graphene layer freely hanging in air. If graphene lies on a dielectric substrate metalized from the back side the intensity of the third harmonic emitted from such a structure can be more than two orders of magnitude larger, see Ref. \cite{Savostianova15}. 

\subsection{Absorption saturation\label{Kerr}}

\begin{figure}
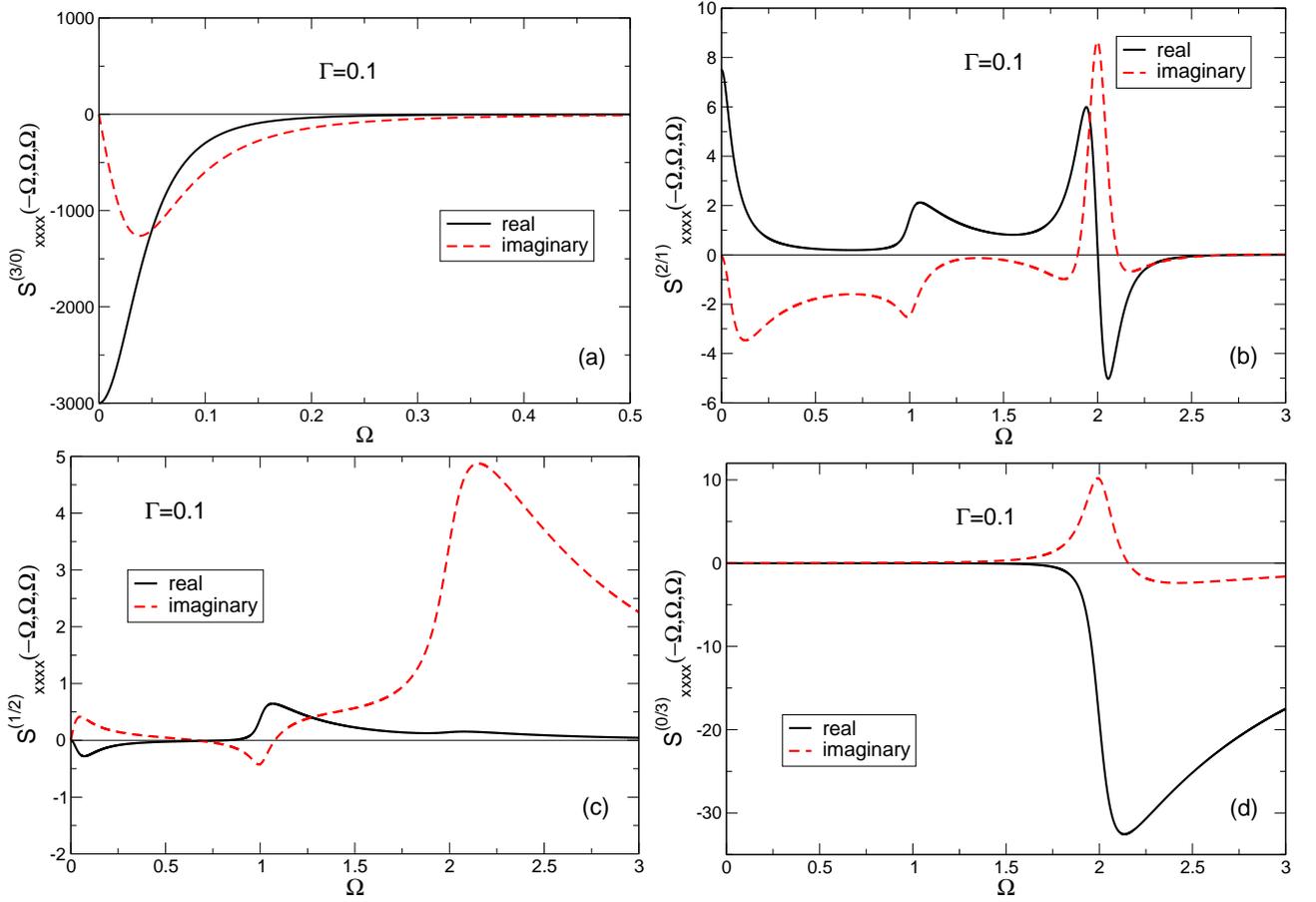

\includegraphics[width=8.5cm]{fig8a.eps}
\includegraphics[width=8.5cm]{fig8b.eps}
\includegraphics[width=8.5cm]{fig8c.eps}
\includegraphics[width=8.5cm]{fig8d.eps}
\caption{Four contributions to the third order conductivity $\sigma_{xxxx}^{(3)} (-\omega,\omega,\omega)/\sigma_{0}^{(3)}$: (a) $(3/0)$ contribution, (b) $(2/1)$ contribution, (c) $(1/2)$ contribution, (d) $(0/3)$ contribution. The effective relaxation rate is $\Gamma=0.1$. \label{fig:Wcontribs}}
\end{figure}

The third-order response of graphene at the frequency of the incident wave $\omega$ is given by the terms in Eq. (\ref{induced3WcurrentFull}) proportional to $e^{-i\omega t}$. Using the symmetry of $\sigma^{(3)}_{xxxx}$ we get
\be
j_{x,\omega}^{(3)}(t)=
\frac {3}8E_0^3
 \sigma^{(3)}_{xxxx}(-\omega,\omega,\omega)
e^{-i\omega t} 
+ \textrm{c.c.}
.\label{inducedWcurrent}
\ee
Figure \ref{fig:Wcontribs} shows four contributions to the third order conductivity ${\cal S}_{xxxx}^{(3)} (-\Omega,\Omega,\Omega)$. One sees that, at low frequencies, $\Omega\lesssim\Gamma$, the largest contribution is, again, the classical one, $(3/0)$. It is of the same order of magnitude as the function ${\cal S}_{xxxx}^{(3)}(\Omega,\Omega,\Omega)$, compare with Figure \ref{fig:3Wcontribs}. At higher frequencies, $\Omega\gtrsim 1$, the largest values of ${\cal S}_{xxxx}^{(3)} (-\Omega,\Omega,\Omega)$ are achieved at the condition $\Omega\approx 2$ which corresponds, again, to the second-order pole at $\Omega_1+\Omega_2+\Omega_3= 2$. The $(2/1)$, $(1/2)$, and $(0/3)$ contributions are of the same order of magnitude and are much larger than the corresponding contributions to ${\cal S}_{xxxx}^{(3)} (\Omega,\Omega,\Omega)$. The term ${\cal S}_{xxxx}^{(2/1)} (-\Omega,\Omega,\Omega)$ has the second-order pole at $\Omega\simeq 2$, see Figure \ref{fig:Wcontribs}(b); the $(1/2)$ and $(0/3)$ terms have logarithmic singularities near this point. 

\begin{figure}
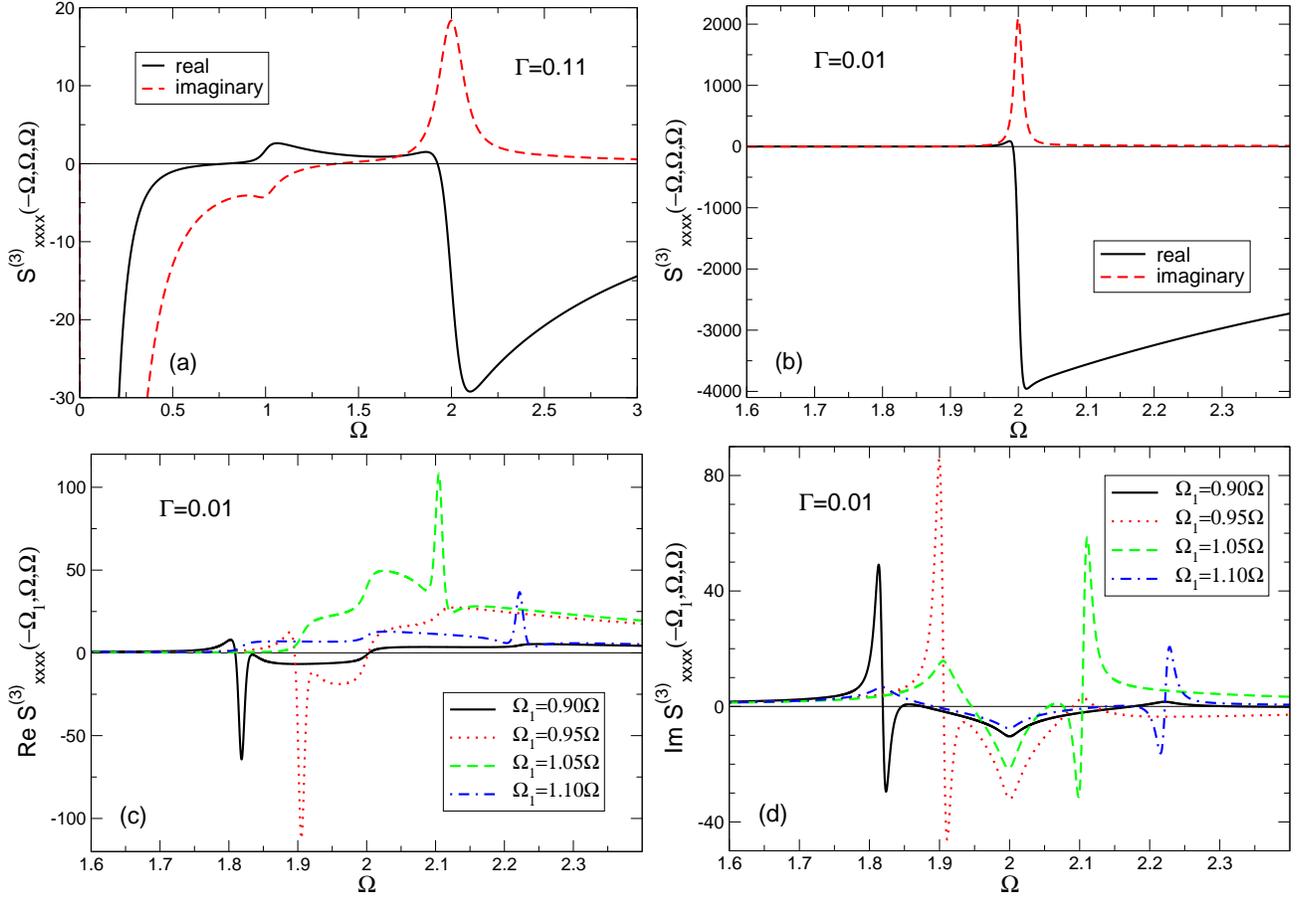

\includegraphics[width=8.5cm]{fig9a.eps}
\includegraphics[width=8.5cm]{fig9b.eps}\\
\includegraphics[width=8.5cm]{fig9c.eps}
\includegraphics[width=8.5cm]{fig9d.eps}
\caption{The third-order dimensionless conductivity ${\cal S}_{xxxx}^{(3)} (-\Omega,\Omega,\Omega)$ as a function of the frequency $\Omega$ (a) under the conditions of Figure 4(b) from Ref. \cite{Cheng15} ($\Gamma=0.11$) and (b) in the vicinity of the most important feature at $\Omega\simeq 2$ and $\Gamma=0.01$.  The real (c) and imaginary (d) parts of the function ${\cal S}_{xxxx}^{(3)} (-\Omega_1,\Omega,\Omega)$ for $\Omega_1=0.9\Omega$, $0.95\Omega$, $1.05\Omega$ and $1.1\Omega$. One sees that the large (amplitude $\simeq 4000$) resonant singularity in (a) is a consequence of the crossing and interaction of three smaller (amplitudes $\simeq 100$) resonances: the second-order pole at $2\Omega-\Omega_1=2$, and two logarithmic singularities at $\Omega=2$ and $\Omega_1=2$.
\label{fig:-W1WW}}
\end{figure}

Figure \ref{fig:-W1WW}(a) shows the total third-order dimensionless conductivity ${\cal S}_{xxxx}^{(3)} (-\Omega,\Omega,\Omega)$ as a function of the frequency $\Omega$ under the conditions of Figure 4(b) from Ref. \cite{Cheng15} (at $\Gamma=0.11$ which corresponds to the relaxation rate 33 meV and the Fermi energy 0.3 eV used in Ref. \cite{Cheng15}). The figures agree with each other. In Figure \ref{fig:-W1WW}(b) we show the vicinity of the most important feature at $\Omega\simeq 2$ and at $\Gamma=0.01$. 
The frequency dependence of the real and imaginary parts of ${\cal S}_{xxxx}^{(3)} (-\Omega,\Omega,\Omega)$ is quite unusual. This is a consequence of the crossing and interaction of three resonant singularities, a second order pole and two logarithmic singularities. This is illustrated in Figures \ref{fig:-W1WW}(b,c). In these graphs we show the real (Figure \ref{fig:-W1WW}(b)) and imaginary (Figure \ref{fig:-W1WW}(c)) parts of the function ${\cal S}_{xxxx}^{(3)} (-\Omega_1,\Omega,\Omega)$ where the first argument $-\Omega_1$ slightly differs from $-\Omega$. Consider, for example, the black curves in Figures \ref{fig:-W1WW}(b,c) which correspond to $\Omega_1=0.9\Omega$. One sees that these curves have two weak logarithmic singularities (a step-like behavior in the real part and a weak resonance in the imaginary part) at the frequencies $\Omega=2$ and $\Omega\approx 2.22$ (the latter corresponds to $\Omega_1=0.9\Omega=2$). In addition, they have a strong second-order pole at $\Omega\approx 1.82$, which corresponds to the three-photon absorption frequency $2\Omega-\Omega_1=2$, i.e., $\Omega=2/1.1\approx 1.82$. All these resonances have a small amplitude $(\lesssim 40)$. When $\Omega_1$ approaches $\Omega$, see the red dotted curves for $\Omega_1=0.95\Omega$, the pole is shifted to the point $\Omega=2/1.05\approx 1.9$ and the logarithmic singularity $\Omega_1=2$ to the point $\Omega=2/0.95\approx 2.1$. Since all three resonances get closer their amplitude increases. If $\Omega_1=\Omega$, Figure \ref{fig:-W1WW}(a), all three singularities meet in one point $\Omega=2$ and three small resonances (with the amplitudes $\lesssim 100$) create a large resonant structure with the amplitude $\simeq 2000 - 4000$. When $\Omega_1$ becomes larger than $\Omega$ ($=1.05$, $1.1$; green and blue curves in Figure \ref{fig:-W1WW}(b,c)) the three resonances move away from each other and their amplitudes are reduced again. 

\begin{figure}
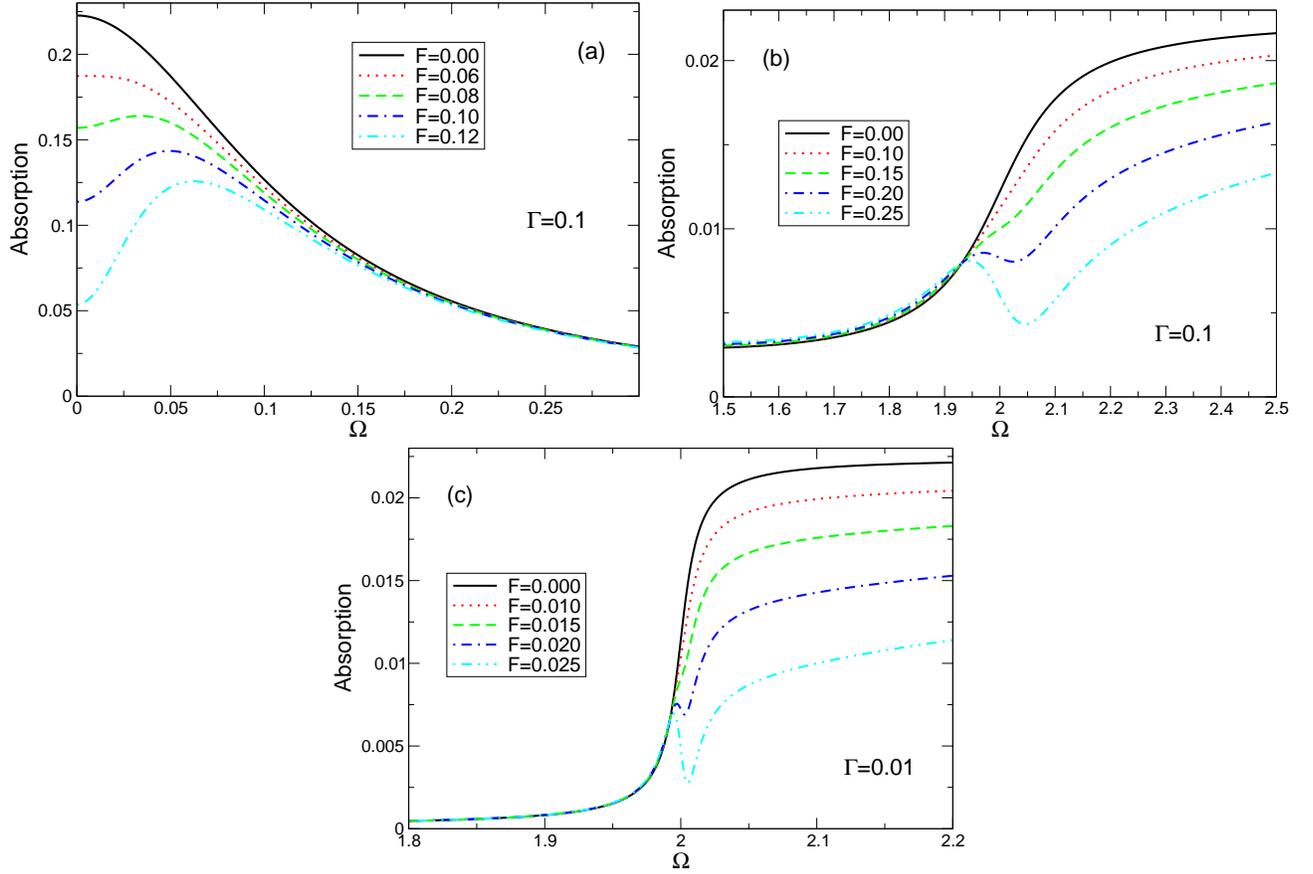

\includegraphics[width=8.5cm]{fig10a.eps}
\includegraphics[width=8.5cm]{fig10b.eps}
\includegraphics[width=8.5cm]{fig10c.eps}
\caption{The dimensionless absorption coefficient (\ref{absorp}) in a single (isolated) graphene layer at (a) low and (b) high frequencies, $\Gamma=0.1$, and (c) high frequencies at $\Gamma=0.01$. The curves are labeled by the value of the dimensionless electric field $F=eE_0/E_Fk_F$. \label{fig:satabs}}
\end{figure}

The third-order current contribution (\ref{inducedWcurrent}) can be added to the linear-response current, Eq. (\ref{sigma1}), to get the electric-field dependent response of graphene at the frequency $\omega$. In up to the third order in the electric field  we have
\be
j_{x,\omega}(t)\approx 
\frac 12E_0
\left[\sigma_{xx}^{(1)}(\omega)
+
\frac 34E_0^2
\sigma_{xxxx}^{(3)}(-\omega,\omega,\omega)\right]
e^{-i\omega t}+\textrm{c.c.}\equiv 
\frac 12
\sigma_{xx}(\omega,E_0)E_0
e^{-i\omega t}+\textrm{c.c.}.
\ee
The corresponding field-dependent dynamic conductivity is then
\ba 
\sigma_{xx}(\omega,E_0)&=& \sigma_{xx}^{(1)}(\omega)
+\frac 34E_0^2
\sigma_{xxxx}^{(3)}(-\omega,\omega,\omega)
\nonumber \\ &=&
\sigma_{0}^{(1)}\left[
{\cal S}^{(1)}_{xx}(\Omega)
+\frac {3}{4\pi}\left(\frac {eE_0}{ E_Fk_F}\right)^2
{\cal S}^{(3)}_{xxxx}(-\Omega,\Omega,\Omega)\right]
\label{field-depdt-conduct}
\ea
where the functions ${\cal S}^{(1)}_{\alpha\beta}(\Omega)$ and ${\cal S}^{(3)}_{\alpha\beta\gamma\delta}(\Omega_1,\Omega_2,\Omega_3)$ are defined in (\ref{sigma1dimless}) -- (\ref{inter-cond}) and (\ref{sigma3resultfulldimless}), (\ref{FuncS30} -- (\ref{FuncS03}), respectively. 
The dimensionless field parameter $F=eE_0/ E_Fk_F$ in (\ref{field-depdt-conduct}) is the work of the field, performed on 2D electrons at the length $\sim k_F^{-1}$, divided by their average (Fermi) energy. Figure \ref{fig:satabs} shows the absorption coefficient 
\be 
A(\omega,E_0)=\frac {4\pi \sigma_{xx}'(\omega,E_0)/c}{|1+2\pi\sigma_{xx}(\omega,E_0)/c|^2}
\label{absorp}
\ee
of an isolated graphene layer with the field-dependent conductivity (\ref{field-depdt-conduct}) at low ($\Omega\ll 1$) and high ($\Omega\simeq 2$) frequencies ($\sigma'$ is the real part of $\sigma$; notice that the self-consistent-field effect is taken into account here). The absorption is seen to be substantially suppressed at the external electric field parameter exceeding a certain value which depends on the scattering rate $\Gamma$. If $\Gamma\simeq 0.1$, the absorption is noticeably suppressed at $F\lesssim 0.05-0.1$. At the electron density $n_s\simeq 10^{12}$ cm$^{-2}$ this corresponds to the electric fields 
\be 
E_{0}\simeq (0.05-0.1)\frac { \hbar v_F \pi n_s}{e}
\simeq (10-20)\textrm{ kV/cm.}
\ee
If $\Gamma$ is ten times smaller, Figure \ref{fig:satabs}(c), the required electric field is reduced by, roughly, one order of magnitude. It is worth noting that at small values of $\Gamma$ the resonance near $\Omega\simeq 2$ is very narrow which can be used for a precise control of the saturable absorption of infrared-laser mirrors, cf. \cite{Popa10,Popa11}. 

\section{Summary and Conclusions\label{sec:conclusion}}

We have presented a quantum theory of the third-order nonlinear electrodynamic response of graphene. The obtained results can be used for analysis of a large number of different nonlinear phenomena, such as the harmonics generation, the four wave mixing, the current induced sum and difference generation, and other.  

The results show that there are two main frequency ranges where the nonlinear effects in graphene are especially large and therefore interesting for applications. The first range corresponds to low (microwave, terahertz) frequencies described by the condition $\Omega\lesssim \Gamma$. The second range corresponds to the three-photon absorption-edge resonance at high (infrared) frequencies $(\Omega_1+\Omega_2+\Omega_3-2)\lesssim\Gamma$. There also exist two-photon and one-photon absorption-edge singularities which are, however, much weaker. The positions of high-frequency resonances depend on the electron density and can therefore be electrically tuned in a broad frequency range, thus opening interesting opportunities for electrically tunable device applications.

The meaning of the relaxation parameter $\Gamma$ may be different in these frequency ranges. At low frequencies $\Gamma$ corresponds to the \textit{intra-band} (Drude) scattering rate, $\Gamma\simeq \hbar/\tau$. In currently available graphene samples the scattering time $\tau$ is about $0.1-1$ ps, so that at low frequencies the strong nonlinear phenomena in graphene are to be expected in the technologically important range of microwave and terahertz frequencies $f\lesssim 10$ THz. At infrared frequencies the width of the three-photon absorption resonance is determined by the \textit{inter-band} relaxation  parameter $\Gamma$, which may differ from the intra-band one. We do not analyze the resonance linewidths further in this work since it would go far beyond its scope. The $\Gamma$ parameters introduced here should just be treated as phenomenological quantities to be extracted from the experiment. Their dependence on the electron density, temperature, excitation power (the electron heating effects) may be investigated in future when enough experimental data will be accumulated. 

Our work removes contradictions between results of different theories which have been published so far. This work corrects our results published in Ref. \cite{Mikhailov14c} and is now in agreement with the relaxation-free theory of Ref. \cite{Cheng14a} (with the reservation concerning the second-order-pole contribution (\ref{SwwwApprox23}) discussed in Section \ref{sec:THG}) and with the full theory of Ref. \cite{Cheng15} whose analytical formulas have been recently corrected in Ref. \cite{Cheng16err}. 

A comparison with experimental data is quite difficult at the present stage. There are several reasons for that. First, most of the experiments have been done at a single (fixed) frequency or in a narrow frequency range, often at the optical frequencies. The value of the Fermi energy and the effective scattering rate $\gamma$ is usually unknown in such publications, which impedes the quantitative comparison with the theory. Second, in experiments graphene usually lies on a substrate and is sometimes covered by another dielectric material, while the theories treat a single graphene layer freely hanging in air. It was shown in Ref. \cite{Savostianova15} that even in a simple structure with graphene lying on a dielectric substrate with the metalized back side the third-order response can be both several order of magnitude larger and several orders of magnitude smaller than in the isolated graphene layer. A precise knowledge of graphene and substrate properties is thus needed for a quantitative comparison of the theory and experiments. Third, many nonlinear experiments are made in the pulsed-excitation regime, with the power density level $\sim$GW/cm$^2$, which lies far beyond the applicability area of the perturbation theory used in our work and in other papers \cite{Cheng14a,Cheng15}. The nonlinear electrodynamics of graphene is thus still at the initial stage, and further experimental and theoretical studies of this interesting area of the fundamental and applied physics is highly desirable. 

\acknowledgments

The author thanks Yuri Kivshar and Mikhail Glazov for interesting discussions of theoretical issues, Nadja Savostianova for assistance, and Costantino de Angelis, Shakil Awan, Giulio Cerullo, J\'er\^ome Faist, Andrea Ferrari, Jonathan Finley, Alexander Grigorenko, Samuel Ver Hoeye, Junichiro Kono, Antonio Lombardo, Geoffrey Nash, Daniel Popa and Hua Qin for interest and useful discussions related to possible experimental observations of the predicted effects. I would also like to thank an anonymous referee for the help in identifying mistakes in the first version of the manuscript. The work was supported by the European Union under the Program Graphene Flagship (No. CNECT-ICT-604391).

\appendix

\section{Some useful formulas\label{app:usefulformulas}}

Some definitions and formulas used in the main text are presented here. 

The two-dimensional Levi-Civita symbol is defined as
\be 
\epsilon_{\alpha\beta}=
\left(
\begin{array}{cc}
0 & 1 \\ -1 & 0 \\
\end{array}
\right)\label{Levi}.
\ee
It satisfies the following relations useful by calculating the integrals in Appendix \ref{app:conduct}: $ 
\epsilon_{\alpha\mu} \epsilon_{\beta\nu} =
\delta_{\alpha\beta}\delta_{\mu\nu}- \delta_{\alpha\nu}\delta_{\mu\beta}$,
 $ 
\epsilon_{\alpha\mu} \epsilon_{\beta\nu} \delta_{\mu\nu}=\epsilon_{\alpha\nu} \epsilon_{\beta\nu} = \delta_{\alpha\beta}$,
\be 
\epsilon_{\alpha\lambda}
\epsilon_{\beta\mu}
\epsilon_{\gamma\nu}
\epsilon_{\delta\sigma}
\Big(\delta_{\lambda\mu}\delta_{\nu \sigma} +\delta_{\lambda\nu}\delta_{\mu\sigma} 
+\delta_{\lambda\sigma}\delta_{\mu\nu}\Big) =
\delta_{\alpha\beta}\delta_{\gamma\delta}
+\delta_{\alpha\gamma} \delta_{\beta\delta}
+\delta_{\alpha\delta} \delta_{\beta\gamma} .\label{4-eps-delta-identity}
\ee

The integration over the polar angle in the ${\bf k}$-space is performed with the help of the formulas
\be 
\langle k_\alpha k_\beta\rangle\equiv \frac 1{2\pi}\int_{-\pi}^\pi k_\alpha k_\beta d\phi=\frac {k^2}2 \delta_{\alpha\beta},\label{average2}
\ee

\be 
\langle k_\alpha k_\beta k_\gamma k_\delta\rangle \equiv \frac 1{2\pi}\int_{-\pi}^\pi k_\alpha k_\beta k_\gamma k_\delta d\phi= \frac {k^4}8
\Big(
\delta_{\alpha\beta}\delta_{\gamma \delta}+
\delta_{\alpha\gamma }\delta_{\beta\delta}+
\delta_{\alpha\delta}\delta_{\beta\gamma}  
\Big) .\label{average4}
\ee

\section{Derivation of formulas (\ref{three-currents}) -- (\ref{inter3}) \label{app:deriv-current}}

To simplify the current expression (\ref{current3}) we first perform the following permutations of the integration variables. In the last line of Eq. (\ref{current3}) we replace $({\bf q}_2\omega_2 {\bf r}_2) \leftrightarrow ({\bf q}_1\omega_1 {\bf r}_1)$ and then $({\bf q}_3\omega_3 {\bf r}_3) \leftrightarrow ({\bf q}_2\omega_2 {\bf r}_2)$. This gives: 
\ba 
{\bf j}^{(3)}({\bf r}_0,t)&=&
\int_{-\infty}^\infty d\omega_1 
\int_{-\infty}^\infty d\omega_2  
\int_{-\infty}^\infty d\omega_3 
\frac {e^4}{2S}\sum_{{\bf \tilde q}{\bf q}_1{\bf q}_2{\bf q}_3} 
\phi_{{\bf q}_1\omega_1}\phi_{{\bf q}_2\omega_2}\phi_{{\bf q}_3\omega_3}
\sum_{\lambda\lambda'} 
\frac {\langle\lambda'|\{{\bf \hat v},e^{-i{\bf \tilde q}\cdot{\bf r}}\}_+|\lambda\rangle e^{i{\bf \tilde q}\cdot{\bf r}_0-i(\omega_1 +\omega_2+\omega_3) t}}
{E_{\lambda'}-E_{\lambda}+\hbar(\omega_1 +\omega_2+\omega_3) +i\hbar \gamma}
\nonumber \\ &\times&
\Bigg[
\sum_{\lambda''\lambda'''}
\frac{\langle\lambda| e^{i{\bf q}_3\cdot{\bf r}_3}|\lambda'''\rangle
\langle\lambda'''|e^{i{\bf q}_2\cdot{\bf r}_2} |\lambda''\rangle 
\langle\lambda''| e^{i{\bf q}_1\cdot{\bf r}_1}  |\lambda'\rangle }
{E_{\lambda'}-E_{\lambda'''} + \hbar (\omega_1 +\omega_2) +i\hbar\gamma }
\frac {f_{\lambda'}-f_{\lambda''}}
{E_{\lambda'}-E_{\lambda''}+\hbar\omega_1+i\hbar\gamma} 
\nonumber \\ &-&
\sum_{\lambda''\lambda'''}
\frac{\langle\lambda| e^{i{\bf q}_3\cdot{\bf r}_3}|\lambda'''\rangle
\langle\lambda'''|e^{i{\bf q}_2\cdot{\bf r}_2} |\lambda''\rangle 
\langle\lambda''| e^{i{\bf q}_1\cdot{\bf r}_1}  |\lambda'\rangle }
{E_{\lambda'}-E_{\lambda'''} + \hbar (\omega_1 +\omega_2) +i\hbar\gamma }
\frac {f_{\lambda''}-f_{\lambda'''}}
{E_{\lambda''}-E_{\lambda'''}+\hbar\omega_2+i\hbar\gamma} 
\nonumber \\ &-&
\sum_{\lambda''\lambda'''}
\frac{\langle\lambda| e^{i{\bf q}_2\cdot{\bf r}_2}|\lambda'''\rangle
\langle\lambda'''|e^{i{\bf q}_1\cdot{\bf r}_1} |\lambda''\rangle 
\langle\lambda''| e^{i{\bf q}_3\cdot{\bf r}_3}  |\lambda'\rangle }
{E_{\lambda''}-E_{\lambda} + \hbar (\omega_1 +\omega_2) +i\hbar\gamma }
\frac {f_{\lambda''}-f_{\lambda'''}}
{E_{\lambda''}-E_{\lambda'''}+\hbar\omega_1+i\hbar\gamma} 
\nonumber \\ &+&
\sum_{\lambda''\lambda'''}
\frac{\langle\lambda| e^{i{\bf q}_2\cdot{\bf r}_2}|\lambda'''\rangle
\langle\lambda'''|e^{i{\bf q}_1\cdot{\bf r}_1} |\lambda''\rangle 
\langle\lambda''| e^{i{\bf q}_3\cdot{\bf r}_3}  |\lambda'\rangle }
{E_{\lambda''}-E_{\lambda} + \hbar (\omega_1 +\omega_2) +i\hbar\gamma }
\frac {f_{\lambda'''}-f_{\lambda}}
{E_{\lambda'''}-E_{\lambda}+\hbar\omega_2+i\hbar\gamma} 
\Bigg] .
\ea 
Then in the second and forth lines in the square parenthesis we replace $({\bf q}_2\omega_2 {\bf r}_2) \leftrightarrow({\bf q}_1\omega_1 {\bf r}_1)$: 
\ba 
{\bf j}^{(3)}({\bf r}_0,t)&=&
\int_{-\infty}^\infty d\omega_1 
\int_{-\infty}^\infty d\omega_2  
\int_{-\infty}^\infty d\omega_3 
\frac {e^4}{2S}\sum_{{\bf \tilde q}{\bf q}_1{\bf q}_2{\bf q}_3} 
\phi_{{\bf q}_1\omega_1}\phi_{{\bf q}_2\omega_2}\phi_{{\bf q}_3\omega_3}
\sum_{\lambda\lambda'} 
\frac {\langle\lambda'|\{{\bf \hat v},e^{-i{\bf \tilde q}\cdot{\bf r}}\}_+|\lambda\rangle e^{i{\bf \tilde q}\cdot{\bf r}_0-i(\omega_1 +\omega_2+\omega_3) t}}
{E_{\lambda'}-E_{\lambda}+\hbar(\omega_1 +\omega_2+\omega_3) +i\hbar \gamma}
\nonumber \\ &\times&
\Bigg[
\sum_{\lambda''\lambda'''}
\frac{\langle\lambda| e^{i{\bf q}_3\cdot{\bf r}_3}|\lambda'''\rangle
\langle\lambda'''|e^{i{\bf q}_2\cdot{\bf r}_2} |\lambda''\rangle 
\langle\lambda''| e^{i{\bf q}_1\cdot{\bf r}_1}  |\lambda'\rangle }
{E_{\lambda'}-E_{\lambda'''} + \hbar (\omega_1 +\omega_2) +i\hbar\gamma }
\frac {f_{\lambda'}-f_{\lambda''}}
{E_{\lambda'}-E_{\lambda''}+\hbar\omega_1+i\hbar\gamma} 
\nonumber \\ &-&
\sum_{\lambda''\lambda'''}
\frac{\langle\lambda| e^{i{\bf q}_3\cdot{\bf r}_3}|\lambda'''\rangle
\langle\lambda'''|e^{i{\bf q}_1\cdot{\bf r}_1} |\lambda''\rangle 
\langle\lambda''| e^{i{\bf q}_2\cdot{\bf r}_2}  |\lambda'\rangle }
{E_{\lambda'}-E_{\lambda'''} + \hbar (\omega_1 +\omega_2) +i\hbar\gamma }
\frac {f_{\lambda''}-f_{\lambda'''}}
{E_{\lambda''}-E_{\lambda'''}+\hbar\omega_1+i\hbar\gamma} 
\nonumber \\ &-&
\sum_{\lambda''\lambda'''}
\frac{\langle\lambda| e^{i{\bf q}_2\cdot{\bf r}_2}|\lambda'''\rangle
\langle\lambda'''|e^{i{\bf q}_1\cdot{\bf r}_1} |\lambda''\rangle 
\langle\lambda''| e^{i{\bf q}_3\cdot{\bf r}_3}  |\lambda'\rangle }
{E_{\lambda''}-E_{\lambda} + \hbar (\omega_1 +\omega_2) +i\hbar\gamma }
\frac {f_{\lambda''}-f_{\lambda'''}}
{E_{\lambda''}-E_{\lambda'''}+\hbar\omega_1+i\hbar\gamma} 
\nonumber \\ &+&
\sum_{\lambda''\lambda'''}
\frac{\langle\lambda| e^{i{\bf q}_1\cdot{\bf r}_1}|\lambda'''\rangle
\langle\lambda'''|e^{i{\bf q}_2\cdot{\bf r}_2} |\lambda''\rangle 
\langle\lambda''| e^{i{\bf q}_3\cdot{\bf r}_3}  |\lambda'\rangle }
{E_{\lambda''}-E_{\lambda} + \hbar (\omega_1 +\omega_2) +i\hbar\gamma }
\frac {f_{\lambda'''}-f_{\lambda}}
{E_{\lambda'''}-E_{\lambda}+\hbar\omega_1+i\hbar\gamma} 
\Bigg] .
\ea 
Now in all denominators we have the same frequency factors, $\omega_1+i\gamma$, $\omega_1+\omega_2+i\gamma$ and $\omega_1+\omega_2+\omega_3+i\gamma$. Substituting the matrix elements (\ref{MEexpexpansion}) at small ${\bf q}$ and taking the sums over the spin indexes and over the momenta ${\bf k}'''$, ${\bf k}''$, ${\bf k}'$ (we use the momentum conservation factors $\delta_{{\bf k}'', {\bf k}'+{\bf q}_1}$, etc.) we get
\ba 
{\bf j}^{(3)}({\bf r}_0,t)&=&
\int_{-\infty}^\infty d\omega_1 
\int_{-\infty}^\infty d\omega_2  
\int_{-\infty}^\infty d\omega_3 
\frac {e^4g_s}{2S}\sum_{{\bf q}_1{\bf q}_2{\bf q}_3} 
\phi_{{\bf q}_1\omega_1}\phi_{{\bf q}_2\omega_2}\phi_{{\bf q}_3\omega_3}
e^{i({\bf q}_3+{\bf q}_1+{\bf q}_2)\cdot{\bf r}_0-i(\omega_1 +\omega_2+\omega_3) t}
\nonumber \\ &\times&
\sum_{l l'{\bf k}} 
\frac {\langle l'{\bf k}|\{{\bf \hat v},e^{-i({\bf q}_3+{\bf q}_1+{\bf q}_2)\cdot{\bf r}}\}_+|l,{\bf k}+{\bf q}_3+{\bf q}_1+{\bf q}_2\rangle }
{E_{ l'{\bf k}}-E_{l,{\bf k}+{\bf q}_3+{\bf q}_1+{\bf q}_2}+\hbar(\omega_1 +\omega_2+\omega_3) +i\hbar \gamma}
\nonumber \\ &\times&
\Bigg[
\sum_{ l'' l'''}
\frac{\left(
\delta_{ll'''}-\frac {(-1)^{l+l'''}}2q_{3\delta}\eta^\delta_{{\bf k}+{\bf q}_2+{\bf q}_1}
\right)
\left(
\delta_{l'''l''}-\frac {(-1)^{l'''+l''}}2q_{2\gamma}\eta^\gamma_{{\bf k}+{\bf q}_1}
\right)
\left(
\delta_{l''l'}-\frac {(-1)^{l''+l'}}2q_{1\beta}\eta^\beta_{{\bf k}}
\right) }
{E_{ l'{\bf k}}-E_{ l''',{\bf k}+{\bf q}_1+{\bf q}_2} + \hbar (\omega_1 +\omega_2) +i\hbar\gamma }
\nonumber \\ &\times&
\frac {f_{ l'{\bf k}}-f_{ l'',{\bf k}+{\bf q}_1}}
{E_{l'{\bf k}}-E_{ l'',{\bf k}+{\bf q}_1}+\hbar\omega_1+i\hbar\gamma} 
\nonumber \\ &-&
\sum_{ l'' l'''}
\frac{\left(
\delta_{ll'''}-\frac {(-1)^{l+l'''}}2q_{3\delta}\eta^\delta_{{\bf k}+{\bf q}_2+{\bf q}_1}
\right)
\left(
\delta_{l'''l''}-\frac {(-1)^{l'''+l''}}2q_{1\beta}\eta^\beta_{{\bf k}+{\bf q}_2}
\right)
\left(
\delta_{l''l'}-\frac {(-1)^{l''+l'}}2q_{2\gamma}\eta^\gamma_{{\bf k}}
\right) }
{E_{ l'{\bf k}}-E_{ l''',{\bf k}+{\bf q}_2+{\bf q}_1} + \hbar (\omega_1 +\omega_2) +i\hbar\gamma }
\nonumber \\ &\times&
\frac {f_{ l'',{\bf k}+{\bf q}_2}-f_{ l''',{\bf k}+{\bf q}_2+{\bf q}_1}}
{E_{ l'',{\bf k}+{\bf q}_2}-E_{ l''',{\bf k}+{\bf q}_2+{\bf q}_1}+\hbar\omega_1+i\hbar\gamma} 
\nonumber \\ &-&
\sum_{ l'' l'''}
\frac{\left(
\delta_{ll'''}-\frac {(-1)^{l+l'''}}2q_{2\gamma}\eta^\gamma_{{\bf k}+{\bf q}_3+{\bf q}_1}
\right)
\left(
\delta_{l'''l''}-\frac {(-1)^{l'''+l''}}2q_{1\beta}\eta^\beta_{{\bf k}+{\bf q}_3}
\right)
\left(
\delta_{l''l'}-\frac {(-1)^{l''+l'}}2q_{3\delta}\eta^\delta_{{\bf k}}
\right) }
{E_{ l'',{\bf k}+{\bf q}_3}-E_{l,{\bf k}+{\bf q}_3+{\bf q}_1+{\bf q}_2} + \hbar (\omega_1 +\omega_2) +i\hbar\gamma }
\nonumber \\ &\times&
\frac {f_{ l'',{\bf k}+{\bf q}_3}-f_{ l''',{\bf k}+{\bf q}_3+{\bf q}_1}}
{E_{ l'',{\bf k}+{\bf q}_3}-E_{ l''',{\bf k}+{\bf q}_3+{\bf q}_1}+\hbar\omega_1+i\hbar\gamma} 
\nonumber \\ &+&
\sum_{ l'' l'''}
\frac{\left(
\delta_{ll'''}-\frac {(-1)^{l+l'''}}2q_{1\beta}\eta^\beta_{{\bf k}+{\bf q}_3+{\bf q}_2}
\right)
\left(
\delta_{l'''l''}-\frac {(-1)^{l'''+l''}}2q_{2\gamma}\eta^\gamma_{{\bf k}+{\bf q}_3}
\right)
\left(
\delta_{l''l'}-\frac {(-1)^{l''+l'}}2q_{3\delta}\eta^\delta_{{\bf k}}
\right) }
{E_{ l'',{\bf k}+{\bf q}_3}-E_{l,{\bf k}+{\bf q}_3+{\bf q}_1+{\bf q}_2} + \hbar (\omega_1 +\omega_2) +i\hbar\gamma }
\nonumber \\ &\times&
\frac {f_{ l''',{\bf k}+{\bf q}_3+{\bf q}_2}-f_{l,{\bf k}+{\bf q}_3+{\bf q}_1+{\bf q}_2}}
{E_{ l''',{\bf k}+{\bf q}_3+{\bf q}_2}-E_{l,{\bf k}+{\bf q}_3+{\bf q}_1+{\bf q}_2}+\hbar\omega_1+i\hbar\gamma} 
\Bigg] .\label{fullcurrent}
\ea 
Now consider different contributions to the current. 

\subsection{The contribution $(3/0)$\label{app:30}}

The purely intra-band contribution $(3/0)$ is obtained from Eq. (\ref{fullcurrent}) if in all matrix elements only the Kronecker-symbol terms 
$\delta_{ll'''}\delta_{l'''l''}\delta_{l''l'}$ are taken into account. Taking the sums over $l'''$, $l''$, and $l'$ we get
\ba 
\underbrace{ j_\alpha^{(3)}({\bf r}_0,t)}_{(3/0)}&=&
\int_{-\infty}^\infty d\omega_1 
\int_{-\infty}^\infty d\omega_2  
\int_{-\infty}^\infty d\omega_3 
\frac {e^4g_s}{2S}\sum_{{\bf q}_1{\bf q}_2{\bf q}_3} 
\phi_{{\bf q}_1\omega_1}\phi_{{\bf q}_2\omega_2}\phi_{{\bf q}_3\omega_3}
e^{i({\bf q}_3+{\bf q}_1+{\bf q}_2)\cdot{\bf r}_0-i(\omega_1 +\omega_2+\omega_3) t}
\nonumber \\ &\times&
\sum_{l {\bf k}} 
\frac {\langle l{\bf k}|\{ \hat v_\alpha,e^{-i({\bf q}_3+{\bf q}_1+{\bf q}_2)\cdot{\bf r}}\}_+|l,{\bf k}+{\bf q}_3+{\bf q}_1+{\bf q}_2\rangle }
{E_{ l{\bf k}}-E_{l,{\bf k}+{\bf q}_3+{\bf q}_1+{\bf q}_2}+\hbar(\omega_1 +\omega_2+\omega_3) +i\hbar \gamma}
\nonumber \\ &\times&
\Bigg[
\frac{1 }
{E_{ l{\bf k}}-E_{ l,{\bf k}+{\bf q}_1+{\bf q}_2} + \hbar (\omega_1 +\omega_2) +i\hbar\gamma }
\frac {f_{ l{\bf k}}-f_{ l,{\bf k}+{\bf q}_1}}
{E_{l{\bf k}}-E_{ l,{\bf k}+{\bf q}_1}+\hbar\omega_1+i\hbar\gamma} 
\nonumber \\ &-&
\frac{1}
{E_{ l{\bf k}}-E_{ l,{\bf k}+{\bf q}_2+{\bf q}_1} + \hbar (\omega_1 +\omega_2) +i\hbar\gamma }
\frac {f_{ l,{\bf k}+{\bf q}_2}-f_{ l,{\bf k}+{\bf q}_2+{\bf q}_1}}
{E_{ l,{\bf k}+{\bf q}_2}-E_{ l,{\bf k}+{\bf q}_2+{\bf q}_1}+\hbar\omega_1+i\hbar\gamma} 
\nonumber \\ &-&
\frac{1}
{E_{ l,{\bf k}+{\bf q}_3}-E_{l,{\bf k}+{\bf q}_3+{\bf q}_1+{\bf q}_2} + \hbar (\omega_1 +\omega_2) +i\hbar\gamma }
\frac {f_{ l,{\bf k}+{\bf q}_3}-f_{ l,{\bf k}+{\bf q}_3+{\bf q}_1}}
{E_{ l,{\bf k}+{\bf q}_3}-E_{ l,{\bf k}+{\bf q}_3+{\bf q}_1}+\hbar\omega_1+i\hbar\gamma} 
\nonumber \\ &+&
\frac{1 }
{E_{ l,{\bf k}+{\bf q}_3}-E_{l,{\bf k}+{\bf q}_3+{\bf q}_1+{\bf q}_2} + \hbar (\omega_1 +\omega_2) +i\hbar\gamma }
\frac {f_{ l,{\bf k}+{\bf q}_3+{\bf q}_2}-f_{l,{\bf k}+{\bf q}_3+{\bf q}_1+{\bf q}_2}}
{E_{ l,{\bf k}+{\bf q}_3+{\bf q}_2}-E_{l,{\bf k}+{\bf q}_3+{\bf q}_1+{\bf q}_2}+\hbar\omega_1+i\hbar\gamma} 
\Bigg] .
\ea
One sees that all the Fermi-function differences are proportional to ${\bf q}_1$ and can be presented as
\be 
f_{ l{\bf k}}-f_{ l,{\bf k}+{\bf q}_1}\approx -q_{1\beta}\frac{\p f_{ l{\bf k}}}{\p k_\beta},
\ee
and similarly for the second, third and forth lines in the square parenthesis. Since the factor $q_{1\beta}$ is obtained, in all other places of this formula ${\bf q}_{1}$ can be set to zero. Then the $(3/0)$ current contribution assumes the form
\ba 
\underbrace{ j_\alpha^{(3)}({\bf r}_0,t)}_{(3/0)}&=&
\int_{-\infty}^\infty d\omega_1 
\int_{-\infty}^\infty d\omega_2  
\int_{-\infty}^\infty d\omega_3 
\frac {e^4g_s}{2S}\sum_{{\bf q}_1{\bf q}_2{\bf q}_3} 
\phi_{{\bf q}_1\omega_1}\phi_{{\bf q}_2\omega_2}\phi_{{\bf q}_3\omega_3}
e^{i({\bf q}_3+{\bf q}_1+{\bf q}_2)\cdot{\bf r}_0-i(\omega_1 +\omega_2+\omega_3) t}
\nonumber \\ &\times&
\sum_{l {\bf k}} 
\frac {\langle l{\bf k}|\{ \hat v_\alpha,e^{-i({\bf q}_3+{\bf q}_2)\cdot{\bf r}}\}_+|l,{\bf k}+{\bf q}_3+{\bf q}_2\rangle }
{E_{ l{\bf k}}-E_{l,{\bf k}+{\bf q}_3+{\bf q}_2}+\hbar(\omega_1 +\omega_2+\omega_3) +i\hbar \gamma}\frac {-q_{1\beta}}
{\hbar\omega_1+i\hbar\gamma} 
\nonumber \\ &\times&
\Bigg[
\frac{1 }
{E_{ l{\bf k}}-E_{ l,{\bf k}+{\bf q}_2} + \hbar (\omega_1 +\omega_2) +i\hbar\gamma }
\frac{\p (f_{ l{\bf k}}
- f_{ l,{\bf k}+{\bf q}_2})}{\p k_\beta}
\nonumber \\ &-&
\frac{1}
{E_{ l,{\bf k}+{\bf q}_3}-E_{l,{\bf k}+{\bf q}_3+{\bf q}_2} + \hbar (\omega_1 +\omega_2) +i\hbar\gamma }
\frac{\p (f_{ l,{\bf k}+{\bf q}_3}- f_{ l,{\bf k}+{\bf q}_3+{\bf q}_2})}{\p k_\beta}
\Bigg] .
\ea 
Now one sees that the terms with the Fermi-function differences in the square parenthesis are proportional to the wavevector ${\bf q}_2$ times the second derivative of the Fermi functions. Omitting ${\bf q}_2$ in the rest of the formula we get
\ba 
\underbrace{ j_\alpha^{(3)}({\bf r}_0,t)}_{(3/0)}&\approx&
\int_{-\infty}^\infty d\omega_1 
\int_{-\infty}^\infty d\omega_2  
\int_{-\infty}^\infty d\omega_3 
\frac {e^4g_s}{2S}\sum_{{\bf q}_1{\bf q}_2{\bf q}_3} 
\phi_{{\bf q}_1\omega_1}\phi_{{\bf q}_2\omega_2}\phi_{{\bf q}_3\omega_3}
e^{i({\bf q}_3+{\bf q}_1+{\bf q}_2)\cdot{\bf r}_0-i(\omega_1 +\omega_2+\omega_3) t}
\nonumber \\ &\times&
\sum_{l {\bf k}} 
\frac {\langle l{\bf k}|\{ \hat v_\alpha,e^{-i{\bf q}_3\cdot{\bf r}}\}_+|l,{\bf k}+{\bf q}_3\rangle }
{E_{ l{\bf k}}-E_{l,{\bf k}+{\bf q}_3}+\hbar(\omega_1 +\omega_2+\omega_3) +i\hbar \gamma}
\frac {-q_{1\beta}}{\hbar\omega_1+i\hbar\gamma} 
\frac{-q_{2\gamma} }{\hbar (\omega_1 +\omega_2) +i\hbar\gamma }
\frac{\p^2 (f_{ l{\bf k}}
-f_{ l,{\bf k}+{\bf q}_3})}{\p k_\beta\p k_\gamma}.
\ea 
Finally, expanding the last Fermi-functions difference at small ${\bf q}_3$ and introducing the Fourier components of the electric field, instead of those of the electric potential, ${\bf E}_{{\bf q}\omega}=-i{\bf q}\phi_{{\bf q}\omega}$, we get Eq. (\ref{intra3}).

\subsection{The contribution $(2/1)$}

The contribution $(2/1)$ consists of the terms containing two Kronecker symbols in the matrix elements in (\ref{fullcurrent}). Collecting the corresponding terms we get
\ba 
&&\underbrace{j_\alpha^{(3)}({\bf r}_0,t)}_{(2/1)}=
\int_{-\infty}^\infty d\omega_1 
\int_{-\infty}^\infty d\omega_2  
\int_{-\infty}^\infty d\omega_3 
\frac {e^4g_s}{2S}\sum_{{\bf q}_1{\bf q}_2{\bf q}_3} 
\phi_{{\bf q}_1\omega_1}\phi_{{\bf q}_2\omega_2}\phi_{{\bf q}_3\omega_3}
e^{i({\bf q}_3+{\bf q}_1+{\bf q}_2)\cdot{\bf r}_0-i(\omega_1 +\omega_2+\omega_3) t}
\nonumber \\ &\times&
\sum_{l l'{\bf k}} 
\frac {\langle l'{\bf k}|\{ \hat v_\alpha,e^{-i({\bf q}_3+{\bf q}_1+{\bf q}_2)\cdot{\bf r}}\}_+|l,{\bf k}+{\bf q}_3+{\bf q}_1+{\bf q}_2\rangle }
{E_{ l'{\bf k}}-E_{l,{\bf k}+{\bf q}_3+{\bf q}_1+{\bf q}_2}+\hbar(\omega_1 +\omega_2+\omega_3) +i\hbar \gamma}
\left(-\frac {(-1)^{l+l'}}2\right)
\nonumber \\ &&
\nonumber \\ &\times&
\Bigg[
\sum_{ l'' l'''}
\frac{
\delta_{ll'''}
\delta_{l'''l''}
q_{1\beta}\eta^\beta_{{\bf k}}
 + 
\delta_{ll'''}
q_{2\gamma}\eta^\gamma_{{\bf k}+{\bf q}_1}
\delta_{l''l'} + q_{3\delta}\eta^\delta_{{\bf k}+{\bf q}_2+{\bf q}_1}
\delta_{l'''l''}\delta_{l''l'}}
{E_{ l'{\bf k}}-E_{ l''',{\bf k}+{\bf q}_1+{\bf q}_2} + \hbar (\omega_1 +\omega_2) +i\hbar\gamma }
\frac {f_{ l'{\bf k}}-f_{ l'',{\bf k}+{\bf q}_1}}
{E_{l'{\bf k}}-E_{ l'',{\bf k}+{\bf q}_1}+\hbar\omega_1+i\hbar\gamma} 
\nonumber \\ &-&
\sum_{ l'' l'''}
\frac{
\delta_{ll'''}
\delta_{l'''l''}
q_{2\gamma}\eta^\gamma_{{\bf k}}
 +
\delta_{ll'''}
q_{1\beta}\eta^\beta_{{\bf k}+{\bf q}_2}
\delta_{l''l'}
+q_{3\delta}\eta^\delta_{{\bf k}+{\bf q}_2+{\bf q}_1}
\delta_{l'''l''}
\delta_{l''l'}}
{E_{ l'{\bf k}}-E_{ l''',{\bf k}+{\bf q}_2+{\bf q}_1} + \hbar (\omega_1 +\omega_2) +i\hbar\gamma }
\frac {f_{ l'',{\bf k}+{\bf q}_2}-f_{ l''',{\bf k}+{\bf q}_2+{\bf q}_1}}
{E_{ l'',{\bf k}+{\bf q}_2}-E_{ l''',{\bf k}+{\bf q}_2+{\bf q}_1}+\hbar\omega_1+i\hbar\gamma} 
\nonumber \\ &-&
\sum_{ l'' l'''}
\frac{
\delta_{ll'''}
\delta_{l'''l''}
q_{3\delta}\eta^\delta_{{\bf k}}
 +
\delta_{ll'''}
q_{1\beta}\eta^\beta_{{\bf k}+{\bf q}_3}
\delta_{l''l'}
+q_{2\gamma}\eta^\gamma_{{\bf k}+{\bf q}_3+{\bf q}_1}
\delta_{l'''l''}
\delta_{l''l'}}
{E_{ l'',{\bf k}+{\bf q}_3}-E_{l,{\bf k}+{\bf q}_3+{\bf q}_1+{\bf q}_2} + \hbar (\omega_1 +\omega_2) +i\hbar\gamma }
\frac {f_{ l'',{\bf k}+{\bf q}_3}-f_{ l''',{\bf k}+{\bf q}_3+{\bf q}_1}}
{E_{ l'',{\bf k}+{\bf q}_3}-E_{ l''',{\bf k}+{\bf q}_3+{\bf q}_1}+\hbar\omega_1+i\hbar\gamma} 
\nonumber \\ &+&
\sum_{ l'' l'''}
\frac{
\delta_{ll'''}
\delta_{l'''l''}
q_{3\delta}\eta^\delta_{{\bf k}}
+
\delta_{ll'''}
q_{2\gamma}\eta^\gamma_{{\bf k}+{\bf q}_3}
\delta_{l''l'}+
q_{1\beta}\eta^\beta_{{\bf k}+{\bf q}_3+{\bf q}_2}
\delta_{l'''l''}
\delta_{l''l'} }
{E_{ l'',{\bf k}+{\bf q}_3}-E_{l,{\bf k}+{\bf q}_3+{\bf q}_1+{\bf q}_2} + \hbar (\omega_1 +\omega_2) +i\hbar\gamma }
\frac {f_{ l''',{\bf k}+{\bf q}_3+{\bf q}_2}-f_{l,{\bf k}+{\bf q}_3+{\bf q}_1+{\bf q}_2}}
{E_{ l''',{\bf k}+{\bf q}_3+{\bf q}_2}-E_{l,{\bf k}+{\bf q}_3+{\bf q}_1+{\bf q}_2}+\hbar\omega_1+i\hbar\gamma} 
\Bigg] .
\nonumber \\ 
\label{eq1}
\ea 
Consider different contributions in Eq. (\ref{eq1}). In the first line in the square parenthesis we have the Fermi-functions difference $f_{ l'{\bf k}}-f_{ l'',{\bf k}+{\bf q}_1}$. The prefactor resulting from the matrix elements has two terms containing $\delta_{l''l'}$ and one term which does not contain $\delta_{l''l'}$. The first two correspond to the intra-band contributions, while the last one -- to the inter-band contribution. The same structure can be seen in the second, third and fourth lines of Eq. (\ref{eq1}). Combining all intra- and all inter-band terms together and removing one of the Kronecker symbols in the intra-band sums we get
\ba 
&&\underbrace{j_\alpha^{(3)}({\bf r}_0,t)}_{(2/1)}=
\int_{-\infty}^\infty d\omega_1 
\int_{-\infty}^\infty d\omega_2  
\int_{-\infty}^\infty d\omega_3 
\frac {e^4g_s}{2S}\sum_{{\bf q}_1{\bf q}_2{\bf q}_3} 
\phi_{{\bf q}_1\omega_1}\phi_{{\bf q}_2\omega_2}\phi_{{\bf q}_3\omega_3}
e^{i({\bf q}_3+{\bf q}_1+{\bf q}_2)\cdot{\bf r}_0-i(\omega_1 +\omega_2+\omega_3) t}
\nonumber \\ &\times&
\sum_{l l'{\bf k}} 
\frac {\langle l'{\bf k}|\{ \hat v_\alpha,e^{-i({\bf q}_3+{\bf q}_1+{\bf q}_2)\cdot{\bf r}}\}_+|l,{\bf k}+{\bf q}_3+{\bf q}_1+{\bf q}_2\rangle }
{E_{ l'{\bf k}}-E_{l,{\bf k}+{\bf q}_3+{\bf q}_1+{\bf q}_2}+\hbar(\omega_1 +\omega_2+\omega_3) +i\hbar \gamma}
\left(-\frac {(-1)^{l+l'}}2\right)
\nonumber \\ &\times&
\Bigg\{
\Bigg[
\sum_{  l'''}
\frac{
\delta_{ll'''}q_{2\gamma}\eta^\gamma_{{\bf k}+{\bf q}_1}
 + 
q_{3\delta}\eta^\delta_{{\bf k}+{\bf q}_2+{\bf q}_1}
\delta_{l'''l''}}
{E_{ l'{\bf k}}-E_{ l''',{\bf k}+{\bf q}_1+{\bf q}_2} + \hbar (\omega_1 +\omega_2) +i\hbar\gamma }
\frac {f_{ l'{\bf k}}-f_{ l',{\bf k}+{\bf q}_1}}
{E_{l'{\bf k}}-E_{ l',{\bf k}+{\bf q}_1}+\hbar\omega_1+i\hbar\gamma} 
\nonumber \\ &-&
\sum_{ l'' }
\frac{
\delta_{ll''}
q_{2\gamma}\eta^\gamma_{{\bf k}}
+q_{3\delta}\eta^\delta_{{\bf k}+{\bf q}_2+{\bf q}_1}
\delta_{l''l'}}
{E_{ l'{\bf k}}-E_{ l'',{\bf k}+{\bf q}_2+{\bf q}_1} + \hbar (\omega_1 +\omega_2) +i\hbar\gamma }
\frac {f_{ l'',{\bf k}+{\bf q}_2}-f_{ l'',{\bf k}+{\bf q}_2+{\bf q}_1}}
{E_{ l'',{\bf k}+{\bf q}_2}-E_{ l'',{\bf k}+{\bf q}_2+{\bf q}_1}+\hbar\omega_1+i\hbar\gamma} 
\nonumber \\ &-&
\sum_{ l''}
\frac{
\delta_{ll''}
q_{3\delta}\eta^\delta_{{\bf k}}
+q_{2\gamma}\eta^\gamma_{{\bf k}+{\bf q}_3+{\bf q}_1}
\delta_{l''l'}}
{E_{ l'',{\bf k}+{\bf q}_3}-E_{l,{\bf k}+{\bf q}_3+{\bf q}_1+{\bf q}_2} + \hbar (\omega_1 +\omega_2) +i\hbar\gamma }
\frac {f_{ l'',{\bf k}+{\bf q}_3}-f_{ l'',{\bf k}+{\bf q}_3+{\bf q}_1}}
{E_{ l'',{\bf k}+{\bf q}_3}-E_{ l'',{\bf k}+{\bf q}_3+{\bf q}_1}+\hbar\omega_1+i\hbar\gamma} 
\nonumber \\ &+&
\sum_{ l'' }
\frac{
\delta_{ll''}
q_{3\delta}\eta^\delta_{{\bf k}}
+
q_{2\gamma}\eta^\gamma_{{\bf k}+{\bf q}_3}
\delta_{l''l'} }
{E_{ l'',{\bf k}+{\bf q}_3}-E_{l,{\bf k}+{\bf q}_3+{\bf q}_1+{\bf q}_2} + \hbar (\omega_1 +\omega_2) +i\hbar\gamma }
\frac {f_{ l,{\bf k}+{\bf q}_3+{\bf q}_2}-f_{l,{\bf k}+{\bf q}_3+{\bf q}_1+{\bf q}_2}}
{E_{ l,{\bf k}+{\bf q}_3+{\bf q}_2}-E_{l,{\bf k}+{\bf q}_3+{\bf q}_1+{\bf q}_2}+\hbar\omega_1+i\hbar\gamma} 
\Bigg] 
\nonumber \\ &+&
\Bigg[
\sum_{ l'' l'''}
\frac{
\delta_{ll'''}
\delta_{l'''l''}
q_{1\beta}\eta^\beta_{{\bf k}}
}
{E_{ l'{\bf k}}-E_{ l''',{\bf k}+{\bf q}_1+{\bf q}_2} + \hbar (\omega_1 +\omega_2) +i\hbar\gamma }
\frac {f_{ l'{\bf k}}-f_{ l'',{\bf k}+{\bf q}_1}}
{E_{l'{\bf k}}-E_{ l'',{\bf k}+{\bf q}_1}+\hbar\omega_1+i\hbar\gamma} 
\nonumber \\ &-&
\sum_{ l'' l'''}
\frac{ 
\delta_{ll'''}
q_{1\beta}\eta^\beta_{{\bf k}+{\bf q}_2}
\delta_{l''l'}
}
{E_{ l'{\bf k}}-E_{ l''',{\bf k}+{\bf q}_2+{\bf q}_1} + \hbar (\omega_1 +\omega_2) +i\hbar\gamma }
\frac {f_{ l'',{\bf k}+{\bf q}_2}-f_{ l''',{\bf k}+{\bf q}_2+{\bf q}_1}}
{E_{ l'',{\bf k}+{\bf q}_2}-E_{ l''',{\bf k}+{\bf q}_2+{\bf q}_1}+\hbar\omega_1+i\hbar\gamma} 
\nonumber \\ &-&
\sum_{ l'' l'''}
\frac{ 
\delta_{ll'''}
q_{1\beta}\eta^\beta_{{\bf k}+{\bf q}_3}
\delta_{l''l'}
}
{E_{ l'',{\bf k}+{\bf q}_3}-E_{l,{\bf k}+{\bf q}_3+{\bf q}_1+{\bf q}_2} + \hbar (\omega_1 +\omega_2) +i\hbar\gamma }
\frac {f_{ l'',{\bf k}+{\bf q}_3}-f_{ l''',{\bf k}+{\bf q}_3+{\bf q}_1}}
{E_{ l'',{\bf k}+{\bf q}_3}-E_{ l''',{\bf k}+{\bf q}_3+{\bf q}_1}+\hbar\omega_1+i\hbar\gamma} 
\nonumber \\ &+&
\sum_{ l'' l'''}
\frac{ 
q_{1\beta}\eta^\beta_{{\bf k}+{\bf q}_3+{\bf q}_2}
\delta_{l'''l''}
\delta_{l''l'} }
{E_{ l'',{\bf k}+{\bf q}_3}-E_{l,{\bf k}+{\bf q}_3+{\bf q}_1+{\bf q}_2} + \hbar (\omega_1 +\omega_2) +i\hbar\gamma }
\frac {f_{ l''',{\bf k}+{\bf q}_3+{\bf q}_2}-f_{l,{\bf k}+{\bf q}_3+{\bf q}_1+{\bf q}_2}}
{E_{ l''',{\bf k}+{\bf q}_3+{\bf q}_2}-E_{l,{\bf k}+{\bf q}_3+{\bf q}_1+{\bf q}_2}+\hbar\omega_1+i\hbar\gamma} 
\Bigg] 
\Bigg\}.
\ea 
Now one sees that each of the intra-band terms contains the Fermi-functions difference of the type $(f_{ l'',{\bf k}+{\bf q}_3}-f_{ l'',{\bf k}+{\bf q}_3+{\bf q}_1})\approx -q_{1\beta}\p f_{ l'',{\bf k}+{\bf q}_3}/\p k_\beta$ and all the inter-band terms have the prefactor $q_{1\beta}$ originating from the matrix elements. One can therefore set ${\bf q}_{1}=0$ in the rest of the formula, which simplifies the second energy-denominators, e.g., $E_{ l'',{\bf k}+{\bf q}_2}-E_{ l'',{\bf k}+{\bf q}_2+{\bf q}_1}+\hbar\omega_1+i\hbar\gamma\to E_{ l'',{\bf k}+{\bf q}_2}-E_{ l'',{\bf k}+{\bf q}_2}+\hbar\omega_1+i\hbar\gamma =\hbar\omega_1+i\hbar\gamma$. Then, taking the sums over $l''$ and $l'''$ in the inter-band contributions we present the result in the form
\ba 
\underbrace{j_\alpha^{(3)}({\bf r}_0,t)}_{(2/1)}&=&
\int_{-\infty}^\infty d\omega_1 
\int_{-\infty}^\infty d\omega_2  
\int_{-\infty}^\infty d\omega_3 
\frac {e^4g_s}{2S}\sum_{{\bf q}_1{\bf q}_2{\bf q}_3} 
\phi_{{\bf q}_1\omega_1}\phi_{{\bf q}_2\omega_2}\phi_{{\bf q}_3\omega_3}
e^{i({\bf q}_3+{\bf q}_1+{\bf q}_2)\cdot{\bf r}_0-i(\omega_1 +\omega_2+\omega_3) t}
\nonumber \\ &\times&
\sum_{l l'{\bf k}} 
\frac {\langle l'{\bf k}|\{ \hat v_\alpha,e^{-i({\bf q}_3+{\bf q}_2)\cdot{\bf r}}\}_+|l,{\bf k}+{\bf q}_3+{\bf q}_2\rangle }
{E_{ l'{\bf k}}-E_{l,{\bf k}+{\bf q}_3+{\bf q}_2}+\hbar(\omega_1 +\omega_2+\omega_3) +i\hbar \gamma}
\left(-\frac {(-1)^{l+l'}}2\right)
\nonumber \\ &\times&
\Bigg\{
\frac {-q_{1\beta}}
{\hbar\omega_1+i\hbar\gamma} 
\Bigg[
\sum_{  l''}
\frac{
\delta_{ll''}q_{2\gamma}\eta^\gamma_{{\bf k}}
 + 
q_{3\delta}\eta^\delta_{{\bf k}+{\bf q}_2}
\delta_{l''l'}}
{E_{ l'{\bf k}}-E_{ l'',{\bf k}+{\bf q}_2} + \hbar (\omega_1 +\omega_2) +i\hbar\gamma }
\frac{\p (f_{ l'{\bf k}}- f_{ l'',{\bf k}+{\bf q}_2})}{\p k_\beta}
\nonumber \\ &-&
\sum_{ l'' }
\frac{
\delta_{ll''}
q_{3\delta}\eta^\delta_{{\bf k}}
+q_{2\gamma}\eta^\gamma_{{\bf k}+{\bf q}_3}
\delta_{l''l'}}
{E_{ l'',{\bf k}+{\bf q}_3}-E_{l,{\bf k}+{\bf q}_3+{\bf q}_2} + \hbar (\omega_1 +\omega_2) +i\hbar\gamma }
\frac{\p (f_{ l'',{\bf k}+{\bf q}_3}- f_{ l,{\bf k}+{\bf q}_3+{\bf q}_2})}{\p k_\beta}
\Bigg]_{=A} 
\nonumber \\ &+&
q_{1\beta}\Bigg[
\frac{1}
{E_{ l'{\bf k}}-E_{ l,{\bf k}+{\bf q}_2} + \hbar (\omega_1 +\omega_2) +i\hbar\gamma }
\left({\cal K}_{l'l}^\beta({\bf k})-{\cal K}_{l'l}^\beta({\bf k}+{\bf q}_2) \right)
\nonumber \\ &-&
\frac{ 1}
{E_{ l',{\bf k}+{\bf q}_3}-E_{l,{\bf k}+{\bf q}_3+{\bf q}_2} + \hbar (\omega_1 +\omega_2) +i\hbar\gamma } 
\left({\cal K}_{l'l}^\beta({\bf k}+{\bf q}_3)-{\cal K}_{l'l}^\beta({\bf k}+{\bf q}_2+{\bf q}_3) \right)
\Bigg]_{=B}
\Bigg\},\label{AB}
\ea 
where the function ${\cal K}_{l'l}^\beta({\bf k})$ is defined as 
\be 
K_{l'l}^\beta({\bf k})=\eta^\beta_{{\bf k}}
\frac {f_{ l'{\bf k}}-f_{ l{\bf k}}}
{E_{l'{\bf k}}-E_{ l{\bf k}}+\hbar\omega_1+i\hbar\gamma}.
\ee
The terms in the first square parenthesis in (\ref{AB}) denoted as $A$ are transformed as follows. First, we take the sums over $l''$ and set ${\bf q}_2=0$ or ${\bf q}_3=0$ in the terms which have already the prefactors $q_{2\gamma}$ or $q_{3\delta}$:
\ba 
A&=&\left(
\frac{q_{2\gamma}
\eta^\gamma_{{\bf k}}}
{E_{ l'{\bf k}}-E_{ l{\bf k}} + \hbar (\omega_1 +\omega_2) +i\hbar\gamma }
\frac{\p (f_{ l'{\bf k}}- f_{ l{\bf k}})}{\p k_\beta}
-
\frac{q_{2\gamma}
\eta^\gamma_{{\bf k}+{\bf q}_3}}
{E_{ l',{\bf k}+{\bf q}_3}-E_{l,{\bf k}+{\bf q}_3} + \hbar (\omega_1 +\omega_2) +i\hbar\gamma }
\frac{\p (f_{ l',{\bf k}+{\bf q}_3}- f_{ l,{\bf k}+{\bf q}_3})}{\p k_\beta}
\right) 
\nonumber \\ &-&
\left(
\frac{q_{3\delta}
\eta^\delta_{{\bf k}}}
{E_{ l{\bf k}}-E_{l,{\bf k}+{\bf q}_2} + \hbar (\omega_1 +\omega_2) +i\hbar\gamma }
\frac{\p (f_{ l{\bf k}}- f_{ l,{\bf k}+{\bf q}_2})}{\p k_\beta} 
-\frac{q_{3\delta}
\eta^\delta_{{\bf k}+{\bf q}_2}}
{E_{ l'{\bf k}}-E_{ l',{\bf k}+{\bf q}_2} + \hbar (\omega_1 +\omega_2) +i\hbar\gamma }
\frac{\p (f_{ l'{\bf k}}- f_{ l',{\bf k}+{\bf q}_2})}{\p k_\beta}
\right).
\nonumber \\ \label{termA}
\ea
Now consider the first line in Eq. (\ref{termA}) which has already the prefactor $q_{2\gamma}$. One sees that the second term there is the same function as the first term but with the argument ${\bf k}+{\bf q}_3$. It can be expanded at ${\bf q}_3\to 0$ which gives the prefactor $q_{3\delta}$. In the second line the differences of the Fermi functions can be expanded at ${\bf q}_2\to 0$; in the rest of the second line we can then set ${\bf q}_2= 0$. So we get:
\be 
A\approx -q_{2\gamma}q_{3\delta}\frac{\p}{\p k_\delta}\left(
\frac{
\eta^\gamma_{{\bf k}}}
{E_{ l'{\bf k}}-E_{ l{\bf k}} + \hbar (\omega_1 +\omega_2) +i\hbar\gamma }
\frac{\p (f_{ l'{\bf k}}- f_{ l{\bf k}})}{\p k_\beta}
\right) 
+
q_{2\gamma}q_{3\delta}
\frac{
\eta^\delta_{{\bf k}}}
{\hbar (\omega_1 +\omega_2 +i\gamma) }
\frac{\p^2 ( f_{ l{\bf k}}- f_{ l'{\bf k}})}{\p k_\beta\p k_\gamma}.
\ee

The terms in the second square parenthesis in (\ref{AB}) denoted as $B$ are transformed as follows. First, we expand the difference of the ${\cal K}$-functions at small ${\bf q}_2$ and set ${\bf q}_2=0$ in the rest of the formula:
\ba 
B&\approx&-q_{2\gamma}
\Bigg(
\frac{1}
{E_{ l'{\bf k}}-E_{ l{\bf k}} + \hbar (\omega_1 +\omega_2) +i\hbar\gamma }
\frac{\p}{\p k_\gamma}\left({\cal K}_{l'l}^\beta({\bf k})\right)
\nonumber \\ &-&
\frac{ 1}
{E_{ l',{\bf k}+{\bf q}_3}-E_{l,{\bf k}+{\bf q}_3} + \hbar (\omega_1 +\omega_2) +i\hbar\gamma } 
\frac{\p}{\p k_\gamma}\left({\cal K}_{l'l}^\beta({\bf k}+{\bf q}_3) \right)\Bigg).\label{b13}
\ea
Now we see that the expression in the second line is the same function as in the first line but with the argument shifted by ${\bf q}_3$. Expanding (\ref{b13}) at ${\bf q}_3\to 0$ we then get
\ba 
B&\approx&q_{2\gamma}q_{3\delta}\frac{\p}{\p k_\delta}
\Bigg(
\frac{1}
{E_{ l'{\bf k}}-E_{ l{\bf k}} + \hbar (\omega_1 +\omega_2) +i\hbar\gamma }
\frac{\p}{\p k_\gamma}\left(\eta^\beta_{{\bf k}}
\frac {f_{ l'{\bf k}}-f_{ l{\bf k}}}
{E_{l'{\bf k}}-E_{ l{\bf k}}+\hbar\omega_1+i\hbar\gamma}\right)
\Bigg).
\ea
One sees that both the $A$ and the $B$ terms differ from zero only if $l'=\bar l$, i.e. $A,B\propto \delta_{l'\bar l}$. Substituting the terms $A$ and $B$ in (\ref{AB}), taking the sum over $l'$ and introducing the Fourier components of the electric field as was described in Section \ref{app:30} we get the expression (\ref{intra2inter}).

\subsection{The contribution $(1/2)$}

The contribution $(1/2)$ consists of the terms containing one Kronecker symbol in the matrix elements in (\ref{fullcurrent}). The corresponding terms are
\ba 
\underbrace{j_\alpha^{(3)}({\bf r}_0,t)}_{(1/2)}&=&
\int_{-\infty}^\infty d\omega_1 
\int_{-\infty}^\infty d\omega_2  
\int_{-\infty}^\infty d\omega_3 
\frac {e^4g_s}{2S}\sum_{{\bf q}_1{\bf q}_2{\bf q}_3} 
\phi_{{\bf q}_1\omega_1}\phi_{{\bf q}_2\omega_2}\phi_{{\bf q}_3\omega_3}
e^{i({\bf q}_3+{\bf q}_1+{\bf q}_2)\cdot{\bf r}_0-i(\omega_1 +\omega_2+\omega_3) t}
\nonumber \\ &\times&
\sum_{l l'{\bf k}} 
\frac {\langle l'{\bf k}|\{\hat v_\alpha,e^{-i({\bf q}_3+{\bf q}_1+{\bf q}_2)\cdot{\bf r}}\}_+|l,{\bf k}+{\bf q}_3+{\bf q}_1+{\bf q}_2\rangle }
{E_{ l'{\bf k}}-E_{l,{\bf k}+{\bf q}_3+{\bf q}_1+{\bf q}_2}+\hbar(\omega_1 +\omega_2+\omega_3) +i\hbar \gamma}
\frac{(-1)^{l+l'}}4
\nonumber \\ &\times&
\Bigg[
\sum_{ l'' l'''}
\frac{\delta_{ll'''}
q_{1\beta}q_{2\gamma}
\eta^\beta_{{\bf k}}\eta^\gamma_{{\bf k}}
 }
{E_{ l'{\bf k}}-E_{ l'''{\bf k}} + \hbar (\omega_1 +\omega_2) +i\hbar\gamma }
\frac {f_{ l'{\bf k}}-f_{ l''{\bf k}}}
{E_{l'{\bf k}}-E_{ l''{\bf k}}+\hbar\omega_1+i\hbar\gamma} 
\nonumber \\ &+&
\sum_{ l'' l'''}
\frac{
q_{1\beta}q_{3\delta}\eta^\delta_{{\bf k}+{\bf q}_2}
\delta_{l'''l''}
\eta^\beta_{{\bf k}}
 }
{E_{ l'{\bf k}}-E_{ l''',{\bf k}+{\bf q}_2} + \hbar (\omega_1 +\omega_2) +i\hbar\gamma }
\frac {f_{ l'{\bf k}}-f_{ l''{\bf k}}}
{E_{l'{\bf k}}-E_{ l''{\bf k}}+\hbar\omega_1+i\hbar\gamma} 
\nonumber \\ &+&
\sum_{ l'' l'''}
\frac{
q_{2\gamma}q_{3\delta}\eta^\delta_{{\bf k}+{\bf q}_1}
\eta^\gamma_{{\bf k}+{\bf q}_1}
\delta_{l''l'} }
{E_{ l'{\bf k}}-E_{ l''',{\bf k}+{\bf q}_1} + \hbar (\omega_1 +\omega_2) +i\hbar\gamma }
\frac {f_{ l'{\bf k}}-f_{ l'',{\bf k}+{\bf q}_1}}
{E_{l'{\bf k}}-E_{ l'',{\bf k}+{\bf q}_1}+\hbar\omega_1+i\hbar\gamma} 
\nonumber \\ &&
\nonumber \\ &-&
\sum_{ l'' l'''}
\frac{
\delta_{ll'''}
q_{1\beta}q_{2\gamma}
\eta^\beta_{{\bf k}}\eta^\gamma_{{\bf k}}}
{E_{ l'{\bf k}}-E_{ l'''{\bf k}} + \hbar (\omega_1 +\omega_2) +i\hbar\gamma }
\frac {f_{ l''{\bf k}}-f_{ l'''{\bf k}}}
{E_{ l''{\bf k}}-E_{ l'''{\bf k}}+\hbar\omega_1+i\hbar\gamma} 
\nonumber \\ &-&
\sum_{ l'' l'''}
\frac{
q_{2\gamma}q_{3\delta}
\eta^\delta_{{\bf k}+{\bf q}_1}
\delta_{l'''l''}
\eta^\gamma_{{\bf k}}}
{E_{ l'{\bf k}}-E_{ l''',{\bf k}+{\bf q}_1} + \hbar (\omega_1 +\omega_2) +i\hbar\gamma }
\frac {f_{ l''{\bf k}}-f_{ l''',{\bf k}+{\bf q}_1}}
{E_{ l''{\bf k}}-E_{ l''',{\bf k}+{\bf q}_1}+\hbar\omega_1+i\hbar\gamma} 
\nonumber \\ &-&
\sum_{ l'' l'''}
\frac{
q_{1\beta}q_{3\delta}
\eta^\delta_{{\bf k}+{\bf q}_2}
\eta^\beta_{{\bf k}+{\bf q}_2}
\delta_{l''l'}}
{E_{ l'{\bf k}}-E_{ l''',{\bf k}+{\bf q}_2} + \hbar (\omega_1 +\omega_2) +i\hbar\gamma }
\frac {f_{ l'',{\bf k}+{\bf q}_2}-f_{ l''',{\bf k}+{\bf q}_2}}
{E_{ l'',{\bf k}+{\bf q}_2}-E_{ l''',{\bf k}+{\bf q}_2}+\hbar\omega_1+i\hbar\gamma} 
\nonumber \\ &&
\nonumber \\ &-&
\sum_{ l'' l'''}
\frac{
\delta_{ll'''}
q_{1\beta}q_{3\delta}
\eta^\beta_{{\bf k}}
\eta^\delta_{{\bf k}} }
{E_{ l''{\bf k}}-E_{l,{\bf k}+{\bf q}_2} + \hbar (\omega_1 +\omega_2) +i\hbar\gamma }
\frac {f_{ l''{\bf k}}-f_{ l'''{\bf k}}}
{E_{ l''{\bf k}}-E_{ l'''{\bf k}}+\hbar\omega_1+i\hbar\gamma} 
\nonumber \\ &-&
\sum_{ l'' l'''}
\frac{
q_{2\gamma}q_{3\delta}
\eta^\gamma_{{\bf k}+{\bf q}_1}
\delta_{l'''l''}
\eta^\delta_{{\bf k}}}
{E_{ l''{\bf k}}-E_{l,{\bf k}+{\bf q}_1} + \hbar (\omega_1 +\omega_2) +i\hbar\gamma }
\frac {f_{ l''{\bf k}}-f_{ l''',{\bf k}+{\bf q}_1}}
{E_{ l''{\bf k}}-E_{ l''',{\bf k}+{\bf q}_1}+\hbar\omega_1+i\hbar\gamma} 
\nonumber \\ &-&
\sum_{ l'' l'''}
\frac{
q_{1\beta}q_{2\gamma}
\eta^\gamma_{{\bf k}+{\bf q}_3}
\eta^\beta_{{\bf k}+{\bf q}_3}
\delta_{l''l'} }
{E_{ l'',{\bf k}+{\bf q}_3}-E_{l,{\bf k}+{\bf q}_3} + \hbar (\omega_1 +\omega_2) +i\hbar\gamma }
\frac {f_{ l'',{\bf k}+{\bf q}_3}-f_{ l''',{\bf k}+{\bf q}_3}}
{E_{ l'',{\bf k}+{\bf q}_3}-E_{ l''',{\bf k}+{\bf q}_3}+\hbar\omega_1+i\hbar\gamma} 
\nonumber \\ &&
\nonumber \\ &+&
\sum_{ l'' l'''}
\frac{
\delta_{ll'''}
q_{2\gamma}q_{3\delta}
\eta^\gamma_{{\bf k}}
\eta^\delta_{{\bf k}} }
{E_{ l''{\bf k}}-E_{l,{\bf k}+{\bf q}_1} + \hbar (\omega_1 +\omega_2) +i\hbar\gamma }
\frac {f_{ l''',{\bf k}}-f_{l,{\bf k}+{\bf q}_1}}
{E_{ l'''{\bf k}}-E_{l,{\bf k}+{\bf q}_1}+\hbar\omega_1+i\hbar\gamma} 
\nonumber \\ &+&
\sum_{ l'' l'''}
\frac{
q_{1\beta}q_{3\delta}
\eta^\beta_{{\bf k}+{\bf q}_2}
\delta_{l'''l''}
\eta^\delta_{{\bf k}} }
{E_{ l''{\bf k}}-E_{l,{\bf k}+{\bf q}_2} + \hbar (\omega_1 +\omega_2) +i\hbar\gamma }
\frac {f_{ l''',{\bf k}+{\bf q}_2}-f_{l,{\bf k}+{\bf q}_2}}
{E_{ l''',{\bf k}+{\bf q}_2}-E_{l,{\bf k}+{\bf q}_2}+\hbar\omega_1+i\hbar\gamma} 
\nonumber \\ &+&
\sum_{ l'' l'''}
\frac{
q_{1\beta}q_{2\gamma}
\eta^\beta_{{\bf k}+{\bf q}_3}
\eta^\gamma_{{\bf k}+{\bf q}_3}
\delta_{l''l'} }
{E_{ l'',{\bf k}+{\bf q}_3}-E_{l,{\bf k}+{\bf q}_3} + \hbar (\omega_1 +\omega_2) +i\hbar\gamma }
\frac {f_{ l''',{\bf k}+{\bf q}_3}-f_{l,{\bf k}+{\bf q}_3}}
{E_{ l''',{\bf k}+{\bf q}_3}-E_{l,{\bf k}+{\bf q}_3}+\hbar\omega_1+i\hbar\gamma} 
\Bigg] ,\label{12}
\ea 
where we have set ${\bf q}_1=0$, ${\bf q}_2=0$ or ${\bf q}_3=0$ in the rest of the terms if they already contained the corresponding $q$-prefactors. In each of the twelve terms in the square parenthesis in (\ref{12}) we have to sum over $l'''$ and $l''$. This can be done as follows. Consider, for example, the first line in the square parenthesis in (\ref{12}). The sum over $l'''$ is taken due to the Kronecker symbol; $l'''$ is then replaced by $l$. To take the sum over $l''$ we notice that $l''$ takes two values, for example, $l'$ and $\bar l'$. As seen from the Fermi-functions difference, $f_{ l'{\bf k}}-f_{ l''{\bf k}}$, the first line vanishes if $l''=l'$. Hence, $l''$ must be equal $\bar l'$. Similarly, one can simplify eight terms out of twelve. 

Now consider the third line. Here $l''=l'$ due to the Kronecker delta. The Fermi-functions difference can then be expanded at ${\bf q}_1\to 0$ and ${\bf q}_1$ can be set to zero in the rest of the formula. Using this way one can simplify the other four terms. Then we obtain: 
\ba 
\underbrace{j_\alpha^{(3)}({\bf r}_0,t)}_{(1/2)}&=&
\int_{-\infty}^\infty d\omega_1 
\int_{-\infty}^\infty d\omega_2  
\int_{-\infty}^\infty d\omega_3 
\frac {e^4g_s}{2S}\sum_{{\bf q}_1{\bf q}_2{\bf q}_3} 
\phi_{{\bf q}_1\omega_1}\phi_{{\bf q}_2\omega_2}\phi_{{\bf q}_3\omega_3}
e^{i({\bf q}_3+{\bf q}_1+{\bf q}_2)\cdot{\bf r}_0-i(\omega_1 +\omega_2+\omega_3) t}
\nonumber \\ &\times&
\sum_{l l'{\bf k}} 
\frac {\langle l'{\bf k}|\{\hat v_\alpha,e^{-i({\bf q}_3+{\bf q}_1+{\bf q}_2)\cdot{\bf r}}\}_+|l,{\bf k}+{\bf q}_3+{\bf q}_1+{\bf q}_2\rangle }
{E_{ l'{\bf k}}-E_{l,{\bf k}+{\bf q}_3+{\bf q}_1+{\bf q}_2}+\hbar(\omega_1 +\omega_2+\omega_3) +i\hbar \gamma}
\frac{(-1)^{l+l'}}4
\nonumber \\ &\times&
\Bigg[q_{1\beta}q_{2\gamma}\Bigg(
\frac{
\eta^\beta_{{\bf k}}\eta^\gamma_{{\bf k}} }
{E_{ l'{\bf k}}-E_{ l{\bf k}} + \hbar (\omega_1 +\omega_2) +i\hbar\gamma }
\frac {f_{ l'{\bf k}}-f_{\bar l'{\bf k}}}
{E_{l'{\bf k}}-E_{\bar l'{\bf k}}+\hbar\omega_1+i\hbar\gamma} 
\nonumber \\ &-&
\frac{\eta^\beta_{{\bf k}+{\bf q}_3} 
\eta^\gamma_{{\bf k}+{\bf q}_3}}
{E_{ l',{\bf k}+{\bf q}_3}-E_{l,{\bf k}+{\bf q}_3} + \hbar (\omega_1 +\omega_2) +i\hbar\gamma }
\frac {f_{ l',{\bf k}+{\bf q}_3}-f_{ \bar l',{\bf k}+{\bf q}_3}}
{E_{ l',{\bf k}+{\bf q}_3}-E_{\bar  l',{\bf k}+{\bf q}_3}+\hbar\omega_1+i\hbar\gamma}\Bigg) 
\nonumber \\ &-&q_{1\beta}q_{2\gamma}\Bigg(
\frac{
\eta^\beta_{{\bf k}}\eta^\gamma_{{\bf k}}}
{E_{ l'{\bf k}}-E_{ l{\bf k}} + \hbar (\omega_1 +\omega_2) +i\hbar\gamma }
\frac {f_{\bar l{\bf k}}-f_{ l{\bf k}}}
{E_{ \bar l{\bf k}}-E_{ l{\bf k}}+\hbar\omega_1+i\hbar\gamma} 
\nonumber \\ &-&
\frac{
\eta^\beta_{{\bf k}+{\bf q}_3}
\eta^\gamma_{{\bf k}+{\bf q}_3} }
{E_{ l',{\bf k}+{\bf q}_3}-E_{l,{\bf k}+{\bf q}_3} + \hbar (\omega_1 +\omega_2) +i\hbar\gamma }
\frac {f_{ \bar l,{\bf k}+{\bf q}_3}-f_{l,{\bf k}+{\bf q}_3}}
{E_{ \bar l,{\bf k}+{\bf q}_3}-E_{l,{\bf k}+{\bf q}_3}+\hbar\omega_1+i\hbar\gamma}\Bigg) 
\nonumber \\ &+&
\frac{q_{1\beta}q_{3\delta}\eta^\delta_{{\bf k}+{\bf q}_2} }
{E_{ l'{\bf k}}-E_{\bar l',{\bf k}+{\bf q}_2} + \hbar (\omega_1 +\omega_2) +i\hbar\gamma }\Bigg(
\eta^\beta_{{\bf k}}
\frac {f_{ l'{\bf k}}-f_{\bar l'{\bf k}}}
{E_{l'{\bf k}}-E_{\bar l'{\bf k}}+\hbar\omega_1+i\hbar\gamma} 
\nonumber \\ &-&
\eta^\beta_{{\bf k}+{\bf q}_2}
\frac {f_{ l',{\bf k}+{\bf q}_2}-f_{\bar  l',{\bf k}+{\bf q}_2}}
{E_{ l',{\bf k}+{\bf q}_2}-E_{ \bar l',{\bf k}+{\bf q}_2}+\hbar\omega_1+i\hbar\gamma} \Bigg)
\nonumber \\ &-&
\frac{q_{1\beta}q_{3\delta}
\eta^\delta_{{\bf k}} }
{E_{ \bar l{\bf k}}-E_{l,{\bf k}+{\bf q}_2} + \hbar (\omega_1 +\omega_2) +i\hbar\gamma }
\Bigg(\eta^\beta_{{\bf k}}\frac {f_{ \bar l{\bf k}}-f_{ l{\bf k}}}
{E_{ \bar l{\bf k}}-E_{ l{\bf k}}+\hbar\omega_1+i\hbar\gamma} 
\nonumber \\ &-&
\eta^\beta_{{\bf k}+{\bf q}_2}
\frac {f_{ \bar l,{\bf k}+{\bf q}_2}-f_{l,{\bf k}+{\bf q}_2}}
{E_{ \bar l,{\bf k}+{\bf q}_2}-E_{l,{\bf k}+{\bf q}_2}+\hbar\omega_1+i\hbar\gamma} \Bigg)
\nonumber \\ &+&
\frac{-q_{1\beta}q_{2\gamma}q_{3\delta}\eta^\delta_{{\bf k}}
\eta^\gamma_{{\bf k}} }{\hbar\omega_1+i\hbar\gamma} 
\Bigg(
\sum_{  l''}
\frac{1}
{E_{ l'{\bf k}}-E_{ l''{\bf k}} + \hbar (\omega_1 +\omega_2) +i\hbar\gamma }
\frac {\p (f_{ l'{\bf k}}- f_{ l''{\bf k}})}{\p k_\beta }
\nonumber \\ &-&
\sum_{ l'' }
\frac{1}
{E_{ l''{\bf k}}-E_{l{\bf k}} + \hbar (\omega_1 +\omega_2) +i\hbar\gamma }
\frac {\p (f_{ l''{\bf k}}- f_{ l{\bf k}}) }
{\p k_\beta} \Bigg)
\Bigg] ;\label{12a}
\ea 
in (\ref{12a}) we have also rearranged terms and collected them in groups according to their $q$-prefactors.

Now one sees that the function in the second line differs from the function in the first line by only the shifted (by ${\bf q}_3$) argument (${\bf k}\to {\bf k}+{\bf q}_3$). It can therefore be expanded in ${\bf q}_3$ and simplified. The same can be done in all other lines except the two last ones which still contain the sums over $l''$. These sums can now be also easily taken since in the last line $l''$ should be equal to $\bar l$, while in the last but one line -- to $\bar l'$. Making all these transformations we get
\ba 
\underbrace{j_\alpha^{(3)}({\bf r}_0,t)}_{(1/2)}&=&
\int_{-\infty}^\infty d\omega_1 
\int_{-\infty}^\infty d\omega_2  
\int_{-\infty}^\infty d\omega_3 
\frac {e^4g_s}{2S}\sum_{{\bf q}_1{\bf q}_2{\bf q}_3} 
\phi_{{\bf q}_1\omega_1}\phi_{{\bf q}_2\omega_2}\phi_{{\bf q}_3\omega_3}
e^{i({\bf q}_3+{\bf q}_1+{\bf q}_2)\cdot{\bf r}_0-i(\omega_1 +\omega_2+\omega_3) t}
\nonumber \\ &\times&
\sum_{l l'{\bf k}} 
\frac {\langle l'{\bf k}|\{\hat v_\alpha,e^{-i({\bf q}_3+{\bf q}_1+{\bf q}_2)\cdot{\bf r}}\}_+|l,{\bf k}+{\bf q}_3+{\bf q}_1+{\bf q}_2\rangle }
{E_{ l'{\bf k}}-E_{l,{\bf k}+{\bf q}_3+{\bf q}_1+{\bf q}_2}+\hbar(\omega_1 +\omega_2+\omega_3) +i\hbar \gamma}
\frac{(-1)^{l+l'}}4
\nonumber \\ &\times&
\Bigg[-q_{1\beta}q_{2\gamma}q_{3\delta} \frac{\p}{\p k_\delta}
\Bigg(
\frac{
\eta^\beta_{{\bf k}}\eta^\gamma_{{\bf k}} }
{E_{ l'{\bf k}}-E_{ l{\bf k}} + \hbar (\omega_1 +\omega_2) +i\hbar\gamma }
\frac {f_{ l'{\bf k}}-f_{\bar l'{\bf k}}}
{E_{l'{\bf k}}-E_{\bar l'{\bf k}}+\hbar\omega_1+i\hbar\gamma} 
\Bigg) 
\nonumber \\ &+&
q_{1\beta}q_{2\gamma}q_{3\delta} \frac{\p}{\p k_\delta}
\Bigg(
\frac{
\eta^\beta_{{\bf k}}\eta^\gamma_{{\bf k}}}
{E_{ l'{\bf k}}-E_{ l{\bf k}} + \hbar (\omega_1 +\omega_2) +i\hbar\gamma }
\frac {f_{\bar l{\bf k}}-f_{ l{\bf k}}}
{E_{ \bar l{\bf k}}-E_{ l{\bf k}}+\hbar\omega_1+i\hbar\gamma} 
\Bigg) 
\nonumber \\ &+&
\frac{-q_{1\beta}q_{2\gamma}q_{3\delta}\eta^\delta_{{\bf k}+{\bf q}_2} }
{E_{ l'{\bf k}}-E_{\bar l',{\bf k}+{\bf q}_2} + \hbar (\omega_1 +\omega_2) +i\hbar\gamma }\frac{\p}{\p k_\gamma}
\Bigg(
\eta^\beta_{{\bf k}}
\frac {f_{ l'{\bf k}}-f_{\bar l'{\bf k}}}
{E_{l'{\bf k}}-E_{\bar l'{\bf k}}+\hbar\omega_1+i\hbar\gamma} 
\Bigg)
\nonumber \\ &-&
\frac{-q_{1\beta}q_{2\gamma}q_{3\delta}
\eta^\delta_{{\bf k}} }
{E_{ \bar l{\bf k}}-E_{l,{\bf k}+{\bf q}_2} + \hbar (\omega_1 +\omega_2) +i\hbar\gamma }\frac{\p}{\p k_\gamma}
\Bigg(\eta^\beta_{{\bf k}}\frac {f_{ \bar l{\bf k}}-f_{ l{\bf k}}}
{E_{ \bar l{\bf k}}-E_{ l{\bf k}}+\hbar\omega_1+i\hbar\gamma} 
\Bigg)
\nonumber \\ &+&
\frac{-q_{1\beta}q_{2\gamma}q_{3\delta}\eta^\delta_{{\bf k}}
\eta^\gamma_{{\bf k}} }{\hbar\omega_1+i\hbar\gamma} 
\Bigg(
\frac{1}
{E_{ l'{\bf k}}-E_{ \bar l'{\bf k}} + \hbar (\omega_1 +\omega_2) +i\hbar\gamma }
\frac {\p (f_{ l'{\bf k}}- f_{ \bar l'{\bf k}})}{\p k_\beta }
\nonumber \\ &-&
\frac{1}
{E_{ \bar l{\bf k}}-E_{l{\bf k}} + \hbar (\omega_1 +\omega_2) +i\hbar\gamma }
\frac {\p (f_{ \bar l{\bf k}}- f_{ l{\bf k}}) }
{\p k_\beta} \Bigg)
\Bigg] .\label{12b}
\ea 
Now we have the required three $q$-prefactors $q_{1\beta}q_{2\gamma}q_{3\delta}$ and can set ${\bf q}_1={\bf q}_2={\bf q}_3={\bf 0}$ in the rest of the formula. Then it can be rewritten as 
\ba 
\underbrace{j_\alpha^{(3)}({\bf r}_0,t)}_{(1/2)}&=&
\int_{-\infty}^\infty d\omega_1 
\int_{-\infty}^\infty d\omega_2  
\int_{-\infty}^\infty d\omega_3 
\frac {ie^4g_s}{2S}\sum_{{\bf q}_1{\bf q}_2{\bf q}_3} 
E_{{\bf q}_1\omega_1}^\beta E_{{\bf q}_2\omega_2}^\gamma E_{{\bf q}_3\omega_3}^\delta
e^{i({\bf q}_3+{\bf q}_1+{\bf q}_2)\cdot{\bf r}_0-i(\omega_1 +\omega_2+\omega_3) t}
\nonumber \\ &\times&
\sum_{l l'{\bf k}} 
\frac {\langle l'{\bf k}|\hat v_\alpha|l{\bf k}\rangle }
{E_{ l'{\bf k}}-E_{l{\bf k}}+\hbar(\omega_1 +\omega_2+\omega_3) +i\hbar \gamma}
\frac{(-1)^{l+l'}}2
\nonumber \\ &\times&
\Bigg[\frac{\p}{\p k_\delta}
\Bigg(
\frac{
\eta^\beta_{{\bf k}}\eta^\gamma_{{\bf k}} }
{E_{ l'{\bf k}}-E_{ l{\bf k}} + \hbar (\omega_1 +\omega_2) +i\hbar\gamma }
\left(
\frac {f_{ l'{\bf k}}-f_{\bar l'{\bf k}}}
{E_{l'{\bf k}}-E_{\bar l'{\bf k}}+\hbar\omega_1+i\hbar\gamma} -\frac {f_{\bar l{\bf k}}-f_{ l{\bf k}}}
{E_{ \bar l{\bf k}}-E_{ l{\bf k}}+\hbar\omega_1+i\hbar\gamma} \right)
\Bigg) 
\nonumber \\ &+&
\frac{\eta^\delta_{{\bf k}} }
{E_{ l'{\bf k}}-E_{\bar l'{\bf k}} + \hbar (\omega_1 +\omega_2) +i\hbar\gamma }\frac{\p}{\p k_\gamma}
\Bigg(
\eta^\beta_{{\bf k}}
\frac {f_{ l'{\bf k}}-f_{\bar l'{\bf k}}}
{E_{l'{\bf k}}-E_{\bar l'{\bf k}}+\hbar\omega_1+i\hbar\gamma} 
\Bigg)
\nonumber \\ &-&
\frac{
\eta^\delta_{{\bf k}} }
{E_{ \bar l{\bf k}}-E_{l,{\bf k}} + \hbar (\omega_1 +\omega_2) +i\hbar\gamma }\frac{\p}{\p k_\gamma}
\Bigg(\eta^\beta_{{\bf k}}\frac {f_{ \bar l{\bf k}}-f_{ l{\bf k}}}
{E_{ \bar l{\bf k}}-E_{ l{\bf k}}+\hbar\omega_1+i\hbar\gamma} 
\Bigg)
\nonumber \\ &+&
\frac{\eta^\delta_{{\bf k}}
\eta^\gamma_{{\bf k}} }{\hbar\omega_1+i\hbar\gamma} 
\Bigg(
\frac{1}
{E_{ l'{\bf k}}-E_{ \bar l'{\bf k}} + \hbar (\omega_1 +\omega_2) +i\hbar\gamma }
\frac {\p (f_{ l'{\bf k}}- f_{ \bar l'{\bf k}})}{\p k_\beta }
\nonumber \\ &-&
\frac{1}
{E_{ \bar l{\bf k}}-E_{l{\bf k}} + \hbar (\omega_1 +\omega_2) +i\hbar\gamma }
\frac {\p (f_{ \bar l{\bf k}}- f_{ l{\bf k}}) }
{\p k_\beta} \Bigg)
\Bigg] .\label{12c}
\ea 

Now we can take the sum over $l'$. One sees that, if $l'=\bar l$ all the expression in the square parenthesis vanishes. Hence $l'$ should be equal to $l$ and we get
\ba 
\underbrace{j_\alpha^{(3)}({\bf r}_0,t)}_{(1/2)}&=&
\int_{-\infty}^\infty d\omega_1 
\int_{-\infty}^\infty d\omega_2  
\int_{-\infty}^\infty d\omega_3 
\frac {ie^4g_s}{2S}
E_{\omega_1}^\beta ({\bf r}_0)E_{\omega_2}^\gamma({\bf r}_0) E_{\omega_3}^\delta({\bf r}_0)
e^{-i(\omega_1 +\omega_2+\omega_3) t}
\nonumber \\ &\times&
\sum_{l {\bf k}} 
\frac {\langle l{\bf k}|\hat v_\alpha|l{\bf k}\rangle }
{\hbar(\omega_1 +\omega_2+\omega_3) +i\hbar \gamma}
\Bigg[\frac{\p}{\p k_\delta}
\Bigg(
\frac{
\eta^\beta_{{\bf k}}\eta^\gamma_{{\bf k}} }
{ \hbar (\omega_1 +\omega_2) +i\hbar\gamma }
\frac {f_{ l{\bf k}}-f_{\bar l{\bf k}}}
{E_{l{\bf k}}-E_{\bar l{\bf k}}+\hbar\omega_1+i\hbar\gamma} 
\Bigg) 
\nonumber \\ &+&
\frac{\eta^\delta_{{\bf k}} }
{E_{ l{\bf k}}-E_{\bar l{\bf k}} + \hbar (\omega_1 +\omega_2) +i\hbar\gamma }\frac{\p}{\p k_\gamma}
\Bigg(
\eta^\beta_{{\bf k}}
\frac {f_{ l{\bf k}}-f_{\bar l{\bf k}}}
{E_{l{\bf k}}-E_{\bar l{\bf k}}+\hbar\omega_1+i\hbar\gamma} 
\Bigg)
\nonumber \\ &+&
\frac{\eta^\delta_{{\bf k}}
\eta^\gamma_{{\bf k}} }{\hbar\omega_1+i\hbar\gamma} 
\Bigg(
\frac{1}
{E_{ l{\bf k}}-E_{ \bar l{\bf k}} + \hbar (\omega_1 +\omega_2) +i\hbar\gamma }
\frac {\p (f_{ l{\bf k}}- f_{ \bar l{\bf k}})}{\p k_\beta }
\Bigg)
\Bigg] .\label{12d}
\ea 
Since the electric field, and hence the current, does not depend on the coordinate, the expression (\ref{12d}) gives the formula (\ref{intrainter2}).

\subsection{The contribution $(0/3)$}

The $(0/3)$ contribution is the only one term in Eq. (\ref{fullcurrent}) which contains no Kronecker symbols. This does not allow one to immediately take some of the sums. On the other hand, the $(0/3)$ contribution has all three $q$-prefactors, $q_{1\beta}q_{2\gamma}q_{3\delta}$, so that one can set ${\bf q}_1={\bf q}_2={\bf q}_3={\bf 0}$ in the rest of the formula and to replace the Fourier components of the potential by those of the electric field. So we get straightforwardly: 
\ba 
\underbrace{ j^{(3)}_\alpha(t)}_{(0/3)}&=&
\int_{-\infty}^\infty d\omega_1 
\int_{-\infty}^\infty d\omega_2  
\int_{-\infty}^\infty d\omega_3 
\frac {ie^4g_s}{8S}
E^{\beta}_{\omega_1}
E^{\gamma}_{\omega_2}
E^{\delta}_{\omega_3}
e^{-i(\omega_1 +\omega_2+\omega_3) t}
\nonumber \\ &\times&
\sum_{l l'{\bf k}} 
\frac {\langle l'{\bf k}| \hat v_\alpha|l{\bf k}\rangle }
{E_{ l'{\bf k}}-E_{l{\bf k}}+\hbar(\omega_1 +\omega_2+\omega_3) +i\hbar \gamma}
(-1)^{l+l'} 
\eta^\beta_{{\bf k}}\eta^\gamma_{{\bf k}}\eta^\delta_{{\bf k}}
\nonumber \\ &\times&
\Bigg[
\sum_{ l'' l'''}
\frac{1 }
{E_{ l'{\bf k}}-E_{ l''{\bf k}} + \hbar (\omega_1 +\omega_2) +i\hbar\gamma }
\frac {f_{ l'{\bf k}}-f_{ l'''{\bf k}}}
{E_{l'{\bf k}}-E_{ l'''{\bf k}}+\hbar\omega_1+i\hbar\gamma} 
\nonumber \\ &-&
\sum_{ l'' l'''}
\frac{1}
{E_{ l'{\bf k}}-E_{ l'''{\bf k}} + \hbar (\omega_1 +\omega_2) +i\hbar\gamma }
\frac {f_{ l''{\bf k}}-f_{ l'''{\bf k}}}
{E_{ l''{\bf k}}-E_{ l'''{\bf k}}+\hbar\omega_1+i\hbar\gamma} 
\nonumber \\ &-&
\sum_{ l'' l'''}
\frac{1}
{E_{ l''{\bf k}}-E_{l{\bf k}} + \hbar (\omega_1 +\omega_2) +i\hbar\gamma }
\frac {f_{ l''{\bf k}}-f_{ l'''{\bf k}}}
{E_{ l''{\bf k}}-E_{ l'''{\bf k}}+\hbar\omega_1+i\hbar\gamma} 
\nonumber \\ &+&
\sum_{ l'' l'''}
\frac{1 }
{E_{ l''{\bf k}}-E_{l{\bf k}} + \hbar (\omega_1 +\omega_2) +i\hbar\gamma }
\frac {f_{ l'''{\bf k}}-f_{l{\bf k}}}
{E_{ l'''{\bf k}}-E_{l{\bf k}}+\hbar\omega_1+i\hbar\gamma} 
\Bigg]. 
\ea 
Now we take a sum over $l'''$. In the first line in the squared parenthesis $l'''$ should be equal $\bar l'$, in the second line -- $\bar l''$, etc. We get
\ba 
\underbrace{ j^{(3)}_\alpha(t)}_{(0/3)}&=&
\int_{-\infty}^\infty d\omega_1 
\int_{-\infty}^\infty d\omega_2  
\int_{-\infty}^\infty d\omega_3 
\frac {ie^4g_s}{8S}
E^{\beta}_{\omega_1}
E^{\gamma}_{\omega_2}
E^{\delta}_{\omega_3}
e^{-i(\omega_1 +\omega_2+\omega_3) t}
\nonumber \\ &\times&
\sum_{l l'{\bf k}} 
\frac {\langle l'{\bf k}| \hat v_\alpha|l{\bf k}\rangle }
{E_{ l'{\bf k}}-E_{l{\bf k}}+\hbar(\omega_1 +\omega_2+\omega_3) +i\hbar \gamma}
(-1)^{l+l'} 
\eta^\beta_{{\bf k}}\eta^\gamma_{{\bf k}}\eta^\delta_{{\bf k}}
\nonumber \\ &\times&
\Bigg[
\sum_{ l''}
\frac{1 }
{E_{ l'{\bf k}}-E_{ l''{\bf k}} + \hbar (\omega_1 +\omega_2) +i\hbar\gamma }
\frac {f_{ l'{\bf k}}-f_{ \bar l'{\bf k}}}
{E_{l'{\bf k}}-E_{ \bar l'{\bf k}}+\hbar\omega_1+i\hbar\gamma} 
\nonumber \\ &-&
\sum_{ l'' }
\frac{1}
{E_{ l'{\bf k}}-E_{ \bar l''{\bf k}} + \hbar (\omega_1 +\omega_2) +i\hbar\gamma }
\frac {f_{ l''{\bf k}}-f_{ \bar l''{\bf k}}}
{E_{ l''{\bf k}}-E_{ \bar l''{\bf k}}+\hbar\omega_1+i\hbar\gamma} 
\nonumber \\ &-&
\sum_{ l'' }
\frac{1}
{E_{ l''{\bf k}}-E_{l{\bf k}} + \hbar (\omega_1 +\omega_2) +i\hbar\gamma }
\frac {f_{ l''{\bf k}}-f_{ \bar l''{\bf k}}}
{E_{ l''{\bf k}}-E_{ \bar l''{\bf k}}+\hbar\omega_1+i\hbar\gamma} 
\nonumber \\ &+&
\sum_{ l'' }
\frac{1 }
{E_{ l''{\bf k}}-E_{l{\bf k}} + \hbar (\omega_1 +\omega_2) +i\hbar\gamma }
\frac {f_{ \bar l{\bf k}}-f_{l{\bf k}}}
{E_{ \bar l{\bf k}}-E_{l{\bf k}}+\hbar\omega_1+i\hbar\gamma} 
\Bigg] .
\ea 
Now we replace, in the first line in the squared parenthesis, the summation index $l''$ by $\bar l''$; this does not change the result. Then the $(0/3)$ current is written as
\ba 
\underbrace{ j^{(3)}_\alpha(t)}_{(0/3)}&=&
\int_{-\infty}^\infty d\omega_1 
\int_{-\infty}^\infty d\omega_2  
\int_{-\infty}^\infty d\omega_3 
\frac {ie^4g_s}{8S}
E^{\beta}_{\omega_1}
E^{\gamma}_{\omega_2}
E^{\delta}_{\omega_3}
e^{-i(\omega_1 +\omega_2+\omega_3) t}
\nonumber \\ &\times&
\sum_{l l'{\bf k}} 
\frac {\langle l'{\bf k}| \hat v_\alpha|l{\bf k}\rangle }
{E_{ l'{\bf k}}-E_{l{\bf k}}+\hbar(\omega_1 +\omega_2+\omega_3) +i\hbar \gamma}
(-1)^{l+l'} 
\eta^\beta_{{\bf k}}\eta^\gamma_{{\bf k}}\eta^\delta_{{\bf k}}
\nonumber \\ &\times&
\Bigg[
\sum_{ l''}
\frac{1 }
{E_{ l'{\bf k}}-E_{ \bar l''{\bf k}} + \hbar (\omega_1 +\omega_2) +i\hbar\gamma }
\left(
\frac {f_{ l'{\bf k}}-f_{ \bar l'{\bf k}}}
{E_{l'{\bf k}}-E_{ \bar l'{\bf k}}+\hbar\omega_1+i\hbar\gamma} 
-\frac {f_{ l''{\bf k}}-f_{ \bar l''{\bf k}}}
{E_{ l''{\bf k}}-E_{ \bar l''{\bf k}}+\hbar\omega_1+i\hbar\gamma} \right)
\nonumber \\ &-&
\sum_{ l'' }
\frac{1}
{E_{ l''{\bf k}}-E_{l{\bf k}} + \hbar (\omega_1 +\omega_2) +i\hbar\gamma }
\Bigg(
\frac {f_{ l''{\bf k}}-f_{ \bar l''{\bf k}}}
{E_{ l''{\bf k}}-E_{ \bar l''{\bf k}}+\hbar\omega_1+i\hbar\gamma} 
-\frac {f_{ \bar l{\bf k}}-f_{l{\bf k}}}
{E_{ \bar l{\bf k}}-E_{l{\bf k}}+\hbar\omega_1+i\hbar\gamma} \Bigg)
\Bigg] .
\ea 
Now one sees that $l''$ should be equal to $\bar l'$ in the first line and $l$ in the second line; otherwise, the corresponding terms vanish. This gives 
\ba 
\underbrace{ j^{(3)}_\alpha(t)}_{(0/3)}&=&
\int_{-\infty}^\infty d\omega_1 
\int_{-\infty}^\infty d\omega_2  
\int_{-\infty}^\infty d\omega_3 
\frac {ie^4g_s}{8S}
E^{\beta}_{\omega_1}
E^{\gamma}_{\omega_2}
E^{\delta}_{\omega_3}
e^{-i(\omega_1 +\omega_2+\omega_3) t}
\nonumber \\ &\times&
\frac{1 }
{\hbar (\omega_1 +\omega_2) +i\hbar\gamma }
\sum_{l l'{\bf k}} 
\frac {\langle l'{\bf k}| \hat v_\alpha|l{\bf k}\rangle }
{E_{ l'{\bf k}}-E_{l{\bf k}}+\hbar(\omega_1 +\omega_2+\omega_3) +i\hbar \gamma}
(-1)^{l+l'} 
\eta^\beta_{{\bf k}}\eta^\gamma_{{\bf k}}\eta^\delta_{{\bf k}}
\nonumber \\ &\times&
\Bigg[
\left(
\frac {f_{ l'{\bf k}}-f_{ \bar l'{\bf k}}}
{E_{l'{\bf k}}-E_{ \bar l'{\bf k}}+\hbar\omega_1+i\hbar\gamma} 
-\frac {f_{ \bar l'{\bf k}}-f_{  l'{\bf k}}}
{E_{ \bar l'{\bf k}}-E_{  l'{\bf k}}+\hbar\omega_1+i\hbar\gamma} \right)
\nonumber \\ &-&
\Bigg(
\frac {f_{ l{\bf k}}-f_{ \bar l{\bf k}}}
{E_{ l{\bf k}}-E_{ \bar l{\bf k}}+\hbar\omega_1+i\hbar\gamma} 
-\frac {f_{ \bar l{\bf k}}-f_{l{\bf k}}}
{E_{ \bar l{\bf k}}-E_{l{\bf k}}+\hbar\omega_1+i\hbar\gamma} \Bigg)
\Bigg].
\ea 
Finally, in this formula, if $l'=l$ the term in the squared parenthesis vanishes. Thus $l'$ should be equal $\bar l$. Substituting $l'=\bar l$ we get, after some transformations, Eq. (\ref{inter3}). 

\section{Calculation of conductivity\label{app:conduct}}

In this Section we calculate the contributions to the third-order conductivity defined by Eqs. (\ref{sigma3}) and (\ref{intra3}) -- (\ref{inter3}). 

\subsection{The contribution $(3/0)$}

Following the definitions (\ref{sigma3}) and (\ref{intra3}) we present the $(3/0)$ contribution to the third-order conductivity in the form
\be 
\underbrace{\tilde \sigma_{\alpha\beta\gamma\delta}^{(3)}
(\omega_1, \omega_2,\omega_3)}_{(3/0)}=
\frac {{\cal W}_{\alpha\beta\gamma\delta}^{(3/0)}}{(\Omega_1 +\Omega_2+\Omega_3 +i \Gamma)
(\Omega_1 +\Omega_2 +i\Gamma)
(\Omega_1+i\Gamma)} ,
\ee
where the factor ${\cal W}^{(3/0)}_{\alpha\beta\gamma\delta}$ is
\be 
{\cal W}^{(3/0)}_{\alpha\beta\gamma\delta}=\frac {ie^4g_s}{E_F^3 S}
\sum_{ l {\bf k}} 
\langle l {\bf k}| \hat v_\alpha| l  {\bf k} \rangle
\frac{\p^3f_{ l {\bf k}}}{\p k_\beta\p k_\gamma\p k_\delta}.
\label{factorW}
\ee
Substituting in (\ref{factorW}) the matrix element of the velocity (\ref{velocity-on}) and integrating by part we rewrite the function ${\cal W}^{(3/0)}_{\alpha\beta\gamma\delta}$ in the form
\be
{\cal W}^{(3/0)}_{\alpha\beta\gamma\delta}=
\frac {ie^4g_s}{\hbar E_F^3 S}  \sum_{l{\bf k}} 
   \frac{\p f(E_{l{\bf k}})}{\p E}\frac{\p E_{l{\bf k}}}{\p k_\delta}
 \frac{\p^3 E_{l{\bf k}}}{\p k_\alpha\p k_\beta \p k_\gamma }.\label{factorW1}
\ee
Due to the factor $\p f(E_{l{\bf k}})/\p E$ only the vicinity of the Fermi level contributes to the integral. Therefore the integration over the whole Brillouin zone in (\ref{factorW1}) is reduced to the integration over the vicinity of two Dirac points. The both Dirac points contribute equally to the integral which leads to an additional, valley degeneracy factor $g_v=2$. Near the Dirac points we have $ 
E_{l{\bf k}}=(-1)^l\hbar v_Fk,
$ 
Eq. (\ref{energynearDirac}). The differentiation gives 
\be 
 \frac{\p E_{l{\bf k}}}{\p k_\delta } =(-1)^l\hbar v_F\frac {k_\delta}k,
\label{Ederiv1}
\ee
\be 
 \frac{\p^3 E_{l{\bf k}}}{\p k_\alpha \p k_\beta \p k_\gamma } =(-1)^l\hbar v_F\left(-\frac {k_{\alpha}\delta_{\beta\gamma}+k_\beta\delta_{\alpha\gamma}+k_\gamma\delta_{\alpha\beta} }{k^3}+3\frac{k_{\alpha}k_\beta k_\gamma}{k^5}\right).
\label{Ederiv3}
\ee
Substituting (\ref{Ederiv1}) and (\ref{Ederiv3}) back to Eq. (\ref{factorW1}) and calculating the integral leads to the following result (at $T\ll|\mu|$): 
\be 
{\cal W}^{(3/0)}_{\alpha\beta\gamma\delta}=
i
\frac {e^4g_sg_v \hbar v_F^2}{16\pi E_F^4 }  
\Big(
\delta_{\alpha\beta}\delta_{\gamma \delta}+
\delta_{\alpha\gamma }\delta_{\beta\delta}+
\delta_{\alpha\delta}\delta_{\beta\gamma}  
\Big)
\ee
Here the angular integration in the ${\bf k}$-space [${\bf k}=k(\cos\phi,\sin\phi)$] has been performed using Eqs. (\ref{average2}) and (\ref{average4}). With the help of the definitions (\ref{sigma^30}), (\ref{Delta3}) and (\ref{O-definitions}) we get Eq. (\ref{FuncS30}).

\subsection{The contribution $(2/1)$}

According to the definition (\ref{intra2inter}) the $(2/1)$-conductivity  has three contributions,
\ba 
\underbrace{\tilde \sigma_{\alpha\beta\gamma\delta}^{(3)}
(\omega_1, \omega_2,\omega_3)}_{(2/1)}&\equiv&
\underbrace{\tilde \sigma_{\alpha\beta\gamma\delta}^{(3)}
(\omega_1, \omega_2,\omega_3)}_{(2/1),a}+
\underbrace{\tilde \sigma_{\alpha\beta\gamma\delta}^{(3)}
(\omega_1, \omega_2,\omega_3)}_{(2/1),b}+
\underbrace{\tilde \sigma_{\alpha\beta\gamma\delta}^{(3)}
(\omega_1, \omega_2,\omega_3)}_{(2/1),c}
\nonumber \\ &=&
\frac {ie^4g_s}{2S}
\sum_{l {\bf k}} 
\frac {\langle \bar l{\bf k}| \hat v_\alpha|l{\bf k}\rangle }
{E_{\bar l{\bf k}}-E_{l{\bf k}}+\hbar(\omega_1 +\omega_2+\omega_3) +i\hbar \gamma}
\Bigg\{
\frac {1}{\hbar^2(\omega_1+i\gamma)(\omega_1 +\omega_2 +i\gamma) }\eta^\delta_{{\bf k}}
\frac{\p^2 (f_{l{\bf k}}- f_{ \bar l{\bf k}})}
{\p k_\beta\p k_\gamma}
\nonumber \\ &+&
\frac {1}{\hbar(\omega_1+i\gamma)} 
\frac{\p}{\p k_\delta}
\Bigg(
\frac{\eta^\gamma_{{\bf k}}}
{E_{\bar l{\bf k}}-E_{ l{\bf k}} + \hbar (\omega_1 +\omega_2) +i\hbar\gamma }
\frac{\p (f_{ l{\bf k}}- f_{\bar l{\bf k}})}{\p k_\beta}
\Bigg)
\nonumber \\ &+&
\frac{\p }{\p k_\delta}
\Bigg[
\frac{1}
{E_{\bar l{\bf k}}-E_{ l{\bf k}} + \hbar (\omega_1 +\omega_2) +i\hbar\gamma }\frac{\p }{\p k_\gamma}
\left(\eta^\beta_{{\bf k}}
\frac {f_{  l{\bf k}}-f_{\bar l{\bf k}}}
{E_{\bar l{\bf k}}-E_{ l{\bf k}}+\hbar\omega_1+i\hbar\gamma} \right)
\Bigg] 
\Bigg\},\label{3lines}
\ea 
corresponding to the three lines in Eq. (\ref{3lines}). Consider them one after another.

The first-line term in Eq. (\ref{3lines}) can be written as
\be 
\underbrace{\tilde \sigma_{\alpha\beta\gamma\delta}^{(3)}
(\omega_1, \omega_2,\omega_3)}_{(2/1),a} =
\frac 1{4O_1O_{12}}
\frac {ie^4g_s}{2E_F^2 S}
\sum_{l {\bf k}} 
\frac {\langle \bar l{\bf k}| \hat v_\alpha|l{\bf k}\rangle }
{E_{\bar l{\bf k}}-E_{l{\bf k}}+\hbar(\omega_1 +\omega_2+\omega_3 +i \gamma)}
\eta^\delta_{{\bf k}}
\frac{\p^2 (f_{l{\bf k}}- f_{ \bar l{\bf k}})}
{\p k_\beta\p k_\gamma}.
\ee
Substituting the inter-band matrix element of the velocity (\ref{velocity-off}) and taking the sum over $l$ we obtain
\ba 
\underbrace{\tilde \sigma_{\alpha\beta\gamma\delta}^{(3)}
(\omega_1, \omega_2,\omega_3)}_{(2/1),a} &=&
\frac 1{4O_1O_{12}}
\frac {ie^4g_s}{4\hbar E_F^2 S}
\sum_{ {\bf k}} 
\Bigg(
\frac {2E_{{\bf k}}  }
{2E_{{\bf k}}+\hbar(\omega_1 +\omega_2+\omega_3 +i \gamma)}
-\frac {2E_{{\bf k}}  }
{2E_{{\bf k}}-\hbar(\omega_1 +\omega_2+\omega_3 +i \gamma)}
\Bigg)\nonumber \\ &\times&
\eta^\alpha_{{\bf k}}\eta^\delta_{{\bf k}}
\frac{\p^2 (f_{1{\bf k}}- f_{ 2{\bf k}})}
{\p k_\beta\p k_\gamma}, \label{21a}
\ea 
where $E_{{\bf k}}=E_{2{\bf k}}$. The Fermi-functions difference in (\ref{21a}), $(f_{1{\bf k}}- f_{ 2{\bf k}})$, is proportional to $\Theta(k-k_F)$ at $T=0$, therefore the integrand in (\ref{21a}) contains the derivative of the delta-function $\delta'(k-k_F)$, only the vicinity of the two Dirac points contributes to the integral, and the integration over $dk$ can be easily done. Performing the angular integration with the help of formulas (\ref{average2}) and (\ref{average4}) we get  
\ba 
\underbrace{\tilde \sigma_{\alpha\beta\gamma\delta}^{(3)}
(\omega_1, \omega_2,\omega_3)}_{(2/1),a} &=&
\sigma_0^{(3)} \frac {i}{4O_1O_{12}}
\Bigg[
\delta_{\alpha\delta}\delta_{\beta\gamma}  
\Bigg(
\frac { 1  }{ 1- O_{123}}
-\frac { 1  }{ 1+ O_{123}}
\Bigg)
\nonumber \\ &&
+ O_{123}\Big(\delta_{\alpha\delta} \delta_{\beta \gamma} - \Delta_{\alpha \beta \gamma \delta}/4\Big)
\Bigg(
\frac { 1 }{ (1+ O_{123})^2}
+\frac { 1 }{ (1- O_{123})^2}
\Bigg)
\Bigg].
\ea 
This gives the first $(a)$ contribution to ${\cal S}^{(2/1)}_{\alpha\beta\gamma\delta}(\Omega_1,\Omega_2,\Omega_3)$. 

The second-line term in Eq. (\ref{3lines}) has the form 
\ba 
\underbrace{\tilde \sigma_{\alpha\beta\gamma\delta}^{(3)}
(\omega_1, \omega_2,\omega_3)}_{(2/1),b}&=&
\frac {ie^4g_s}{8O_1\hbar E_F S}
\sum_{l {\bf k}} 
\frac {(E_{\bar l{\bf k}}-E_{l{\bf k}})\eta_{\bf k}^\alpha }
{E_{\bar l{\bf k}}-E_{l{\bf k}}+\hbar(\omega_1 +\omega_2+\omega_3 +i \gamma)}
\nonumber \\ &\times&
\frac{\p}{\p k_\delta}
\Bigg(
\frac{\eta^\gamma_{{\bf k}}}
{E_{\bar l{\bf k}}-E_{ l{\bf k}} + \hbar (\omega_1 +\omega_2 +i\gamma) }
\frac{\p (f_{ l{\bf k}}- f_{\bar l{\bf k}})}{\p k_\beta}
\Bigg)
\ea 
Calculating it similarly we get
\ba 
\underbrace{\tilde \sigma_{\alpha\beta\gamma\delta}^{(3)}
(\omega_1, \omega_2,\omega_3)}_{(2/1),b}&=&
\sigma_0^{(3)}\frac {-i}{ 4O_1}
\Bigg\{\Bigg(
 \delta_{\alpha \beta}\delta_{\gamma\delta}
 -\Delta_{\alpha \beta \gamma \delta} /4
+\frac {\delta_{\alpha\gamma} \delta_{\beta \delta} -
\Delta_{\alpha \beta \gamma \delta} /4}{ (1+O_{123})}
\Bigg)
\frac{1}
{( 1 + O_{12})(1+O_{123})} \nonumber \\ &&
+\Bigg(
 \delta_{\alpha \beta}\delta_{\gamma\delta}
 -\Delta_{\alpha \beta \gamma \delta} /4
+\frac {\delta_{\alpha\gamma} \delta_{\beta \delta} -
\Delta_{\alpha \beta \gamma \delta} /4}{ (1-O_{123})}
\Bigg)
\frac{1}
{( 1 - O_{12})(1-O_{123})} 
\Bigg\}.
\ea
This gives the second $(b)$ contribution to ${\cal S}^{(2/1)}_{\alpha\beta\gamma\delta}(\Omega_1,\Omega_2,\Omega_3)$. 

The third-line term in Eq. (\ref{3lines}),
\ba 
\underbrace{\tilde \sigma_{\alpha\beta\gamma\delta}^{(3)}
(\omega_1, \omega_2,\omega_3)}_{(2/1),c} &=&
\frac {ie^4g_s}{4\hbar S}
\sum_{l {\bf k}} 
\frac {(E_{\bar l{\bf k}}-E_{l{\bf k}})\eta^\alpha_{\bf k} }
{E_{\bar l{\bf k}}-E_{l{\bf k}}+\hbar(\omega_1 +\omega_2+\omega_3) +i\hbar \gamma}\nonumber \\ &\times&
\frac{\p }{\p k_\delta}
\Bigg[
\frac{1}
{E_{\bar l{\bf k}}-E_{ l{\bf k}} + \hbar (\omega_1 +\omega_2) +i\hbar\gamma }\frac{\p }{\p k_\gamma}
\left(\eta^\beta_{{\bf k}}
\frac {f_{  l{\bf k}}-f_{\bar l{\bf k}}}
{E_{\bar l{\bf k}}-E_{ l{\bf k}}+\hbar\omega_1+i\hbar\gamma} \right)
\Bigg] ,
\ea 
can be reduced, after a long integration, to the form
\ba 
\underbrace{\tilde \sigma_{\alpha\beta\gamma\delta}^{(3)}
(\omega_1, \omega_2,\omega_3)}_{(2/1),c}&=&
\sigma_0^{(3)} \frac {i}4
\Bigg[
(\delta_{\alpha\beta} \delta_{\gamma \delta} -\Delta_{\alpha\beta\gamma \delta}/4) 
\Big({\cal J}_1(O_1,O_{12},O_{123})-{\cal J}_1(-O_1,-O_{12},-O_{123})\Big)
\nonumber \\ &-&
\left( \delta_{\alpha \gamma} \delta_{\beta\delta} - 
 \Delta_{\alpha \beta \gamma \delta}/4 \right)
\Big({\cal J}_2(O_1,O_{12},O_{123})-{\cal J}_2(-O_1,-O_{12},-O_{123})\Big)
\Bigg], \label{21c}
\ea
where the integrals ${\cal J}_1(a,b,c)$ and ${\cal J}_2(a,b,c)$ are defined in (\ref{IntI1}) -- (\ref{IntI2}). The formula (\ref{21c}) gives the last $(c)$ contribution to ${\cal S}^{(2/1)}_{\alpha\beta\gamma\delta}(\Omega_1,\Omega_2,\Omega_3)$. Combining all three contributions we get after some transformations Eq. (\ref{FuncS21}).

\subsection{The contribution $(1/2)$}

According to the definition (\ref{intrainter2}) the $(1/2)$-conductivity also has three contributions,
\ba 
\underbrace{\tilde \sigma_{\alpha\beta\gamma\delta}^{(3)}
(\omega_1, \omega_2,\omega_3)}_{(1/2)}&=&
\underbrace{\tilde \sigma_{\alpha\beta\gamma\delta}^{(3)}
(\omega_1, \omega_2,\omega_3)}_{(1/2),a}+\underbrace{\tilde \sigma_{\alpha\beta\gamma\delta}^{(3)}
(\omega_1, \omega_2,\omega_3)}_{(1/2),b}+\underbrace{\tilde \sigma_{\alpha\beta\gamma\delta}^{(3)}
(\omega_1, \omega_2,\omega_3)}_{(1/2),c}
\nonumber \\ &=&
\frac {1}{O_{123}}
\frac {ie^4g_s}{4E_FS}
\sum_{l {\bf k}} 
\langle l{\bf k}| \hat v_\alpha|l{\bf k}\rangle 
\Bigg[
\frac{1}{\hbar (\omega_1 +\omega_2) +i\hbar\gamma }
\frac{\p }{\p k_\delta}
\Bigg(
\eta^\beta_{{\bf k}}\eta^\gamma_{{\bf k}}
\frac {f_{ l{\bf k}}-f_{ \bar l{\bf k}}}
{E_{l{\bf k}}-E_{ \bar l{\bf k}}+\hbar\omega_1+i\hbar\gamma}  \Bigg)
\nonumber \\ &+&
\frac{\eta^\delta_{{\bf k}} }
{E_{ l{\bf k}}-E_{ \bar l,{\bf k}} + \hbar (\omega_1 +\omega_2) +i\hbar\gamma }
\frac{\p}{\p k_\gamma}
\Bigg(\eta^\beta_{{\bf k}}
\frac {f_{ l{\bf k}}-f_{ \bar l{\bf k}}}
{E_{l{\bf k}}-E_{ \bar l{\bf k}}+\hbar\omega_1+i\hbar\gamma} 
\Bigg)
\nonumber \\ &+&
\frac{\eta^\gamma_{{\bf k}} \eta^\delta_{{\bf k}}}{\hbar\omega_1+i\hbar\gamma}
\frac{1}
{E_{ l{\bf k}}-E_{ \bar l{\bf k}} + \hbar (\omega_1 +\omega_2) +i\hbar\gamma }
\frac{\p (f_{ l{\bf k}}- f_{ \bar l{\bf k}})}{\p k_\beta}
\Bigg]
,\label{3linesA}
\ea 
corresponding to the three lines in Eq. (\ref{3linesA}). The first one $(a)$ has the form 
\be 
\underbrace{\tilde \sigma_{\alpha\beta\gamma\delta}^{(3)}
(\omega_1, \omega_2,\omega_3)}_{(1/2),a}=
\frac {1}{O_{123}}
\frac {ie^4g_s}{4E_FS}
\sum_{l {\bf k}} 
\frac{\langle l{\bf k}| \hat v_\alpha|l{\bf k}\rangle }{\hbar (\omega_1 +\omega_2) +i\hbar\gamma }
\frac{\p }{\p k_\delta}
\Bigg(
\eta^\beta_{{\bf k}}\eta^\gamma_{{\bf k}}
\frac {f_{ l{\bf k}}-f_{ \bar l{\bf k}}}
{E_{l{\bf k}}-E_{ \bar l{\bf k}}+\hbar\omega_1+i\hbar\gamma}  \Bigg).
\ee
Calculating it like in the previous cases we get:
\be 
\underbrace{\tilde \sigma_{\alpha\beta\gamma\delta}^{(3)}
(\omega_1, \omega_2,\omega_3)}_{(1/2),a}=
\sigma_0^{(3)}
\frac {i\Delta_{\alpha \beta \gamma\delta} }
{8O_{123}O_{12}O_1}
\Bigg(\frac 1{2O_1}
\ln\frac{1+O_1}{1-O_1}- 1
\Bigg).
\ee

The second $(b)$ contribution is 
\be
\underbrace{\tilde \sigma_{\alpha\beta\gamma\delta}^{(3)}
(\omega_1, \omega_2,\omega_3)}_{(1/2),b}=
\frac {1}{O_{123}}
\frac {ie^4g_s}{4E_FS}
\sum_{l {\bf k}} 
\frac{\langle l{\bf k}| \hat v_\alpha|l{\bf k}\rangle 
\eta^\delta_{{\bf k}} }
{E_{ l{\bf k}}-E_{ \bar l,{\bf k}} + \hbar (\omega_1 +\omega_2) +i\hbar\gamma }
\frac{\p}{\p k_\gamma}
\Bigg(\eta^\beta_{{\bf k}}
\frac {f_{ l{\bf k}}-f_{ \bar l{\bf k}}}
{E_{l{\bf k}}-E_{ \bar l{\bf k}}+\hbar\omega_1+i\hbar\gamma} 
\Bigg).
\ee 
Calculating it we get 
\ba 
\underbrace{\tilde \sigma_{\alpha\beta\gamma\delta}^{(3)}
(\omega_1, \omega_2,\omega_3)}_{(1/2),b}&=&
\sigma_0^{(3)}
\frac {-i}{4O_{123}}
\Bigg\{
\Big(\delta_{\alpha\beta}\delta_{\gamma\delta} 
-\Delta_{\alpha \beta \gamma \delta}/4 \Big)
\Big(
{\cal J}_3(O_{1},O_{12})+{\cal J}_3(-O_{1},-O_{12})
\Big)
\nonumber \\ &-&
\Big(\delta_{\alpha  \gamma}\delta_{\beta\delta} 
-\Delta_{\alpha \beta \gamma \delta}/4 \Big)
\Big(
{\cal J}_4(O_{1},O_{12})+{\cal J}_4(-O_{1},-O_{12})
\Big)
\Bigg\}
\ea
where the integrals ${\cal J}_3(a,b)$ and ${\cal J}_4(a,b)$ are defined in Eqs. (\ref{IntI3}) -- (\ref{IntI4}). 

The last $(c)$ contribution corresponding to the third line in (\ref{3linesA}) can be written as
\be
\underbrace{\tilde \sigma_{\alpha\beta\gamma\delta}^{(3)}
(\omega_1, \omega_2,\omega_3)}_{(1/2),c}=
\frac {1}{O_1O_{123}}
\frac {ie^4g_s}{4E_F^2S}
\sum_{l {\bf k}} 
\frac{\langle l{\bf k}| \hat v_\alpha|l{\bf k}\rangle 
\eta^\gamma_{{\bf k}} \eta^\delta_{{\bf k}}}
{E_{ l{\bf k}}-E_{ \bar l{\bf k}} + \hbar (\omega_1 +\omega_2) +i\hbar\gamma }
\frac{\p (f_{ l{\bf k}}- f_{ \bar l{\bf k}})}{\p k_\beta}.
\ee 
After the integration we obtain 
\be
\underbrace{\tilde \sigma_{\alpha\beta\gamma\delta}^{(3)}
(\omega_1, \omega_2,\omega_3)}_{(1/2),c}=
\sigma_0^{(3)}
\frac {i}{4O_{123}O_1}
\Bigg(
\frac{1} { 1 + O_{12} }
- \frac{1} {  1 -  O_{12} }
\Bigg) 
\Big(\delta_{\alpha \beta}  
\delta_{\gamma\delta}-
 \Delta_{\alpha \beta \gamma \delta}/4 
\Big).
\ee 
Combining all three contributions we get after some transformations Eq. (\ref{FuncS12}). 

The last, $(0/3)$, contribution is calculated similarly. 

\section{Special cases for the integrals ${\cal J}_n$\label{specialJn}}

The formula (\ref{IntI1abc}) -- (\ref{IntI5ab}) provide results for the integrals ${\cal J}_1(a,b,c)$ -- ${\cal J}_5(a,b)$ at all values of the arguments $a$, $b$ and $c$ but in the ``degenerate'' cases $a=b$, $a=c$, etc. it is useful to have their simplified explicit expressions. Here they are. 

\subsection{Integral  ${\cal J}_1(a,b,c)$ }

In the special ``degenerate'' cases we have:
\ba 
{\cal J}_1(a,a,c)&=&
\frac{1}{a^2 c }
+\frac{c-2 a}{a^2 (a-c)^2 (a+1)}
+\frac{4 c-a}{c (a-c)^3 (c+1)}
-\frac{1}{2(a-c)^2 (a+1)^2}
-\frac{1}{(a-c)^2(c+1)^2}
\nonumber \\ &-& 
\frac{-3a^3+12 a^2 c-8 a c^2+2 c^3}{a^3 (a-c)^4 }\ln(a+1)
-\frac{2 a-5 c}{c (a-c)^4}\ln (c+1),
\ea
\ba 
{\cal J}_1(a,b,a)&=&
\frac{1}{a^2 b }
+\frac{1}{a (a-b)^2 (a+1)}
-\frac{1}{(a-b)^3 (b+1)}
+\frac{2 a-b}{2a (a-b)^2 (a+1)^2}
+\frac{2}{3(a-b)(a+1)^3}
\nonumber \\ &-&
\frac{-a^3+2 a^2 b-3 a b^2+b^3}{a^3 (a-b)^4 }\ln(a+1)
-\frac{a^2-3 a b+3 b^2}{b^2 (a-b)^4 }\ln(b+1),
\ea
\ba 
{\cal J}_1(a,b,b)&=&
\frac{1}{a b^2 }
+\frac{a-4 b}{b (a-b)^3(b+1)}
+\frac{4 b-a}{2b (a-b)^2 (b+1)^2}
-\frac{1}{(a-b) (b+1)^3}
\nonumber \\ &-&
\frac{-a^2-3 a b+b^2}{a^2 (a-b)^4 }\ln(a+1)
-\frac{a^3-3 a^2 b+2a b^2+3 b^3}{b^3 (a-b)^4 }\ln(b+1),
\ea
\ba 
{\cal J}_1(a,a,a)&=&
\frac{1}{a^3}
+\frac{1}{a^3 (a+1)}
-\frac{3}{4(a+1)^4}
-\frac{1}{3a (a+1)^3}
-\frac{2}{a^4 }\ln(a+1).
\ea

\subsection{Integral  ${\cal J}_2(a,b,c)$ }

For the second integral in the special cases we have:

\be 
{\cal J}_2(a,a,c)=
\frac{1}{2a (a-c) (a+1)^2}
+\frac{2 a-c}{a^2 (a-c)^2 (a+1)}
-\frac{3 a^2-3 a c+c^2}{a^3 (a-c)^3 }\ln(a+1)
+\frac{1}{c (a-c)^3 }\ln(c+1),
\ee
\be
{\cal J}_2(a,b,a)=
-\frac{1}{a (a-b)^2 (a+1)}
-\frac{1}{b (a-b)^2 (b+1)}
-\frac{b-3 a}{a^2 (a-b)^3 }\ln(a+1)
-\frac{3 b-a}{b^2 (a-b)^3 }\ln(b+1),
\ee

\be 
{\cal J}_2(a,b,b)=
\frac{2 b-a}{b^2(a-b)^2 (b+1)}
-\frac{1}{2b (a-b) (b+1)^2}
-\frac{1}{a (a-b)^3 }\ln(a+1)
-\frac{-a^2+3 a b-3 b^2}{b^3 (a-b)^3 }\ln(b+1),
\ee

\be 
{\cal J}_2(a,a,a)=
-\frac{1}{a^3 (a+1)}
-\frac{1}{2a^2(a+1)^2}
-\frac{1}{3a (a+1)^3}+\frac{1}{a^4 }\ln (a+1).
\ee

\subsection{Integrals  ${\cal J}_3(a,b)-{\cal J}_5(a,b)$ }

For these integrals the number of special cases is small and we get:
\be 
{\cal J}_3(a,a)=
-\frac{2}{a^3 }\ln(1+a)
+\frac{1}{a^2 }
+\frac{1}{a^2 (1+a)},
\ee
\be
{\cal J}_4(a,a)=
-\frac{1}{2a (1+a)^2}
-\frac{1}{a^2 (1+a)}
+\frac{1}{a^3 }\ln (1+a)
\ee
\be 
{\cal J}_5(a,a)=-{\cal J}_5(a,-a)=
\frac{4}{a^2 }
+\frac{2}{a^2 (1-a^2)}
+\frac{3}{a^3 }\ln\frac{(1-a)}{(1+a)}.
\ee


\end{document}